\documentclass[a4paper,12pt]{article}
\usepackage[margin=1in]{geometry}
 \usepackage{amsmath}
\numberwithin{equation}{subsection}
\usepackage{amsthm}
\usepackage{comment}
    \usepackage{array}
\usepackage[table]{xcolor}
\usepackage{upgreek}
    \usepackage{bbm}
\usepackage{chngcntr}
\usepackage[normalem]{ulem} 
\usepackage{natbib}
\usepackage{lipsum}
    \usepackage{placeins}
\usepackage{indentfirst}

\usepackage{appendix}      
\usepackage{tocloft}       
\usepackage{chngcntr} 
\usepackage{booktabs}
\usepackage{rotating}
    \usepackage{amssymb}
    \usepackage{float}
\usepackage{booktabs,algorithm,algorithmicx,algpseudocode,subcaption,xcolor}
\usepackage{enumitem}
    \usepackage{pdflscape}
    \usepackage{subcaption}
    \usepackage{caption}
    \usepackage{graphicx}
    \usepackage{tabularx}  
\usepackage{authblk} 
\counterwithin{figure}{section}
\counterwithin{table}{section}

\newcommand{\X}{\mathbf{X}_{ijt}}
\newcommand{\x}{\mathbf{X}}
\newcommand{\E}{\mathbb{E}}
\newcommand{\F}{\mathcal{F}}
\newcommand{\R}{\mathbb{R}}
\newcommand{\1}{\mathbbm{1}}
\newcommand{\e}{\mathcal{E}}

\newtheorem{prop}{Proposition}
\newtheorem{corollary}{Corollary}
\newtheorem{assumption}{Assumption}

\newtheorem{lemma}{Lemma}
\newtheorem{definition}{Definition}
\newtheorem{remark}{Remark}

\title{Sharp Hybrid Confidence Bands for Partially Identified Treatment Effects under Tail Uncertainty with an Application to Workforce Gender Diversity and Firm Performance}

\author[1]{Grace Lordan}
\author[1,2]{Kaveh S. Nobari}

\affil[1]{The Inclusion Initiative, London School of Economics and Political Science, UK}
\affil[2]{Data Science Institute, London School of Economics and Political Science, UK}

\date{\today}
\newcommand{\bm}{\mathbf}

\begin{document}

\maketitle

\begin{abstract}
Manski's nonparametric bounds partially identify the average treatment effects (ATEs) under minimal assumptions, yielding an interval-valued estimand with endpoints that depend on the outcome support - typically treated as known or fixed. In many empirical settings, however, credible bounds on the outcome support are often unavailable and outcomes may be heavy-tailed, so common empirical implementations that rely on ad-hoc truncation or observed extrema can compromise finite-sample coverage. We develop concATE, a hybrid confidence band for interval-identified ATEs that explicitly accounts for tail uncertainty without imposing parametric assumptions. The inference method combines a distribution-free concentration bound for the outcome distribution based on the Dvoretzky-Kiefer-Wolfowitz inequality with the asymptotic delta-method inference for smooth mean components, and allocates size across bound endpoints using Bonferroni’s inequality to guarantee joint coverage. We further extend concATE to a group-sequential procedure that controls the family-wise error rate using Pocock correction. Applying the method to panel data on 901 listed firms (2015Q2--2022Q1), we find that senior-level gender diversity has a statistically significant positive effect on firm value (Tobin’s $Q$) only after crossing substantial representation thresholds: in Growth \& Innovation sectors, significance emerges at approximately 55\% female leadership, while in Defensive sectors it appears only beyond about 60\%.
\end{abstract}



\section{Introduction\label{sec:introduction}}

Estimating the causal effects of a treatment in the presence of partially observed counterfactuals, where for each unit only one of the two potential outcomes is observed, i.e., the one corresponding to the treatment status actually realised - is a central problem within the causal inference literature. When only one of two potential outcomes is observed, the average treatment effect (ATE hereafter) is not point-identified without imposing additional assumptions. A large literature therefore studies partial identification, characterising causal effects estimands as sets rather than points \citep[see~Ch.10 of][for an overview of the partial identification literature]{honore2017advances}. In the canonical case of a scalar ATE, this yields an interval-identified parameter, whose lower and upper endpoints depend on the observed outcomes with minimal restrictions on the outcome support \citep{manski1990nonparametric,manski2003partial}. A critical but largely unexamined restriction in this literature is that the outcome support - the range of values the potential outcomes can take - is treated as known or fixed. In many empirical settings, however, credible global support bounds are unavailable, and outcomes may be heavy-tailed, so that common implementations relying on ad-hoc truncation or observed sample extrema can compromise finite-sample coverage \citep[see][for an adaptive truncation example]{d3fc5a28-c5bb-3e85-a388-e8132ffff151}. In this paper, we propose concATE, a hybrid confidence band for interval-identified ATEs that treats the latent outcome support as an inferential object rather than an identification assumption. The method combines finite-sample, distribution-free control of tail uncertainty via the Dvoretzky-Kiefer-Wolfowitz inequality with asymptotic delta-method inference for the smooth components of the Manski bound endpoints, and we show that the resulting tail-bracketing procedure achieves the minimax optimal rate among all distribution-free confidence procedures over the class of distributions with finite mean.

The problem of inference for set-identified causal effects has been studied extensively. \citet{imbens2004confidence} and \citet{stoye2009more} develop confidence intervals for partially identified scalar parameters by exploiting the asymptotic normality of the estimated bound endpoints. While \citet{imbens2004confidence} provide sharp inference for partially identified parameters, they assume the identification region is consistently estimable. More generally, a broad literature develops inference for identified sets and functions thereof, including approaches based on moment inequalities, set estimation, and resampling methods \citep[see][among others]{chernozhukov2007estimation,rosen2008confidence,romano2008inference,andrews2010inference,andrews2009validity,bugni2017inference}. Related work studies intersection and projection bounds, providing asymptotic inference for parameters defined as extrema over index sets \citep{chernozhukov2013intersection,kaido2019confidence}. Surveys of these developments emphasize the richness of available asymptotic tools for partially identified models that guarantee asymptotic validity, but not finite-sample guarantees \citep{honore2017advances,MOLINARI2020355}.

While these studies provide tools for large-sample inference on partially identified estimators, they typically assume a fixed class of data-generating processes in which the outcome support adheres to regularity conditions that ensure the asymptotic validity of estimated bound endpoints. The moment inequality framework, for instance, takes population moment conditions as given: for Manski bounds on the ATE with known support, the moment inequalities are well-defined, and the inference problem reduces to accounting for sampling uncertainty in the estimated moments. Support uncertainty is categorically different. It concerns whether the bounding constants entering Manski's identified region, which depend on the outcome support, are themselves correctly specified. When the support is unknown or unbounded, the standard moment inequalities may be misspecified, and no amount of asymptotic refinement addresses this. In practice, applied researchers confronting this problem often substitute observed sample extrema or ad-hoc quantile truncations for the unknown support bounds. As our Monte Carlo results in Section \ref{sec:montecarlo} demonstrate, while plug-in methods using sample extrema achieve valid coverage, they do so at the cost of substantially wider intervals, often 2--4 times the width of the concATE band, reflecting the systematic overshoot of sample extrema relative to the true population support.

The recent literature has also aimed at addressing the issue of exact finite-sample inference in partially identified models. \citet{li2022finite} consider the general class of incomplete structural models, though their paradigm requires the parametric specification of latent variable distribution and that the support restriction is known a priori. Furthermore, their approach relies on simulation-based test inversion with known latent distributions. \citet{song2024categorical} construct finite-sample confidence intervals for partially identified ATE under categorical IV models, using the Kullback-Leibler divergence concentration bounds, but this is restricted to categorical variables. \citet{cox2023simple} provide exact finite-sample size control, but only in DGPs arising from normal models. In the point-identified setting, \citet{armstrong2021finite} derive finite-sample minimax optimal confidence intervals for the ATE under unconfoundedness with known outcome support, though their framework requires point-identification and known bounds on the outcome.

This paper studies finite-sample and hybrid inference procedures for interval-identified ATEs when the outcome support and its corresponding data-generating process are unknown. We propose concATE, a hybrid concentration-driven confidence band that explicitly accounts for tail uncertainty without imposing parametric assumptions on the distribution of the unobserved counterfactuals. Our approach separates uncertainty arising from smooth functional components\footnote{In the Hadamard differentiability sense.} of the Manski bound endpoints, such as treated and untreated means, from uncertainty about unobserved outcome tails. For the latter, we exploit the Dvoretzky-Kiefer-Wolfowitz inequality \citep{dvoretzky1956asymptotic} to construct a finite-sample, distribution-free envelope for the latent outcome distribution. This envelope induces, with controlled probability, a confidence set for the Manski identified interval. For the smooth components, we employ standard delta-method asymptotic approximations, and we combine the two sources of uncertainty using Bonferroni's inequality to obtain a simultaneous confidence band that guarantees joint coverage of the lower and upper ATE bounds. While many existing confidence regions for partially identified parameters are only pointwise asymptotically valid and may fail in finite samples, we construct confidence bands with explicit finite-sample coverage guarantees for the tail components and asymptotic guarantees for the smooth components, yielding a hybrid band that remains valid under weak distributional assumptions.

We further extend concATE and its finite-sample counterpart to an optimal stopping setting, in which a group-sequential procedure is employed to identify the first threshold at which the ATE becomes non-zero while controlling the family-wise error rate. This extension is motivated by empirical contexts featuring threshold or “tipping point” effects and provides a principled approach to post-selection inference under partial identification.

To demonstrate the utility of our approach, we apply it to the question: Does greater senior-level gender workforce diversity causally improve firm performance? This question has taken on renewed importance as many firms have invested heavily in Diversity, Equity, and Inclusion (DEI) initiatives, yet establishing causality is difficult due to non-random adoption of diversity practices. A rich body of literature in management and economics has examined the links between top management team composition and organisational outcomes. The foundational “upper echelons” theory of \citet{hambrick1984upper} posits that a firm’s strategies and performance reflect the backgrounds of its senior executives. Consistent with this view, numerous studies document associations between executive attributes and firm outcomes such as innovation and financial performance. For example, prior research finds that gender-diverse boards tend to exhibit improved internal governance (e.g., better oversight and attendance) although the average impact on firm profitability or market value is mixed \citep{adams2009women}. A comprehensive meta-analysis by \citet{post2015women} reports that female board representation is positively related to accounting returns, especially in societies with greater gender parity, but the correlation with market-based performance metrics is weaker. More recent work has begun to address endogeneity in this relationship: \citet{safiullah2022gender}, analyzing Spain’s Gender Equality Act, use GMM techniques and find that while gender-diverse boards outperform on accounting measures, they can underperform on market valuation measures, suggesting investors may respond differently than internal metrics. Similarly, a study of Russian firms by \citet{kuzmina2021gender} finds that gender-diverse boards are associated with higher profitability and market value, with the benefits particularly pronounced during economic downturns. Beyond gender, other aspects of diversity have been linked to innovation outcomes: \citet{ostergaard2011does} show that employee gender and educational diversity positively predict firm innovation, and in a study of London firms, \citet{nathan2013cultural} find cultural diversity in management boosts product innovation and entrepreneurship. Field experiments also echo these benefits. \citet{hoogendoorn2013impact} conducted a randomized experiment with startup teams and found that gender-balanced teams outperformed male-dominated teams in terms of sales and profits.

An intriguing hypothesis in this literature is the existence of non-linear effects, or “critical mass” thresholds, in the diversity-performance relationship. Sociologist Rosabeth Kanter’s classic work on tokenism \citep{kanter1977some,kanter1987men} theorised that women in extreme minority (a “token” few) face marginalisation, whereas once a minority group reaches a substantial share of the team, dynamics shift and their influence grows disproportionately. Kanter’s typology categorises group gender composition into skewed (up to $\sim$15\% women), tilted ($\sim$20-35\% women), and balanced ($\sim$40-50\% women) categories, proposing that performance benefits might emerge when moving from skewed to tilted or balanced distributions. Subsequent studies have sought empirical evidence of such tipping points. For instance, \citet{torchia2011women} find that having at least three women directors (roughly a critical mass on many boards) is associated with a jump in innovation outputs, consistent with moving beyond token representation. \citet{ali2011gender} report an inverted U-shaped relationship between female representation and firm performance in certain contexts, suggesting that the strongest returns may occur at intermediate diversity levels before tapering off. These studies, while suggestive, largely report correlations or rely on linear/quadratic models that may not capture the true causal threshold. Our study contributes to this literature by using a robust, partially identified approach to formally test for causal tipping points. By refraining from imposing a specific functional form, we let the data reveal whether and where increasing diversity has a statistically reliable positive effect on firm value.

Our empirical analysis uses a panel of 901 publicly listed firms observed quarterly from 2015 Q2 to 2022 Q1. We focus on Tobin’s $Q$ (the ratio of market value to the replacement cost of assets) as the outcome of interest, which is a standard proxy for a firm’s growth opportunities and innovative performance. Originally introduced by \citet{tobin1969general} and later expounded in Tobin’s subsequent work \citep{tobin1978monetary}, the $Q$-ratio captures market expectations of future returns. A value above 1 indicates that the firm’s market valuation exceeds book value, signalling strong investment incentives \citep{brainard1968pitfalls,tobin1976asset}. For each quarter, we define the “treatment” as whether the firm’s top management team or board exceeds a given diversity threshold. In separate analyses, we consider thresholds for the percentage of women in senior leadership (e.g., 30\%, 40\%, 50\%, etc.), reflecting the critical mass levels discussed above. We then estimate the nonparametric bounds on the ATE of diversity at each threshold using our concATE procedure. This approach does not assume that firms with different diversity levels are comparable on unobservables; instead, it provides an interval estimate for the possible causal effect, given the observable data, without invoking full identification. In contrast to most prior studies that report point estimates after making identification assumptions, our results will highlight the range of plausible causal impacts of diversity on Tobin’s $Q$, emphasising what can be learned with minimal assumptions.

Our findings yield informative insights. In broad terms, the concATE analysis suggests that senior-level gender diversity has a significantly positive causal effect on Tobin's $Q$, but only after a certain threshold of representation is achieved. In innovation-driven sectors (such as technology and healthcare, where overall growth opportunities are high), we find that once female representation in leadership surpasses roughly 55\%, the lower bound of the ATE becomes positive, and the confidence band excludes zero. The estimated effect size grows as diversity increases, with particularly strong gains evident as teams approach and exceed gender balance (around 50\% female). This provides causal empirical support for the notion of a ``tipping point'' around high diversity levels in dynamic industries. One interpretation is that innovation-oriented firms, facing fast-moving and competitive markets, have strong incentives to harness the benefits of workforce diversity. Such firms may actively invest in inclusive cultures and leadership practices that allow diverse perspectives to be heard and integrated, thereby capturing value from diversity once a basic critical mass is present.

By contrast, in more traditional or cyclically oriented industries, the data suggest that a higher critical mass (on the order of 60\% female representation) is needed before we detect a reliably positive impact on firm value. Below that level, the confidence bands include zero, indicating we cannot rule out no effect in those sectors (but we can rule out negative effects). This stringent ``tipping point'' in traditional industries may reflect a lack of inclusion---when women remain a small minority, they may not experience the psychological safety needed to freely voice their insights or challenge prevailing viewpoints. Indeed, diversity alone (without an inclusive environment) can lead to friction or marginalization of minority members, whereas inclusion actively involves and values those members. This aligns with prior research suggesting that diversity must be complemented by inclusion to unlock its benefits: simply adding a few token individuals from underrepresented groups often fails to improve outcomes unless the organizational climate empowers those individuals to participate fully \citep{roberson2006disentangling,nishii2013benefits,josten2025makes}. Alternatively, it is possible that the intrinsic gains to diversity are lower in these contexts. Importantly, these conclusions are drawn with rigorous uncertainty quantification. The concATE bands allow us to assert, for example, that at 95\% confidence a firm in a growth industry with a gender-balanced leadership enjoys an ATE on Tobin's $Q$ that is positive (bounded away from zero), whereas at lower diversity levels the effect cannot be distinguished from zero. Such results illustrate how our methodological innovation can uncover nuanced causal relationships that might be obscured or misstated by conventional point estimation approaches.

The remainder of the paper is organised as follows. Section \ref{sec:framework} formalises the problem and presents the theoretical framework for nonparametric identification, extending Manski’s bounds to our context. Section \ref{sec:estimation} discusses the estimation procedure of nonparametric bounds in practice. Section \ref{sec:inference} details the construction of finite-sample concentration-inequality-driven inference and the hybrid concATE confidence band, and extends it to sequential threshold analysis, including theoretical guarantees. Section \ref{sec:optimality} derives the minimax lower and upper bounds for the convergence rate of the hybrid concATE approach and compares it with a few distribution-free methods, and Section \ref{sec:montecarlo} reports results from a Monte Carlo simulation that compares the finite-sample performance of concATE to traditional methods under a variety of data-generating processes. Section \ref{sec:results} then presents the empirical findings from our panel of firms, highlighting the estimated diversity tipping points and their interpretation. Finally, Section \ref{sec:conclusion} concludes with a discussion of implications for research and policy, and potential extensions of our framework.

\section{Framework\label{sec:framework}}
 In this section, inspired by the noted shortcomings in causal inference methodologies rigorously discussed by \citet{angrist2009mostly}, we extend the theoretical framework of \citet{manski1990nonparametric, manski2003partial} to derive nonparametric bounds on the “average” diversity treatment effect.

\subsection{Nonparametric Bounds\label{sec:nonparametricbounds}}
Let us denote the potential outcomes for firm $i$ in sector $j$ in quarter $t$ by $Y_{ijt}^{(0)}$ and $Y_{ijt}^{(1)}$, corresponding to the scenarios of no diversity efforts (no treatment) and with diversity efforts (treatment), respectively. Regardless of whether firm $i$ actually adopts diversity, $Y_{ijt}^{(0)}$ represents the hypothetical (counterfactual) outcome had the firm not exercised any diversity efforts, and $Y_{ijt}^{(1)}$ represents the outcome if the firm did adopt diversity. In essence, the question we seek to investigate is whether these potential outcomes differ (i.e., whether diversity efforts affect $Y_{ijt}$).

For simplicity of exposition, assume that $Y_{ijt}^{(k)}\in\mathbb{R}$ for  
$k\in\{0,1\}$ and define the treatment indicator
\begin{equation}\label{eq:indicator}
Z_{ijt}(\tau):=\1\left\{\mathcal{D}\ge\tau\right\},
\end{equation}
where $\1\{\cdot\}$ is the indicator function, $\mathcal{D}$ is a
diversity signal, and $\tau$ is a threshold chosen by the investigator.
Let $\X=(X_{ijt}^{1},\dots,X_{ijt}^{d})^{\top}\in\R^{d}$
denote a $(d\times1)$ vector of covariates.
Our goal is to learn the conditional treatment effect
$Y_{ijt}^{(1)}-Y_{ijt}^{(0)}\mid\X$.
Following the notation of \citet{angrist2009mostly}, the \emph{observed} outcome
can be written in terms of \emph{potential} outcomes as
\begin{align}
Y_{ijt} &=
\begin{cases}
Y_{ijt}^{(1)}, & \text{if } Z_{ijt}(\tau)=1,\\[4pt]
Y_{ijt}^{(0)}, & \text{if } Z_{ijt}(\tau)=0,
\end{cases}\\[6pt]
&= Y_{ijt}^{(0)} 
   +\left(Y_{ijt}^{(1)}-Y_{ijt}^{(0)}\right)Z_{ijt}(\tau).
\label{eq:observedoutcome}
\end{align}

Because only one potential outcome is ever observed for a given firm--quarter
$(i,j,t)$, a naïve comparison of conditional means by treatment status is
\begin{equation}
\delta(\x):=
\E\left[Y_{ijt}\mid \X,Z_{ijt}(\tau)=1\right]-
\E\left[Y_{ijt}\mid \X,Z_{ijt}(\tau)=0\right].
\label{eq:averagecomparison}
\end{equation}
Substituting \eqref{eq:observedoutcome} into \eqref{eq:averagecomparison} gives
\begin{align}
\begin{split}\label{eq:selectionbias}
\delta(\x):=&\underbrace{\E\left[Y_{ijt}^{(1)}-Y_{ijt}^{(0)}
       \mid \X,Z_{ijt}(\tau)=1\right]}_{\rho(\x)}
\\[2pt]
&\quad+\underbrace{\E\left[Y_{ijt}^{(0)}
       \mid \X,Z_{ijt}(\tau)=1\right]
       -\E\left[Y_{ijt}^{(0)}
       \mid \X,Z_{ijt}(\tau)=0\right]}_{\mathcal{B}(\x)},
\end{split}
\end{align}
where $\rho(\x)$ is the (conditional) treatment effect and
$\mathcal{B}(\x)$ is the selection bias. \noindent
As a corollary, the unconditional mean‐comparison parameter
$\delta$ of \citet{angrist2009mostly} is obtained by integrating
$\delta(\x)$ over the (marginal) distribution of $\x\mid Z(\tau)$, i.e.,
\begin{align}
\begin{split}
\delta
= \int_{\mathbb{R}^d}
& \E\big[Y \mid \x, Z(\tau)=1\big] f_{\x\mid Z=1}(\mathbf{x})
 \, \mathrm{d}\mathbf{x} \nonumber \\
&\quad -
\int_{\mathbb{R}^d} 
  \E\big[Y \mid \x, Z(\tau)=0\big] f_{\x\mid Z=0}(\mathbf{x})  \mathrm{d}\mathbf{x} ,
\end{split}
\end{align}
hence
\begin{equation}\label{eq:condtouncond}
  \delta = \E\left[Y_{ijt}\mid Z_{ijt}(\tau)=1\right]
           - \E\left[Y_{ijt}\mid Z_{ijt}(\tau)=0\right].
\end{equation}

The latter may be non-zero because firms that adopt diversity efforts might do so
precisely when they face innovation shortfalls, either to signal responsiveness
to investors or to diversify their workforce in search of new ideas; in such
cases $\mathcal{B}(\x)<0$. Conversely, if a firm scales up diversity after large innovation gains, aiming to sustain that momentum, then $\mathcal{B}(\x)>0$.

\citet{manski1990nonparametric, manski2003partial} formalise the problem
differently.  For firms characterised by attributes $\mathbf{X}$, define the
difference in expected outcomes as
\begin{align}
\label{eq:treatmenteffect_manski}
\Re(\x)
&= \E\left[Y_{ijt}^{(1)} \mid \X\right]
   -\E\left[Y_{ijt}^{(0)} \mid \X\right] \\[4pt]
&= \E\left[Y_{ijt}^{(1)}-Y_{ijt}^{(0)}
        \mid \X\right]. \notag
\end{align}
Using the law of total expectation, each conditional mean in
\eqref{eq:treatmenteffect_manski} can be decomposed; for example,
\begin{align}
\label{eq:lte_decomp}
\E\left[Y_{ijt}^{(1)} \mid \X\right]
 &= \E\left[Y_{ijt}^{(1)}
        \mid \X,Z_{ijt}(\tau)=1\right]
    \Pr\left(Z_{ijt}=1 \mid \X\right) \notag\\
 &\quad + \E\left[Y_{ijt}^{(1)}
        \mid \X,Z_{ijt}(\tau)=0\right]
    \Pr\left(Z_{ijt}=0 \mid \X\right),
\end{align}
and an analogous expression holds for $k=0$.

Equation \eqref{eq:lte_decomp} makes clear that two latent expectations,
\begin{equation}
\label{eq:latentexpectations}
\E\left[Y_{ijt}^{(1)} \mid \X, Z_{ijt}(\tau)=0\right]
\quad \text{and} \quad
\E\left[Y_{ijt}^{(0)} \mid \X, Z_{ijt}(\tau)=1\right],
\end{equation}
are never observed in the data. Put differently, we do not observe the
innovation outcome a firm \emph{would have} achieved without diversity efforts when it
actually implemented them ($Z_{ijt}(\tau)=1$), nor the outcome \emph{with} diversity
efforts when it did not implement them ($Z_{ijt}(\tau)=0$).

In both scenarios, one can conduct a \textit{randomized experiment}, as noted by \citet{angrist2009mostly}, which coincides with the \textit{mean-independence} assumption in \citet{manski2003partial}. In that case:
\begin{equation}\label{eq:mar}
\E\left[Y_{ijt}^{(k)} \mid \X, Z_{ijt}(\tau)=1\right]
= \E\left[Y_{ijt}^{(k)} \mid \X, Z_{ijt}(\tau)=0\right],
\quad \text{for } k=0,1.
\end{equation}
Then, $\delta(\x) = \rho(\x)$, and the expression in \eqref{eq:treatmenteffect_manski} simplifies to:
\begin{equation}
\Re(\x) = \E[Y_{ijt}^{(1)} \mid \X, Z_{ijt}(\tau)=1]
- \E[Y_{ijt}^{(0)} \mid \X, Z_{ijt}(\tau)=0].
\end{equation}
Hence, under random assignment, $\delta(\x)$ suffers no selection bias, and $\Re(\x)$ is point-identified.

The mean-independence assumption, however, is rather strict. For expositional purposes, consider bounding arguments that replace unknown support limits with quantile-indexed truncations of tail mass:
\begin{equation}
-\infty < L^{(k)} < Q_Y^{(k)}(p) \leq Y_{ijt}^{(k)} \leq Q_Y^{(k)}(p^c) < U^{(k)} < +\infty,
\end{equation}
where $ Q_Y(p) = \inf\{ y : F_Y(y) \geq p \} $, with $ F_Y(\cdot) $ the CDF of $Y$ and $p^c$ is the complement of $p$, i.e., $p^c=1-p$. Suppose further that we are interested in the \emph{unconditional} average treatment effect rather than the effect for each unit. Using the law of iterated expectations:
\begin{align}
\begin{split}\label{eq:condtouncond}
\E_{\x}\left[\Re(\x)\right]
&= \E_{\x}\left[\E\left[Y_{ijt}^{(1)} \mid \X\right]\right]
 - \E_{\x}\left[\E\left[Y_{ijt}^{(0)} \mid \X\right]\right] \\
&= \E\left[Y_{ijt}^{(1)} - Y_{ijt}^{(0)}\right] = \Re.
\end{split}
\end{align}

Now, since some conditional expectations remain latent (as shown in \eqref{eq:lte_decomp}), one may bound them using either known outcome supports $ [L^{(k)},\; U^{(k)}] $ or their quantile-based versions $ [Q_Y(p), Q_Y(1-p)] $. \citet{manski1990nonparametric, manski2003partial} propose the following nonparametric bounds for the treatment effect:
\begin{align}
\begin{split}\label{eq:nonparametricbounds}
\Re \in \bigg[
&\E[Y_{ijt}^{(1)} \mid Z_{ijt}(\tau)=1] \Pr(Z_{ijt}(\tau)=1)
+ L^{(1)} \Pr(Z_{ijt}(\tau)=0) \\
&\quad - U^{(0)} \Pr(Z_{ijt}(\tau)=1)
- \E[Y_{ijt}^{(0)} \mid Z_{ijt}(\tau)=0] \Pr(Z_{ijt}(\tau)=0), \\
&\E[Y_{ijt}^{(1)} \mid Z_{ijt}(\tau)=1] \Pr(Z_{ijt}(\tau)=1)
+ U^{(1)} \Pr(Z_{ijt}(\tau)=0) \\
&\quad - L^{(0)} \Pr(Z_{ijt}(\tau)=1)
- \E[Y_{ijt}^{(0)} \mid Z_{ijt}(\tau)=0] \Pr(Z_{ijt}(\tau)=0)
\bigg]
\end{split}
\end{align}
These expressions illustrate how the identified set depends on tail behaviour.
Replacing $ L^{(k)} $ and $ U^{(k)} $ with quantile-indexed limits, say 
$ Q_Y(p) $ and $ Q_Y(1-p)$, correspond to
a heuristic truncation of tail mass and does not define a new estimand. In essence, using a similar notation to \citet{manski2003partial}, the region $\mathcal{H}[\Re]$ is the \textit{identification region} for $\Re$, where $\mathcal{H}[\Re]$ is defined as the bound in \eqref{eq:nonparametricbounds}. Note that $\mathcal{H}[\Re]$ is only partially identified when $0<\Pr[Z_{ijt}(\tau)=k]<1$ for $k=0,1$, as otherwise, $\mathcal{H}[\Re]$ is simply a singleton. In other words, if, say, $\Pr[Z_{ijt}(\tau)=1]=1$, then both upper and lower bounds coincide with the treated mean, so $\mathcal{H}[\Re]$ collapses.

Throughout, the target parameter remains the original Manski-identified set
$\mathcal{H}[\Re]$; quantile-based bounds are introduced only as a device to
illustrate sensitivity to tail behaviour and will be formally calibrated using
finite-sample concentration inequalities in subsequent sections.

In the following sections, we outline estimation procedures for both the naïve unconditional difference and the nonparametric bounds. We also construct $100(1 - \alpha)\%$ confidence intervals for the bounds using Bonferroni-adjusted intervals as proposed by \citet{horowitz1998censoring}, and derive standard errors via the delta method [see \citet{casella2024statistical}].
  
\subsection{Interpretation of the Bounding Constants\label{sec:interpret}}
For a fixed $\tau$, the bounding constants
\[
L^{(1)} \le \E\left[Y_{ijt}^{(1)} \mid Z_{ijt}(\tau)=0\right] \le U^{(1)},
\quad
L^{(0)} \le \E\left[Y_{ijt}^{(0)} \mid Z_{ijt}(\tau)=1\right] \le U^{(0)},
\]
state that the latent (never-observed) mean outcome a “treated’’ firm
would have realised had it not been treated cannot be lower than
$L^{(1)}$ nor higher than $U^{(1)}$; analogously for an “untreated’’ firm
under treatment.
Without bounding these counterfactual means, the ATE, $\Re$, is
not point-identifiable: any value between $-\infty$ and $\infty$
could be rationalised by suitable (and untestable) choices of
$\E[Y_{ijt}^{(1)}\mid Z_{ijt}(\tau)=0]$ and
$\E[Y_{ijt}^{(0)}\mid Z_{ijt}(\tau)=1]$.
Because our outcome of interest (Tobin’s~$Q$) is in theory unbounded from above, we use quantile-based limits here, e.g.\ $L^{(k)} = 0$ and
$U^{(k)} = Q_Y^{(k)}(0.90)$, purely for illustration; in Sections \ref{subsec:dkw}--\ref{sec:inference} these fixed cut-offs are replaced by data-driven, DKW-calibrated endpoints that deliver valid outer confidence sets for the original Manski identified region: they confine the worst-case counterfactual means to
the central 90\% of the empirical outcome distribution,
ruling out only the most extreme tail behaviour while introducing
minimal additional assumptions. The 10\% tails trimmed was deemed a reasonable balance between realism and conservatism in our firm performance data, which can be heavy-tailed.  In Section \ref{sec:results}, when applying Manski bounds to study the causal impact of workforce gender diversity on Tobin’s~$Q$, we further experiment with tighter bounds (such as the 10\textsuperscript{th}--90\textsuperscript{th} percentiles) to assess how the results are affected.

Under these mild restrictions, the interval in
\eqref{eq:nonparametricbounds}
remains robust to selection on unobservables, yet is now finite, so
if the entire interval lies above (below)~0
we may still conclude a positive (negative) causal effect even when
ignorability fails.
We therefore describe $\mathcal{H}[\Re]$ as a set of
“worst-case bounds’’ for the ATE under no unverifiable assumptions beyond the outcome range.

\subsection{Testing in the Presence of a Random Tipping Point\label{sec:tippingpoint}}

As it is evident from Eq.~\eqref{eq:indicator}--\eqref{eq:observedoutcome},
the sampling composition of the treated and untreated firms depends on the threshold
$\tau$. In theory one could fix $\tau$ and analyse the resulting samples accordingly, though in the context of this work, our
goal is different: we seek the \emph{tipping point} at which the average
treatment effect $\Re$ becomes strictly positive (or negative). Thus,
$\tau=\tau(\omega)$ must be regarded as a \emph{random} stopping time with values in $\mathcal{M}$, where $\omega$ is the realization.

 Formally, let $(\Omega,\mathcal{F},\mathbb{P})$ be the underlying probability
space, with $\{\mathcal{F}_u\}_{u\in\mathcal{M}}$ denoting the filtration
generated by the sequential statistics up to look $u$; for instance,
$\mathcal{F}_u:=\sigma(\Re_v:v\in\mathcal{M},v\le u)$, where $\sigma(.)$ is the $\sigma$-field spanned by the random variables $\{\Re_v:v\leq u\}$. The random
threshold $\tau=\tau(\omega)$, for each $u \in \mathcal{M}$ with
\[
\{\omega: \tau(\omega)\leq u\} \in \mathcal{F}_u,\quad \forall u \in \mathcal{M}.
\]
 is selected by the procedure will be a stopping time with respect
to $\{\mathcal{F}_u\}_{u\in\mathcal{M}}$.

Let $\mathcal{D}_{ijt}$ denote the diversity signal for firm $i$ in sector
$j$ at time $t$ (e.g.\ the percentage of female or non-white executives).
A firm is labelled ``treated'' when $\mathcal{D}_{ijt}\ge\tau$. Rather
than prespecify $\tau$, we examine a grid of meaningful cut-offs,
\[
\tau_m := m, \qquad m\in\mathcal{M},
\]
where in our context
$\mathcal{M}=\{5,10,15,\dots,90,95\}$, and
$\overline{\mathcal{M}}:=\lvert\mathcal{M}\rvert$.
For each $m$ we define
\[
Z_{ijt}(\tau_m):=\mathbbm{1}\{\mathcal{D}_{ijt}\ge \tau_m\}.
\]

For a chosen significance level $\alpha\in(0,1)$ we test
\begin{equation}
\label{eq:null}
H_0:\;0\in\mathcal{H}[\Re_u]\quad \forall u\in\mathcal{M}
\qquad\text{vs.}\qquad
H_1:\;\exists u\in\mathcal{M}\text{ s.t.\ }0\notin\mathcal{H}[\Re_u].
\end{equation}
Following \citet{siegmund2013sequential}, define the stopping rule
\[
\tilde{\tau}
:=
\inf\bigl\{\tau_u:\ u\in\mathcal{M},\ 
\mathcal{H}_*[\Re_u]>0\ \text{or}\ \mathcal{H}^*[\Re_u]<0\bigr\},
\]
with the convention $\inf\varnothing=+\infty$.
We then reject $H_0$ if $\tilde{\tau}\le \tau_{m_1}$. Denote the rejection
event at look $u$ by
\[
S^{\tau_u}
:=
\{\mathcal{H}_*[\Re_u]>0\}\cup\{\mathcal{H}^*[\Re_u]<0\},
\]
where $\mathcal{H}_*[\Re_u]:=\inf \mathcal{H}[\Re_u]$ and
$\mathcal{H}^*[\Re_u]:=\sup \mathcal{H}[\Re_u]$. By construction,
$S^{\tau_u}\in\mathcal{F}_u$, and $\tilde{\tau}$ is a stopping time with
respect to $\{\mathcal{F}_u\}_{u\in\mathcal{M}}$. For a \emph{fixed}
threshold $\tau_u$, the test can be sized at level $\alpha_u$, i.e.
\[
\Pr\left(S^{\tau_u}\mid H_0\right)\le \alpha_u.
\]
For the sequential procedure, the family-wise Type~I error requirement is
\[
\Pr\left(
\exists u\in\{m_0,\dots,m_1\}:\ 
(\tilde{\tau}=\tau_u)\cap S^{\tau_u}
\ \Big\vert\ H_0
\right)\le\alpha.
\]
Since the rejection event is a union over looks,
\[
\{\exists u:\tilde{\tau}=\tau_u\text{ and }S^{\tau_u}\}
=
\bigcup_{u=m_0}^{m_1}
\left\{(\tilde{\tau}=\tau_u)\cap S^{\tau_u}\right\},
\]
Boole's inequality yields
\begin{equation}
\label{eq:BooleFixed}
\Pr\left(
\bigcup_{u=m_0}^{m_1}
\left\{(\tilde{\tau}=\tau_u)\cap S^{\tau_u}\right\}
\ \Big\vert\ H_0
\right)
\le
\sum_{u=m_0}^{m_1}
\Pr\left((\tilde{\tau}=\tau_u)\cap S^{\tau_u}\mid H_0\right)
\le \sum_{u=m_0}^{m_1}\alpha_u.
\end{equation}
where \eqref{eq:BooleFixed} holds, because, $\left\{(\tilde{\tau}=\tau_u)\cap S^{\tau_u}\right\}\subseteq \left\{S^{\tau_u}\right\}$, and
\[
\Pr\left((\tilde{\tau}=\tau_u)\cap S^{\tau_u}\mid H_0\right)\leq \Pr\left( S^{\tau_u}\mid H_0\right)\leq \alpha_u.
\]
Thus, a sufficient condition for family-wise control at level $\alpha$ is
\[
\sum_{u=m_0}^{m_1}\alpha_u=\alpha.
\]
Since all $\overline{\mathcal M}=19$ looks are pre-scheduled, we adopt the
equal-spending rule of \citet{pocock1977group},
\[
\alpha_u=\frac{\alpha}{\overline{\mathcal{M}}},\qquad u=m_0,\dots,m_1.
\]
The corresponding two-sided critical value is
$c_u=\Phi^{-1}(1-\alpha_u/2)\approx 3.007$ for $\alpha=0.05$ and
$\overline{\mathcal M}=19$.\footnote{Alternative allocations include
O'Brien-Fleming \citep{o1979multiple} or the Lan-DeMets spending function
\citep{gordon1983discrete}.}
\subsection{Empirical-Population CDF Divergence and Calibrated Tail Endpoints}
\label{subsec:dkw}

As before, for a \emph{fixed} $\tau$, let $k\in\{0,1\}$ index the untreated and treated groups respectively. For the indices $i,j,t$, denote the population and empirical distribution functions by
\[
F^{(k)}(y)=\Pr\left(Y_{ijt}^{(k)}\le y\right),
\quad
\hat F^{(k)}_{N_k}(y)=\frac{1}{N_k}\sum_{i,j,t}^{N_k}\1\{Y_{ijt}^{(k)} \le y\},
\]
respectively, and define the uniform deviation $\Delta_k:=\sup_{y\in\mathbb R}\lvert \hat F^{(k)}_{N_k}(y)-F^{(k)}(y)\rvert$.
As in Sections \ref{sec:nonparametricbounds}--\ref{sec:tippingpoint}, the problem consists of finding suitable bounding thresholds for the latent conditional expectations $\E[Y_{ijt}^{(0)}\mid Z_{ijt}(\tau)=1]$ and $\E[Y_{ijt}^{(1)}\mid Z_{ijt}(\tau)=0]$. The problem of controlling the family-wise size $\alpha$ for a given level $\alpha\in (0,1)$ using the Bonferroni approach is discussed extensively in Section \ref{sec:inference}. However, in this Section, we discuss how to control for an arbitrary size $\alpha'$ for the uniform divergence of the true CDF $F^{(k)}(y)$ and its empirical counterpart $\hat{F}_{N_k}^{(k)}(y)$.

For group $k$, the Dvoretzky-Kiefer-Wolfowitz inequality (Massart-sharp) \citep{dvoretzky1956asymptotic} implies that for every $c>0$,
\begin{equation}
\Pr\left(\Delta_k>c\right)\le2\exp(-2N_k c^2).
\end{equation}
Selecting $c=\varepsilon_{k}$ with
\begin{equation}
\label{eq:eps-alpha-prime}
\varepsilon_{k}
:=
\sqrt{\frac{\log\left(2/\alpha'\right)}{2N_k}}
\end{equation}
yields the DKW event $\mathcal E_{k}:=\{\Delta_k\le \varepsilon_{k}\}$ with
$\Pr(\mathcal E_{k})\ge 1-\alpha'$. Thus,
\begin{equation}
\Pr\left(\Delta_k\le \varepsilon_{k}\right) \ge 1-\alpha',
\end{equation}
so the construction is jointly valid at the targeted family-wise level.

Uniform control of the CDFs translates into control of tail quantiles. On the event
$\{\Delta_k\le\varepsilon_{k}\}$, for every $p\in[\varepsilon_{k},1-\varepsilon_{k}]$ and noting Lemma \ref{lem:quantilesando} in the Appendix, one has
\begin{equation}
\label{eq:quantile-sandwich}
F^{-1}(p-\varepsilon_{k}) \le \hat F^{-1}_{N_k}(p) \le F^{-1}(p+\varepsilon_{k}).
\end{equation}
where the superscript $(k)$ is dropped from the inverse CDFs hereafter for the ease of exposition. 
Choosing $p=2\varepsilon_{k}$ and $p=1-2\varepsilon_{k}$ (which lie strictly inside the admissible
interval) and writing $Y^{(k)}_{(1)}\le\cdots\le Y^{(k)}_{(N_k)}$ for the order statistics,
\[
F^{-1}(\varepsilon_{k}) \le \hat F^{-1}_{N_k}(2\varepsilon_{k})
=Y^{(k)}_{(\lceil 2\varepsilon_{k} N_k\rceil)},
\quad
\hat F^{-1}_{N_k}(1-2\varepsilon_{k})
=Y^{(k)}_{(\lceil (1-2\varepsilon_{k}) N_k\rceil)}
 \le F^{-1}(1-\varepsilon_{k}).
\]
We therefore define the \emph{DKW-calibrated tail endpoints}
\begin{equation}
\label{eq:LkUk}
L_{k}:=Y^{(k)}_{(\lceil 2\varepsilon_{k} N_k\rceil)},
\quad
U_{k}:=Y^{(k)}_{(\lceil (1-2\varepsilon_{k}) N_k\rceil)}.
\end{equation}
These are ``outward" in the sense that up to $\varepsilon_{k}$ probability mass is allowed to lie
beyond each observed tail; the widening occurs on the probability axis via
\eqref{eq:quantile-sandwich}, and the resulting endpoints are empirical quantiles (order statistics). When population bounds $L^{(k)},U^{(k)}$ are unknown in Sections \ref{sec:nonparametricbounds} and \ref{sec:interpret}, we plug in $L^{(k)}\leftarrow L_{k}$ and $U^{(k)}\leftarrow U_{k}$ for the relevant group $k$.

\begin{figure}[!t]
  \centering
  \includegraphics[width=0.82\textwidth]{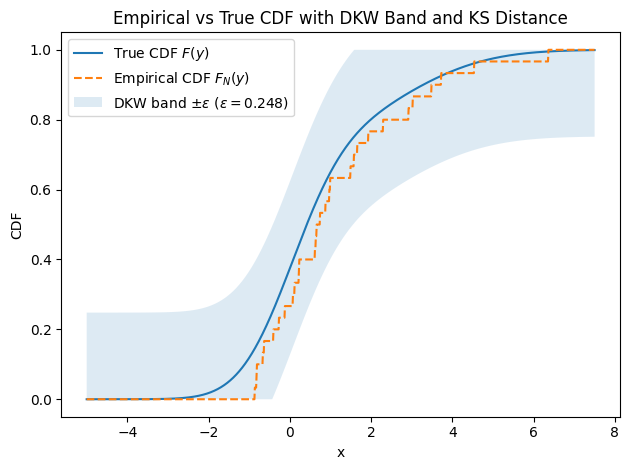}
  \caption{Empirical and population CDFs with a DKW band for $N_k=200$. The shaded region is $F(y)\pm\varepsilon$ with $\varepsilon=\sqrt{\ln(2/\alpha')/(2N_k)}=0.096$ for $\alpha'=0.05$.}
  \label{fig:ecdf-vs-true-n50}
\end{figure}

Figure~\ref{fig:ecdf-vs-true-n50} illustrates the construction for a sample of size $N_k=200$. Taking
$\alpha'=0.05$ gives
\[
\varepsilon_{k}
=\sqrt{\frac{\ln(2/0.05)}{2\cdot 200}}
=0.096,
\]
and the shaded band $F^{(k)}\pm \varepsilon_{k}$ envelopes the empirical curve uniformly. In this case
the indices in \eqref{eq:LkUk} are $\lceil 2\varepsilon_{k} N_k\rceil=39$ and
$\lceil (1-2\varepsilon_{k})N_k\rceil=162$, which are the data-driven tail endpoints replacing unknown
support extremes in the bounds and tests that follow.

\begin{remark}
If $\varepsilon_{k}\le \frac{1}{4}$, the convenient choice
\[
L_{k}=Y^{(k)}_{\left(\lceil 2\varepsilon_{k} N_k\rceil\right)},
\quad
U_{k}=Y^{(k)}_{\left(\lceil (1-2\varepsilon_{k}) N_k\rceil\right)}
\]
ensures $L_{k}\le U_{k}$. When $\varepsilon_{k}> \frac{1}{4}$, pick any tail mass level
$r\in\left(0,\frac{1}{2}-\varepsilon_{k}\right]$ and set
\[
L_{k}=F^{-1}_{N_k}(r+\varepsilon_{k})
=Y^{(k)}_{\left(\lceil (r+\varepsilon_{k})N_k\rceil\right)},
\quad
U_{k}=F^{-1}_{N_k}(1-r-\varepsilon_{k})
=Y^{(k)}_{\left(\lceil (1-r-\varepsilon_{k})N_k\rceil\right)}.
\]
Then $L_{k}\le U_{k}$ by construction, and on the DKW event
$\{\sup_y\lvert\hat F^{(k)}_{N_k}(y)-F^{(k)}(y)\rvert\le \varepsilon_{k}\}$ we have
\[
F^{-1}(r) \le L_{k} \le U_{k} \le F^{-1}(1-r).
\]
For a concrete choice that avoids ties in small samples, one may take
$r=\frac{1}{2}-\varepsilon_{k}-\frac{1}{N_k}$ (or any $o(1)$ slack).
When indices fall outside $\{1,\dots,N_k\}$ due to rounding, truncate them to the boundary.
\end{remark}

\section{Estimation and Identification\label{sec:estimation}}


In Section \ref{sec:framework} we defined the unconditional mean‐comparison parameter
$\delta$ (Eq.~\eqref{eq:condtouncond}) and Manski’s bounds
$\Re$ (Eq.~\eqref{eq:nonparametricbounds}).  We now give their
sample estimates and show how to tighten the bounds via quantiles.  

For a fixed $\tau$, the estimator of $\delta$ can be written as a single weighted sum:
\[
  \hat\delta
  = \sum_{j=1}^{K}\sum_{i=1}^{n^j}\sum_{t=1}^{T}
      Y_{ijt}\,w_{ijt},
\qquad
  w_{ijt}
  = \frac{Z_{ijt}(\tau)}{N_1}
  -\frac{1 - Z_{ijt}(\tau)}{N_0},
\]
where
\[
  N_k =\sum_{i,j,t}\1\{Z_{ijt}(\tau)=k\},
  \quad k\in\{0,1\},
  \quad N=N_0+N_1.
\]
It then follows by the Central Limit Theorem (CLT hereafter) that 
\begin{equation}\label{eq:clt}
\sqrt{N}(\hat{\delta}-\delta)\xrightarrow{\mathcal{D}} N(0,\sigma^2).
\end{equation}

To estimate $\Re$, recall from \eqref{eq:nonparametricbounds} that
$\Re$ involves the four quantities
$\E[Y_{ijt}^{(k)}\mid Z_{ijt}(\tau)=k]$ and $\Pr(Z_{ijt}(\tau)=k)$, $k=0,1$, plus the endpoints
$\{L^{(k)},U^{(k)}\}$.  We estimate them by
\[
  \hat\delta_k
  = \frac{1}{N_k}\sum_{i,j,t}Y_{ijt}\1\{Z_{ijt}(\tau)=k\},
\qquad
  \hat p_k = \frac{N_k}{N},
\]
and in the case of classical \citet{manski1990nonparametric} bounds
\[
  L^{(k)} = \min\{Y_{ijt}:Z_{ijt}(\tau)=k\},
\quad
  U^{(k)} = \max\{Y_{ijt}:Z_{ijt}(\tau)=k\},
\]
noting that $\hat\delta_1-\hat\delta_0=\hat\delta$ and
$\hat p_1=1-\hat p_0$. 

To tighten the raw‐support bounds, as means of illustrating the sensitivity of the bounds, one may replace $L^{(k)},U^{(k)}$ by the sample
$p$- and $(1-p)$-quantiles in each group:
\[
  \hat F^{(k)}(y)
  = \frac{1}{N_k}
    \sum_{i,j,t}\1\{Y_{ijt}\le y,\;Z_{ijt}(\tau)=k\},
\qquad
  \hat Q_Y^{(k)}(p)
  = \inf\{y:\hat F^{(k)}(y)\ge p\},
\]
or equivalently
$\hat Q_Y^{(k)}(p)=Y^{(k)}_{(\lceil pN_k\rceil)}$ when
$Y_{(1)}^{(k)}\le\cdots\le Y_{(N_k)}^{(k)}$ are group‑$k$ order stats.
Finally, in \eqref{eq:nonparametricbounds} substitute
\[
  L^{(k)}\mapsto \hat Q_Y^{(k)}(p),
\qquad
  U^{(k)}\mapsto \hat Q_Y^{(k)}(1-p),
\]
to obtain the quantile‐based bounds.	

\bigskip
In Section \ref{sec:inference} below we describe how to construct 
$(1-\alpha)\%$ confidence bands for $\hat\delta$ and for the nonparametric bounds via the Bonferroni‐adjusted delta‐method.
\section{Inference\label{sec:inference}}

For $u = m_0,\dots,m_1$, constructing $(1-\alpha_u)$ confidence intervals for the
naïve estimator $\hat\delta$ is rather straightforward, because by definition, $\hat\delta$ is simply a
difference of sample means. Recall from \eqref{eq:clt} that
\[
\sqrt{N}(\hat\delta - \delta)\xrightarrow{\mathcal{D}} N(0,\sigma^2).
\]
where
\begin{equation}
\sigma^2(\hat\delta)
  = \frac{\text{Var}(Y_{ijt}\mid Z_{ijt}(\tau_u)=1)}{p_1}
  + \frac{\text{Var}(Y_{ijt}\mid Z_{ijt}(\tau_u)=0)}{p_0}
\end{equation}
is the asymptotic variance parameter and $p_k = \Pr(Z_{ijt}(\tau_u)=k)$. Hereafter, for notational simplicity and without loss of generality, we write $N_k$ instead of $N_k(\tau_u)$ for each look $u$, keeping the dependence on $\tau_u$ implicit.

A consistent estimator of $\sigma^2(\hat\delta)$ is the usual
difference-in-means estimator:
\begin{equation}
\hat\sigma^2(\hat\delta)
  = \frac{\hat\sigma_1^2}{\hat p_1}
  + \frac{\hat\sigma_0^2}{\hat p_0},
\end{equation}
where
\[
\hat\sigma_k^2 
  = \frac{1}{N_k-1}\sum_{Z_{ijt}(\tau)=k}(Y_{ijt}-\bar Y_k)^2,
  \qquad
  \bar Y_k = \frac{1}{N_k}\sum_{Z_{ijt}(\tau)=k} Y_{ijt},
\]
and $\hat p_k = N_k/N$.

Since $\text{Var}(\hat\delta)\approx \sigma^2(\hat\delta)/N$, a
$(1-\alpha_u)$ Wald-type confidence interval for $\delta$ is therefore
\begin{equation}
\hat\delta \pm 
\Phi^{-1}(1-\alpha_u/2)\,
\sqrt{\frac{\hat\sigma^2(\hat\delta)}{N}},
\qquad u = m_0,\dots,m_1,
\end{equation}
where $\Phi^{-1}(\cdot)$ denotes the standard normal quantile function.

Obtaining confidence intervals for the nonparametric bounds is less
straightforward, since the upper and lower bound estimators are nonlinear
functions of the data. We therefore rely on a Bonferroni-adjusted
delta method, as formalized in the following proposition.

\begin{prop}\label{prop:1}

Suppose the latent conditional expectations in Eq.~\eqref{eq:latentexpectations} are within a ``known" bounded interval $[L^{(k)},\;U^{(k)}]$ for $k\in\{0,1\}$. Let us denote $\mathcal{L}(\hat{\theta})$ and $\mathcal{U}(\hat{\theta})$ as the lower and upper bound estimates of the nonparametric bounds respectively, where $\hat{\theta}=(\hat{\delta}_1,\;\hat{\delta}_{0},\;\hat{p}_1,\; \hat{p}_0)^\top$ is a $4\times 1$ vector of estimators. The $100(1-\alpha_u)\%$ confidence interval for the union of the bounds is obtained by:
\begin{equation}
\mathcal{L}(\hat{\theta})\pm \Phi^{-1}(1-\alpha_u/2)\times \text{S.E.}\left(\mathcal{L}(\hat{\theta})\right)\quad\text{and}\quad \mathcal{U}(\hat{\theta})\pm  \Phi^{-1}(1-\alpha_u/2)\times \text{S.E.}\left(\mathcal{U}(\hat{\theta})\right)
\end{equation}
where $\text{Var} \left(\mathcal{L}(\hat{\theta})\right)\approx\nabla \mathcal{L}(\theta)^\top \frac{\Omega_{\hat{\theta}}}{N}\nabla \mathcal{L}(\theta)$ with
\begin{align}
\nabla \mathcal{L}(\theta)=\left(p_1,\;-p_0,\;\delta_1-U^{(0)},\;L^{(1)}-\delta_0\right)^\top\\
\nabla \mathcal{U}(\theta)=\left(p_1,\;-p_0,\;\delta_1-L^{(0)},\;U^{(1)}-\delta_0\right)^\top
\end{align} 
and the covariance matrix of the vector of estimators $\hat{\theta}$ is given explicitly by:
\begin{equation}
\Omega_{\hat{\bm{\theta}}}=\begin{pmatrix}
\text{Var}(\hat{\delta}_{1}) & 0 & 0 & 0 \\
0 & \text{Var}(\hat{\delta}_{0}) & 0 & 0 \\
0 & 0 & \text{Var}(\hat{p}_1) & -\text{Var}(\hat{p}_1) \\
0 & 0 & -\text{Var}(\hat{p}_1) & \text{Var}(\hat{p}_0)
\end{pmatrix},
\end{equation}
\end{prop}

\subsection{Concentration-Driven Confidence Bands for Average Treatment Effects\label{sec:concate}}

A major shortcoming of the nonparametric bounds proposed by 
\citet{manski1990nonparametric,manski2003partial} and introduced in
Section~\ref{sec:framework} is the strong assumption that the latent
conditional expectations in Eq.~\eqref{eq:latentexpectations} lie inside a
known bounded interval
\[
[L^{(k)},\;U^{(k)}], \qquad k\in\{0,1\}.
\]
In practice, these expectations may be unbounded. To address this, we
reformulate the problem probabilistically and study the probability
\[
\Pr\left(\forall u \in \mathcal{M},\;\Re_u \in \mathcal{H}_{\alpha_u}[\Re_u]\right).
\]

We first analyse a stylized setting in which the observations $Y_{ijt}$ are
independent across all indices $i,j,t$.  Although independence may be
reasonable for purely cross-sectional snapshots (for instance, a single
quarter across many sectors) it is unrealistic in panel settings, where serial
correlation is typically present. Accordingly, we later extend our results to
allow for temporal dependence within firm, while maintaining cross-sectional
independence across firms.

For each threshold value $\tau_u$, we define
\[
\delta_k := \mathbb{E}\left[Y^{(k)}_{ijt}\mid Z_{ijt}(\tau_u)=k\right],
\qquad
p_k := \Pr\left(Z_{ijt}(\tau_u)=k\right),
\qquad k\in\{0,1\},
\]
with the dependence on $u$ left implicit to lighten notation. Following
\citet{manski1990nonparametric,manski2003partial}, the identified set for the
tipping-point functional $\Re_u$ is then
\begin{equation}
\Re_u \in 
\left[
\delta_{1}p_1 + L^{(1)}p_0 - U^{(0)}p_1 - \delta_{0}p_0,\;\;
\delta_{1}p_1 + U^{(1)}p_0 - L^{(0)}p_1 - \delta_{0}p_0
\right].
\label{eq:manski}
\end{equation}
The latent cross-terms 
$\delta_{10} := \mathbb{E}[Y^{(1)}_{ijt}\mid Z_{ijt}(\tau_u)=0]$ and
$\delta_{01} := \mathbb{E}[Y^{(0)}_{ijt}\mid Z_{ijt}(\tau_u)=1]$
are replaced by their support bounds $(L^{(k)},\;U^{(k)})$ to obtain the interval
in \eqref{eq:manski}.  In the first setting, we assume that the data exhibit
neither cross-sectional nor serial dependence.

\begin{assumption}[Independent sampling]\label{ass:iid}
The collection ${(Y_{ijt},Z_{ijt})}_{i,j,t}$ consists of i.i.d. draws from a sub-exponential distribution.
\end{assumption}

\begin{prop}[Finite-sample coverage under i.i.d.\ sampling]
\label{prop:iid}
Let $0<\alpha_u<1$ for $u=m_0,\cdots,m_1$ and denote by $N_k=\sum_{i}\mathbbm{1}\{Z_i(\tau_u)=k\}$ the sample size in treatment group $k\in\{0,1\}$ and by  
\[
Y^{(k)}_{(1)}\le\cdots\le Y^{(k)}_{(N_k)}
\]
the order statistics of the observed outcomes in that group. Set
\[
\varepsilon_k:=\sqrt{\frac{\log\left(12/\alpha_u\right)}{2N_k}},
\]
choose any $r_k\in\left(0,\frac12-\varepsilon_k\right]$, and define
\[
L_{\alpha_u}^{(k)}:=Y^{(k)}_{\left(\lceil (r_k+\varepsilon_k)N_k\rceil\right)},
\quad
U_{\alpha_u}^{(k)}:=Y^{(k)}_{\left(\lceil (1-r_k-\varepsilon_k)N_k\rceil\right)},
\quad k=0,1 .
\]
(In particular, if $\varepsilon_k\le \frac14$, one may take $r_k=\varepsilon_k$, which yields
$L_{\alpha_u}^{(k)}=Y^{(k)}_{(\lceil 2\varepsilon_k N_k\rceil)}$
\emph{and}
$U_{\alpha_u}^{(k)}=Y^{(k)}_{(\lceil (1-2\varepsilon_k) N_k\rceil)}$.)%
\par\smallskip
Define the two thresholds
\[
t_{p,k}:=\sqrt{\frac{\log(12/\alpha_u)}{2N}},\qquad t_{\mu,k}:=\min\left\{M_k\sqrt{\frac{\log(12/\alpha_u)}{cN_k}},\frac{M_k}{cN_k}\log\left(\frac{12}{\alpha_u}\right)\right\}.
\]
Let  
$
\hat\mu_k=\frac{1}{N_k}\sum_{Z_i(\tau_u)=k}Y_i
$
and  
$
\hat p_k=\frac{N_k}{N_0+N_1}
$
be the sample means and treatment shares.
Define the random interval
\begin{align}
\begin{split}\label{eq:Halpha}
\mathcal H_{\alpha_u}[\Re_u]
=\left[
      \hat{\mu}_{1}^{-}\hat{p}_{1}^{-}
      +L_{\alpha_u}^{(1)}\hat{p}_{0}^{-}
      -U_{\alpha_u}^{(0)}\hat p_{1}^{+}
      -\hat\mu_{0}^{+}\hat p_{0}^+,
     \; \hat\mu_{1}^+\hat p_{1}^+
      +U_{\alpha_u}^{(1)}\hat p_{0}^+
      -L_{\alpha_u}^{(0)}\hat p_{1}^-
      -\hat\mu_{0}^-\hat p_{0}^-
  \right],
\end{split}
\end{align}
where $\hat{\mu}_{k}^{\pm}=\hat{\mu}_k\pm t_{\mu,k}$ and $\hat{p}_{k}^{\pm}=\hat{p}_k\pm t_{p,k}$.
Then, under assumption \ref{ass:iid}
\begin{equation}\label{eq:coverage}
\Pr\left(\forall u\in \mathcal{M},\;\Re_u\in \mathcal H_{\alpha_u}[\Re_u]\right)\ge1-\alpha,
\end{equation}
\end{prop}

Proposition~\ref{prop:iid} states that the data-driven set $\mathcal H_{\alpha_u}[\Re_u]$ in \eqref{eq:Halpha} is a $100(1-\alpha_u)\%$-level confidence region for the average treatment effect $\Re_u$ under nothing more than i.i.d. sampling. Because $\E[Y_{ijt}^{(1)}\mid Z_{ijt}(\tau_u)=0]$ and $\E[Y_{ijt}^{(0)}\mid Z_{ijt}(\tau_u)=1]$ are latent, point identification is impossible without additional assumptions; the proposition nevertheless guarantees that the random interval constructed from the empirical means, treatment proportions, and DKW-calibrated empirical quantiles will cover the true $\Re_u$ in at least $100(1-\alpha_u)\%$ of repeated samples. Practically, one computes $\mathcal H_{\alpha_u}[\Re_u]$ by (i) splitting the sample into treated and untreated subsamples, (ii) forming the subsample-specific means $\hat\mu_k$ and proportions $\hat{p}_k$, (iii) computing $\varepsilon_k=\sqrt{\log(12/\alpha_u)/(2N_k)}$ and choosing any $r_k\in(0,\frac12-\varepsilon_k]$, then setting the tail endpoints
\[
L_{\alpha_u}^{(k)}=Y^{(k)}_{\left(\lceil (r_k+\varepsilon_k)N_k\rceil\right)},
\quad
U_{\alpha_u}^{(k)}=Y^{(k)}_{\left(\lceil (1-r_k-\varepsilon_k)N_k\rceil\right)},
\]
(in particular, if $\varepsilon_k\le \frac14$ one may take $r_k=\varepsilon_k$, yielding the convenient indices $\lceil 2\varepsilon_k N_k\rceil$ and $\lceil(1-2\varepsilon_k)N_k\rceil$), and (iv) plugging these objects into~\eqref{eq:Halpha}. The resulting band can be used exactly like an ordinary confidence interval: the null hypothesis $H_0: 0\in \Re_u$ is rejected at level $\alpha_u$ whenever $0\notin \mathcal H_{\alpha_u}[\Re_u]$. 

If substantive knowledge implies that the latent outcomes are truncated on one or both tails, the extreme-value inputs in Manski’s bounds can be replaced by the true population limits. Let $\lambda$ (lower) and $\Lambda$ (upper) denote any such known bounds. When both limits are known one sets $L^{(k)}=\lambda$ and $U^{(k)}=\Lambda$ in the plug-in formulas; the resulting $100(1-\alpha)\%$ simultaneous band coincides with Proposition \ref{prop:1} and requires no DKW calibration. When only one tail is known, say $Y\ge \lambda$, but the upper support is unknown, we fix the lower extreme at $\lambda$ while retaining the DKW-calibrated endpoint on the upper side. The next Corollary shows that this hybrid construction preserves the nominal family-wise coverage probability even when the first significant threshold is data-selected.

\begin{corollary}[Finite-sample coverage under i.i.d.\ sampling and truncated distribution]
\label{corol:iidtrunc}
Let $0<\alpha_u<1$ for $u=m_0,\cdots,m_1$ and denote by $N_k=\sum_{i}\mathbbm{1}\{Z_i(\tau_u)=k\}$ the sample size in treatment group $k\in\{0,1\}$ and by  
\[
 \lambda\leq Y^{(k)}_{(1)}\le\cdots\le Y^{(k)}_{(N_k)}
\]
the order statistics of the observed outcomes in that group. Set
\[
\varepsilon_k:=\sqrt{\frac{\log\left(6/\alpha_u\right)}{2N_k}},
\quad
L^{(k)}:=\lambda,
\]
choose any $r_k\in(0,\,1-\varepsilon_k]$, and define the upper endpoint via the one-sided DKW quantile relation as
\[
U_{\alpha_u}^{(k)}:= \hat F^{-1}_{N_k}(1-r_k-\varepsilon_k)
= Y^{(k)}_{\left(\lceil (1-r_k-\varepsilon_k)N_k\rceil\right)},
\qquad k=0,1 .
\]
Define the two thresholds
\[
t_{p,k}:=\sqrt{\frac{\log(12/\alpha_u)}{2N}},\qquad t_{\mu,k}:=\min\left\{M_k\sqrt{\frac{\log(12/\alpha_u)}{cN_k}},\frac{M_k}{cN_k}\log\left(\frac{12}{\alpha_u}\right)\right\}.
\]
Let  
$
\hat\mu_k=\frac{1}{N_k}\sum_{Z_i=k}Y_i
$
and  
$
\hat p_k=\frac{N_k}{N_0+N_1}
$
be the sample means and treatment shares.
Define the random interval
\begin{align}
\begin{split}
\mathcal H_{\alpha_u}[\Re_u]
=\left[
      \hat{\mu}_{1}^{-}\hat{p}_{1}^{-}
      +L^{(1)}\hat{p}_{0}^{-}
      -U_{\alpha_u}^{(0)}\hat p_{1}^{+}
      -\hat\mu_{0}^{+}\hat p_{0}^+,
     \; \hat\mu_{1}^+\hat p_{1}^+
      +U_{\alpha_u}^{(1)}\hat p_{0}^+
      -L^{(0)}\hat p_{1}^-
      -\hat\mu_{0}^-\hat p_{0}^-
  \right],
\end{split}
\end{align}
where $\hat{\mu}_{k}^{\pm}=\hat{\mu}_k\pm t_{\mu,k}$ and $\hat{p}_{k}^{\pm}=\hat{p}_k\pm t_{p,k}$.
Then, under assumption \ref{ass:iid}
\begin{equation}\label{eq:coverage}
\Pr\left(\forall u\in \mathcal{M},\;\Re_u\in \mathcal H_{\alpha_u}[\Re_u]\right)\ge1-\alpha,
\end{equation}
\end{corollary}

From here on, we weaken the i.i.d.\ assumption and allow the data to exhibit weak dependence by assuming each series is a stationary $\upalpha$-mixing process. For example, any stationary AR($1$) model satisfies this condition.

\begin{assumption}[$\upalpha$-mixing sampling]\label{ass:mixing}
The collection $(Y_{ijt},Z_{ijt})_{i,j,t}$ is a strictly stationary $\upalpha$-mixing process in the sense of Definition~\ref{def:mixing}, with mixing coefficients
\[
\upalpha(k)
=\sup_{m\ge1}\,
\upalpha\left(\mathcal B_{1}^{m},\,\mathcal B_{m+k}^{nT}\right),
\]
satisfying
$\upalpha(k)\to0$ as $k\to\infty$ and 
$C_\upalpha=\sum_{k=1}^\infty\upalpha(k)^{1/2}<\infty$.
Moreover, each outcome $Y_{ijt}$ has a uniformly bounded sub-exponential norm,
$\sup_{i,j,t}\|Y_{ijt}\|_{\psi_1}<\infty$.
\end{assumption}

\begin{prop}[Finite-sample coverage under $\upalpha$-mixing sampling]
\label{prop:mixing}
Let $0<\alpha_u<1$ for $u=m_0,\cdots,m_1$ and let $(Y_{ijt},Z_{jt})_{i,j,t}$ be a strictly stationary sequence with 
$Z_i(\tau_u)\in\{0,1\}$, $Y_i\in\R$, and strong-mixing coefficients $\upalpha(r)$ satisfying
\[
C_\upalpha =\sum_{r=1}^\infty \upalpha(r)^{1/2} <\infty.
\]
Write
\[
N_k =\sum_{i=1}^n\mathbbm1\{Z_i(\tau_u)=k\}, 
\quad
\hat p_k =\frac{N_k}{N_0+N_1},
\quad
\hat\mu_k =\frac1{N_k}\sum_{Z_i(\tau_u)=k}Y_i,
\quad
k=0,1.
\]
Define the two thresholds
\[
t_{p,k}
:=(1+4C_\upalpha)\sqrt{\frac{2\log(12/\alpha_u)}{N_k}},
\qquad
t_{\mu,k}
:=\max\left\{t_k^{(1)},t_k^{(2)},t_k^{(3)}\right\},
\]
where the $t_k^{(j)}$ are the unique solutions making each term of
Lemma \ref{lem:bernsteinweak} bounded by $\alpha_u/18$.  Finally, set
\[
\varepsilon_k:= t_{p,k},
\]
choose any $r_k\in\left(0,\frac12-\varepsilon_k\right]$, and define the DKW-calibrated endpoints
\[
L^{(k)}_{\alpha_u}
:= Y^{(k)}_{\left(\lceil (r_k+\varepsilon_k) N_k\rceil\right)},
\quad
U^{(k)}_{\alpha_u}
:= Y^{(k)}_{\left(\lceil (1-r_k-\varepsilon_k) N_k\rceil\right)},
\quad
k=0,1.
\]
(In particular, if $\varepsilon_k\le \frac14$, one may take $r_k=\varepsilon_k$, yielding
$L^{(k)}_{\alpha_u}=Y^{(k)}_{(\lceil 2\varepsilon_k N_k\rceil)}$
and
$U^{(k)}_{\alpha_u}=Y^{(k)}_{(\lceil (1-2\varepsilon_k) N_k\rceil)}$.) 

Then the random interval
\[
\mathcal H_{\alpha_u}[\Re_u]
=
\left[
\hat\mu_1^-\hat p_1^-
+L^{(1)}_{\alpha_u}\hat p_0^-
-U^{(0)}_{\alpha_u}\hat p_1^+
-\hat\mu_0^+\hat p_0^+,\;
\hat\mu_1^+\hat p_1^+
+U^{(1)}_{\alpha_u}\hat p_0^+
-L^{(0)}_{\alpha_u}\hat p_1^-
-\hat\mu_0^-\hat p_0^-
\right],
\]
where $\hat{\mu}_{k}^{\pm}=\hat{\mu}_k\pm t_{\mu,k}$ and $\hat{p}_{k}^{\pm}=\hat{p}_k\pm t_{p,k}$.
Then, under assumption \ref{ass:mixing}
\[
\Pr\left(\forall u \in \mathcal{M}, \;\Re_u\in \mathcal H_{\alpha_u}[\Re_u]\right)\ge1-\alpha.
\]
\end{prop}
Similar to Corollary~\ref{corol:iidtrunc}, the extension of Proposition~\ref{prop:mixing} to the case of one-tail truncated latent conditional expectations simply requires modifying the \emph{endpoints}. Specifically, for the case of lower-tail truncation, replace
\[
L^{(k)} := \lambda,
\quad
U^{(k)}_{\alpha_u} := \hat F^{-1}_{N_k}\left(1 - r_k - \tilde{\varepsilon}_k\right)
= Y^{(k)}_{\left(\lceil (1 - r_k - \tilde{\varepsilon}_k)\,N_k\rceil\right)},
\]
with the fixed lower bound and a one-sided DKW-calibrated upper empirical quantile, where
\[
\tilde{\varepsilon}_k:=(1+4C_{\upalpha})\sqrt{\frac{2\log\!\left(6/\alpha_u\right)}{N_k}},
\qquad
r_k\in\left(0,1-\tilde{\varepsilon}_k\right],
\]
for $k = 0,1$, and the observed order statistics satisfy $\lambda \leq Y_{(1)}^{(k)} \leq \cdots \leq Y_{\left(\lceil (1-r_k-\tilde{\varepsilon}_k)(N_k)\rceil\right)}^{(k)}$.
(When $\tilde{\varepsilon}_k\le \tfrac14$, a convenient choice is $r_k=\tilde{\varepsilon}_k$, yielding the index $\lceil (1-2\tilde{\varepsilon}_k)N_k\rceil$.)

A drawback of the finite‐sample bands in Propositions~\ref{prop:iid} and~\ref{prop:mixing} is that the Bernstein and Hoeffding‐type paddings for sub‐exponential tails depend on multiple nuisance constants (mixing rates, sub‐exponential parameters, etc.), which quickly becomes cumbersome in practice. Moreover, although the sub‐exponential assumption is fairly general, it is still a substantive restriction on the data. In Proposition~\ref{prop:hybrid} we therefore introduce a hybrid confidence band that combines

\begin{itemize}
\item The Dvoretzky-Kiefer-Wolfowitz concentration bound (which requires no tail assumptions beyond finiteness) for the order-statistic endpoints, and

\item The usual asymptotic delta‐method (CLT) for the sample means and proportions.
The DKW inequality controls the uniform deviation $\sup_x \left\lvert F_n(x) - F(x)\right\rvert$ in finite samples without any distributional assumptions on $Y$ (see, e.g., Chapter 3 of \citet{van1996weak}). This hybrid approach preserves the simplicity of the DKW envelope for the nonparametric piece while relying on asymptotic normality only for the low‐dimensional parameters.
\end{itemize}

Although asymptotic normality can be used for smooth mean components, asymptotic quantile inference would require additional tail regularity conditions and delivers only pointwise validity. The DKW inequality instead provides uniform, finite-sample control of tail uncertainty without imposing distributional assumptions, which is essential for constructing joint confidence bands for interval-identified parameters.

\begin{remark}
The sample mean is a smooth (Hadamard-differentiable) functional of the
underlying distribution $P$ and therefore satisfies a $\sqrt{n}$-Central
Limit Theorem under standard moment conditions. In contrast, quantile
estimation involves the functional $T(F)=F^{-1}(p)$, whose asymptotic
behaviour follows by applying the functional delta method to the empirical
distribution $F_n$.

For each fixed $p \in (0,1)$, if $F$ is continuously differentiable at
$q(p)=F^{-1}(p)$ with density $f(q(p))>0$, then the map
$F \mapsto F^{-1}(p)$ is Hadamard differentiable at $F$, with derivative
\[
\dot T_F(h) = -\frac{h(q(p))}{f(q(p))}.
\]
Combining this with the empirical process central limit theorem and the
functional delta method (see \citealp[Ch.~20]{van2000asymptotic}) yields
\[
\sqrt{n}\left(\hat q(p)-q(p)\right)
\;\xrightarrow{D}\;
N\!\left(0,\frac{p(1-p)}{f(q(p))^2}\right).
\]

This approximation is inherently local in $p$, as the derivative depends
on $1/f(q(p))$, which may become large in tail regions where the density
is small. Uniform validity over ranges of $p$ therefore requires additional
regularity conditions, such as densities bounded away from zero or
extreme-value assumptions \citep{van1996weak,de2006extreme}.

To avoid imposing such tail restrictions, we instead exploit the
Dvoretzky-Kiefer-Wolfowitz inequality, which provides finite-sample,
distribution-free control of $\sup_y \lvert F_n(y)-F(y) \rvert$. This yields
probabilistic bracketing of quantiles on the probability scale, allowing
us to control tail uncertainty uniformly without relying on asymptotic
approximations.
\end{remark}

\begin{prop}[$100(1-\alpha)\%$ hybrid confidence band under $\upalpha$-mixing]\label{prop:hybrid}
Let $(Y_{ijt},Z_{ijt})_{i,j,t}$ be strictly stationary with $\upalpha$-mixing
coefficients $\upalpha(r)$ such that
\[
C_\upalpha=\sum_{r=1}^{\infty}\upalpha(r)^{1/2}<\infty .
\]
For
\[
Y^{(k)}_{(1)}\le\cdots\le Y^{(k)}_{(N_k)},
\]
define
\[
\varepsilon_k:=(1+4C_\upalpha)\,
             \sqrt{\frac{2\log(8/\alpha_u)}{N_k}},
\]
choose any $r_k\in\left(0,\tfrac12-\varepsilon_k\right]$, and set the DKW-calibrated empirical endpoints
\[
L_{\alpha_u}^{(k)}:=Y^{(k)}_{\left(\lceil (r_k+\varepsilon_k)N_k\rceil\right)},\quad
U_{\alpha_u}^{(k)}:=Y^{(k)}_{\left(\lceil (1-r_k-\varepsilon_k)N_k\rceil\right)},\qquad k=0,1,
\]
(in particular, if $\varepsilon_k\le \tfrac14$ one may take $r_k=\varepsilon_k$, which yields
$L_{\alpha_u}^{(k)}:=Y^{(k)}_{(\lceil 2\varepsilon_k N_k\rceil)}$
and
$U_{\alpha_u}^{(k)}:=Y^{(k)}_{(\lceil (1-2\varepsilon_k) N_k\rceil)}$).
Let $\mathcal{L}_{\alpha_u}(\hat{\theta})$ and $\mathcal{U}_{\alpha_u}(\hat{\theta})$ denote the lower and upper bound estimators, where $\hat{\theta}=(\hat{\delta}_1,\hat{\delta}_{0},\hat{p}_1, \hat{p}_0)^\top$ and
\[
\sqrt{N}(\hat{\theta}-\theta)\xrightarrow{d}N(0,\Omega_{\theta}),
\]
for some positive semidefinite $4\times4$ matrix $\Omega_{\theta}$. Let $\hat\Omega_{\theta}$ be any consistent estimator of $\Omega_{\theta}$ (e.g.\ a HAC or cluster-robust estimator).

Then a $100(1-\alpha)\%$ confidence band for the union of the bounds is
\[
\mathcal{L}_{\alpha_u}(\hat{\theta})\pm \Phi^{-1}(1-\alpha_u/4)\, \text{S.E.}\!\left(\mathcal{L}_{\alpha_u}(\hat{\theta})\right)\quad\text{and}\quad
\mathcal{U}_{\alpha_u}(\hat{\theta})\pm \Phi^{-1}(1-\alpha_u/4)\, \text{S.E.}\!\left(\mathcal{U}_{\alpha_u}(\hat{\theta})\right),
\]
where
\begin{align*}
\nabla \mathcal{L}(\theta)&=\left(p_1,\;-p_0,\;\delta_1-U_{\alpha_u}^{(0)},\;L_{\alpha_u}^{(1)}-\delta_0\right)^\top,\\
\nabla \mathcal{U}(\theta)&=\left(p_1,\;-p_0,\;\delta_1-L_{\alpha_u}^{(0)},\;U_{\alpha_u}^{(1)}-\delta_0\right)^\top,
\end{align*}
and
\[
\text{S.E.}\left(\mathcal{L}_{\alpha_u}(\hat{\theta})\right)
:=\sqrt{\frac{1}{N}\nabla \mathcal{L}(\hat{\theta})^\top\hat\Omega_{\theta}\nabla \mathcal{L}(\hat{\theta})},\quad
\text{S.E.}\!\left(\mathcal{U}_{\alpha_u}(\hat{\theta})\right)
:=\sqrt{\frac{1}{N}\nabla \mathcal{U}(\hat{\theta})^\top\hat\Omega_{\theta}\nabla \mathcal{U}(\hat{\theta})}.
\]
\end{prop}

Proposition~\ref{prop:hybrid} thus guarantees that our hybrid concATE band achieves the desired confidence level for the ATE bounds even when data are dependent, by using the DKW inequality with appropriate mixing corrections.

Following the same logic as Corollary~\ref{corol:iidtrunc}, the extension of Proposition~\ref{prop:hybrid}
to the case of one-tail truncated latent conditional expectations replaces the endpoints by
\[
L^{(k)} := \lambda,
\qquad
U^{(k)}_{\alpha_u} := \hat F^{(k),-1}_{N_k}\left(1 - r_k - \tilde{\varepsilon}_k\right)
= Y^{(k)}_{\left(\lceil (1 - r_k - \tilde{\varepsilon}_k) N_k\rceil\right)},
\]
where \(\tilde{\varepsilon}_k=(1+4C_{\upalpha})\sqrt{2\log(4/\alpha_u)/N_k}\) and \(r_k\in(0,1-\tilde{\varepsilon}_k]\).

Note that in the case of weakly dependent data, the hybrid concATE approach in Proposition~\ref{prop:hybrid} depends on the nuisance constant $C_{\upalpha}=\sum_{k\geq 1}\upalpha(k)^{1/2}$. For the important special case of a stable Gaussian AR($1$) process, this constant admits a closed-form upper bound.

\begin{lemma}[Mixing constant for a stable Gaussian AR($1$) model]\label{lem:calpha_ar1}
Let $\{Y_t\}$ be a centered stationary Gaussian AR($1$) process, with
\begin{equation*}
Y_t=\theta Y_{t-1}+\varepsilon_t,\quad \lvert\theta\rvert<1,\quad\varepsilon_t\overset{i.i.d.}{\sim}N(0,\sigma^2)
\end{equation*}
Then the strong-mixing coefficients satisfy
\begin{equation}
\upalpha(k) \leq \rho(\F_{-\infty}^0,\F_{k}^{+\infty})=\lvert \theta\rvert^k
\end{equation}
and consequently
\begin{equation}
C_\upalpha := \sum_{k\geq 1}\upalpha(k)^{1/2} \leq \frac{\sqrt{\lvert\theta\rvert}}{\left(1-\sqrt{\lvert\theta\rvert}\right)}.
\end{equation}
\end{lemma}    

To further strengthen these results, we formally derive and discuss the optimality of the concATE approach in the minimax sense in Section \ref{sec:optimality}.

\section{Minimax Optimality of concATE\label{sec:optimality}}

The hybrid concATE approach introduced in Proposition~\ref{prop:hybrid} uses DKW-calibrated order statistics of the treated and untreated outcome variables to derive $\alpha'$-sized lower and upper bands on the set-identified ATE. A natural question is whether a different distribution-free procedure (e.g., bootstrapping) could achieve a tighter worst-case interval while maintaining the same coverage. In what follows, we show that the DKW-calibrated endpoints $L_{\alpha'}^{(k)}$ and $U_{\alpha'}^{(k)}$ of concATE, for $k\in\{0,1\}$, are minimax-rate-optimal \citep[see~Ch. 2 of ][]{tsybakov2008nonparametric}, which implies that no other distribution-free procedure can achieve a faster rate of worst-case excess bracket width, while maintaining valid coverage over the class $\mathcal{F}$ of all outcome distribution with finite mean\footnote{Note that the definition of $\mathcal{F}$ is distinct from the $\sigma$-field definition in Section~\ref{sec:tippingpoint}}. The minimax results concern the tail-bracketing component of the confidence band, since the smooth components, i.e., $\hat{\delta}_k$ and $\hat{p}_k$, rely on the asymptotically valid delta-method, where the CLT applies. 

Before introducing the first set of results, let us denote $\mathcal{F}$ as the class of all Borel probability measures on $\R$ with finite mean. The results presented are for each arm $k\in\{0,1\}$, but for ease of exposition, we drop the subscript (and subscripts) $k$. Furthermore, following the notations in Section \ref{subsec:dkw}, $\alpha'$ is some generic $\alpha'\in(0,1)$ and the thresholds $\tau$ are suppressed. 
  
For $F\in\mathcal{F}$ and a fixed tail probability $r\in\left(0,\tfrac{1}{2}\right)$, define the true tail endpoints 
\[
q_r:=F^{-1}(r),\qquad q_{1-r}:=F^{-1}(1-r),
\]
where $G^{-1}(u)=\inf\{y:G(y)\geq u\}$ as in Lemma~\ref{lem:quantilesando}. In what follows, we find the minimax lower and upper bounds for concATE.

\begin{prop}[Minimax lower bound]
\label{prop:minimaxlow}
Let the pair $(L_{\alpha'},U_{\alpha'})$, measurable with respect to the order statistics $Y_{(1)},\dots,Y_{(N)}$, be called a distribution-free $(1-\alpha')$-bracket over $\mathcal{F}$, if 
\begin{equation}
\inf_{F\in\mathcal{F}}\Pr_{F}\left((L_{\alpha'}\leq q_r) \cap (U_{\alpha'}\geq q_{1-r})\right)\geq 1-\alpha'
\end{equation}
For such a bracket, define its worst-case high-probability excess bracket width $\mathcal{W}_N$, as the smallest value $w\geq 0$, such that
\begin{equation}
\inf_{F\in\mathcal{F}}\Pr_{F}\left(\max\{q_r-L_{\alpha'},0\}+\max\{U_{\alpha'}-q_{1-r},0\}\leq w\right)\geq 1-\alpha'.
\end{equation}
$\mathcal{W}$ measures the maximal width by which any valid bracket overshoots the true tail endpoints with high probability. A smaller $\mathcal{W}$ yields tighter identification regions for $\Re$.
For any distribution-free $(1-\alpha')$-bracket $(L_{\alpha'},U_{\alpha'})$ over $\mathcal{F}$ with $\alpha'\in\left(0,\frac{1}{2}\right)$, there exists a universal constant $c>0$ such that 
\begin{equation}
\mathcal{W}(L_{\alpha'},U_{\alpha'})\geq c\sqrt{\frac{\log(1/\alpha')}{N}}
\end{equation}
\end{prop}

\begin{prop}[Minimax upper bound]
\label{prop:minimaxupp}
Under Assumption~\ref{ass:mixing} and letting $\varepsilon:=(1+4C_{\upalpha})\sqrt{2\log(8/\alpha')/N}$ as per Proposition~\ref{prop:hybrid}, the DKW-calibrated endpoints
\begin{equation}
\label{eq:prop6q}
L_{\alpha'}:=Y_{(\lceil(r+\varepsilon)N\rceil)},\qquad U_{\alpha'}:=Y_{(\lceil(1-r-\varepsilon)N\rceil)},
\end{equation}
form a distribution-free $(1-\alpha')$-bracket over $\mathcal{F}$, satisfying
\begin{equation}
\mathcal{W}\left(L_{\alpha'},U_{\alpha'}\right)\leq C\sqrt{\frac{\log(2/\alpha')}{N}}
\end{equation}
for some constant $C>0$.
\end{prop}

The lower and upper bounds of the the bracket width of the DKW-calibrated tail endpoints in Propositions~\ref{prop:minimaxupp} and~\ref{prop:minimaxlow} lead to the following Corollary. 

\begin{corollary}[Convergence rate of concATE]
Among all distribution-free $(1-\alpha')$-brackets over the class of distributions $\mathcal{F}$, the DKW-calibrated  bracket $(L_{\alpha'},U_{\alpha'})$ achieves the minimax optimal rate
\begin{equation}
\mathcal{W}(L_{\alpha'},U_{\alpha'})=\Theta\left(\sqrt{\frac{\log(1/\alpha')}{N}}\right).
\end{equation}
In essence, no distribution-free procedure, including the nonparametric bootstrap, empirical likelihood, or plug-in sample-extrema, can achieve a strictly faster rate of convergence of $\mathcal{W}$, while maintaining valid coverage over $\mathcal{F}$.
\label{corr:corol2}
\end{corollary}
 The statements from Corollary~\ref{corr:corol2} can be verified as follows: let us first consider the case of plug-in sample extrema. Given the order statistics $Y_{(1)},\dots,Y_{(N)}$, using $Y_{(N)}$ are the upper bound tail endpoint, fails coverage whenever the hypothesised class of distributions $F_{1}$ concentrates mass above the observed sample range with positive probability (e.g., DGP E in Section~\ref{sec:montecarlo}. Hence, the sample maximum does not satisfy the distribution-free bracket condition over $\mathcal{F}$. Similarly, boostrap procedures achieve the rate $O\left(N^{-1/2}\right)$ only under additional regularity condition (e.g., bounded support of Donsker conditions \citep[see Theorem 3.6.1][]{van1996weak}, and can fail for heavy-tailed distributions $F\in \mathcal{F}$.

\section{Monte Carlo Study\label{sec:montecarlo}}

\subsection{Monte Carlo Design}

To study the finite sample behaviour of the hybrid band in
Proposition~\ref{prop:hybrid}, we run a Monte Carlo experiment with
six data generating processes (DGPs). Each design is replicated
$B=2{,}000$ times on a single sector with $n=50$ firms observed for
$T\in\{1,2,5\}$ periods, giving sample sizes $N=nT\in\{50,100,250\}$ respectively.
A single diversity cut-off $\tau^{\circ}=50\%$ is analyzed; hence no
Bonferroni size split is required. The overall two-sided size is fixed
at $\alpha=0.05$, giving the critical values
\[
c_{M}=\Phi^{-1}\!\left(1-\alpha/2\right)\approx1.96
\quad\text{and}\quad
c_{H}=\Phi^{-1}\!\left(1-\alpha/4\right)\approx2.24
\]
correspondingly for the \citet{manski1990nonparametric} and Hybrid (concATE) approaches. 

Within each replication and arm $k\in\{0,1\}$ we compute
\[
\varepsilon_k:=\sqrt{\frac{\log\left(8/\alpha\right)}{2N_k}},
\quad
r_k\in\left(0,\frac12-\varepsilon_k\right],
\]
and set the outward empirical-quantile endpoints
\[
L^{(k)}:=Y^{(k)}_{\left(\lceil (r_k+\varepsilon_k)N_k\rceil\right)},
\quad
U^{(k)}:=Y^{(k)}_{\left(\lceil (1-r_k-\varepsilon_k)N_k\rceil\right)}.
\]
(When $\varepsilon_k\le\frac14$ we take $r_k=\varepsilon_k$, yielding the
convenient indices $\lceil 2\varepsilon_kN_k\rceil$ and
$\lceil(1-2\varepsilon_k)N_k\rceil$; if $\varepsilon_k>\frac14$,
we set $r_k=\frac12-\varepsilon_k-\frac1{N_k}$ to preserve ordering.)
The Manski benchmark uses the usual plug-in with known support when available
(DGP F), and otherwise the sample extrema. The realized outcome is
\[
Y_{it}^{\mathrm{obs}} = Y_{it}^{0} + \Delta D_{it},
\qquad
\Delta = 4,
\]
where $Y_{it}^{0}$ follows the distribution listed below and
$D_{it}\sim\mathrm{Bernoulli}(0.3)$.

For each replication, we also compute: (i) the \citet{imbens2004confidence} confidence interval, which solves for the critical value $c_n$, satisfying $\Phi(c_N+\sqrt{N}\Delta_N)-\Phi(-c_N)=1-\alpha$, where $\Delta_N=(\hat{U}-\hat{L})/\hat{\sigma}_{\max}$; (ii) The \citet{stoye2009more} interval, which sets $c=\max(c_N,z_{\alpha/2})$ to ensure uniform validity; and (iii) percentile bootstrap with $B_{boot}=999$ replications, using a Sieve bootstrap for DGP C, when $T> 1$ \citep[see][]{buhlmann1998sieve}.

\begin{enumerate}[label=\textbf{DGP~\Alph*:}, itemsep=2pt, leftmargin=16mm]

\item \emph{i.i.d.\ Standard Normal}\\
$Y_{it}^{0}\sim N(0,1)$,\quad
$D_{it}\sim\mathrm{Bernoulli}(0.3)$.
\smallskip

\item \emph{Heavy tail (sub-exponential)}\\
$Y_{it}^{0}\sim t_3/\sqrt{3}$ (unit variance),\quad
$D_{it}\sim\mathrm{Bernoulli}(0.3)$.
\smallskip

\item \emph{AR(1) panel with positive selection bias}\\
$Y_{it}^{0}=0.4Y_{i,t-1}^{0}+\varepsilon_{it}$,\quad
$\varepsilon_{it}\stackrel{\text{i.i.d.}}{\sim} N(0,1)$.\\
Treatment probability:
$\Pr(D_{it}=1\mid Y_{it}^{0})
=\mathrm{logit}\!\left(+0.5Y_{it}^{0}+\eta_{it}\right)$,\quad
$\eta_{it}\sim N(0,0.5^{2})$.

\item \emph{Rare-extreme point mass (controlled tail visibility)}\\
Let the per-tail probability be $\pi_N=\lambda/(2N)$ with
$\lambda\in\{0.35,\,0.69,\,1.61\}$ so that
$\Pr(\text{no extreme in the sample})\approx e^{-\lambda}\in\{0.70,\,0.50,\,0.20\}$.
Set
\[
Y_{it}^{0} =
\begin{cases}
- M, & \text{w.p.\ }\pi_N,\\
Z,   & \text{w.p.\ }1-2\pi_N,\quad Z\sim N(0,1),\\
+ M, & \text{w.p.\ }\pi_N,
\end{cases}
\quad M\in\{6,10\},\quad
D_{it}\sim\mathrm{Bernoulli}(0.3).
\]
\textit{Purpose:} directly tunes the probability that finite samples miss the true extremes.

\item \emph{Left-truncated $\chi^{2}$ tail}\\
$Y_{it}^{0} \sim \chi^{2}(3)$,\quad
$D_{it}\sim\mathrm{Bernoulli}(0.3)$.\\
\textit{Note:} for the hybrid band we fix $L^{(k)}=0$ and use only the one-sided
upper empirical quantile $U^{(k)}=\hat F^{-1}_{N_k}\left(1-r_k-\varepsilon_k\right)$.
\smallskip

\item \emph{Uniform support known a priori}\\
$Y_{it}^{0}\sim\mathrm{Uniform}[-5,5]$,\quad
$D_{it}\sim\mathrm{Bernoulli}(0.3)$.\\
\textit{Note:} the estimator is supplied with the true support
$a=-5,b=5$ when forming Manski bounds.

\end{enumerate}

DGP D is specifically constructed so that the finite sample has a tunable probability
of not observing the population extremes, the precise scenario for which the hybrid
band was developed.

\subsection{Simulation Results}

\vspace{-1ex}
\begin{table}[H]
\footnotesize
\centering
\begin{tabular}{lcccccccccc}
\toprule
&\multicolumn{5}{c}{Coverage (\%)} & \multicolumn{5}{c}{Width} \\
\cmidrule(lr){2-6}\cmidrule(lr){7-11}\\
DGP & Manski & IM & Stoye & Bootstrap& Hybrid  &Manski & IM & Stoye & Bootstrap& Hybrid \\
\midrule
\textbf{A}          &&&&&&&&&&\\
$N=50$                &100 &100 & 100 &99.95 & 98.95 & 5.2 & 4.9 &5.2 &4.2 &2.0 \\
$N=100$                 &100 & 100 & 100 & 100 &  99.80 &  5.4 &  5.2  &  5.4  &  4.6  &  1.5 \\
$N=250$                 &100 & 100 & 100 & 100 &100 &     5.6 &    5.5  &  5.6 &   5.2 &   1.7 \\
 &&&&&&&&&&\\
\textbf{B}          &&&&&&&&&&\\
$N=50$                &100 & 100 & 100 & 100 & 99.85 &     5.1 &   4.9 &   5.1 &   4.1  &  1.9\\
$N=100$                 &100 & 100 & 100 & 100 & 100 &       5.9 &   5.7 &   5.9  &  5.1  &  1.4 \\
$N=250$                 &100 & 100 & 100 & 100 & 100  &     7.5 &   7.4 &   7.5 &   7.0 &   1.4 \\
 &&&&&&&&&&\\
\textbf{C}          &&&&&&&&&&\\
$N=50$                &100& 100& 100& 100&  99.75  &     5.6  &  5.3  &  5.6  &  4.3  &  2.3\\
$N=100$                 &100& 100& 100& 100&  97.75  &     5.8 &   5.6 &   5.8  &  5.2  &  1.7 \\
$N=250$                 &100& 100& 100& 100& 100&       6.2  &  6.0 &   6.2 &  14.2 &   2.1 \\
 &&&&&&&&&&\\
\textbf{D}          &&&&&&&&&&\\
$N=50$                &100& 100& 100& 100&  99.30 &     5.4 &   5.2 &   5.4 &   4.4 &   2.1\\
$N=100$                 &100& 100& 100& 100&  99.55  &     5.7 &   5.6 &   5.7 &   5.0 &   1.6 \\
$N=250$                 &100& 100& 100& 100& 100&     8.1 &   7.9 &   8.1 &   7.6 &   1.7 \\
 &&&&&&&&&&\\
\textbf{E}          &&&&&&&&&&\\
$N=50$                &100& 100& 100& 100&  99.45 &    14.1 &  13.6 &  14.1 &  12.9  &  7.9\\
$N=100$                 &100& 100& 100& 100&  99.55 &    14.8  & 14.5 &  14.8 &  13.9&    7.2 \\
$N=250$                 &100& 100& 100& 100& 100&    16.4 &  16.1 &  16.4 &  15.7  &  7.5\\
 &&&&&&&&&&\\
\textbf{F}          &&&&&&&&&&\\
$N=50$                &100& 100& 100&  99.95 &100&    12.6 &  12.2 &  12.6 &  10.5 &  12.6\\
$N=100$                 &100 & 100 & 100 & 100 & 100 &    11.9 &  11.6 &  11.9 &  10.6 &  11.9 \\
$N=250$                 &100& 100& 100&100& 100&    11.2 &  11.0 &  11.2 &  10.5 &  11.2\\
\bottomrule
\end{tabular}
\caption{Monte Carlo coverage and the median width of the 95\% of \citet{manski1990nonparametric} (Manski), \citet{imbens2004confidence} (IM), \citet{stoye2009more} (Stoye), Bootstrap and Hybrid (concATE) bounds at $\tau^{\circ}=50\%$. All methods achieve nominal coverage; the Hybrid band achieves comparable coverage with substantially sharper intervals}
\label{tab:coverage}
\end{table}
Table~\ref{tab:coverage} reports the pointwise Monte Carlo coverage of the identified set for the average treatment effect evaluated at the tipping point $\tau^{\circ} = 50\%$. For each replication, we check whether $\Delta$ lies inside the \citet{manski1990nonparametric}, \citet{imbens2004confidence}, \citet{stoye2009more}, Bootstrap and Hybrid (concATE) bounds, at this single threshold.

At first glance, it is evident that all the bounds are valid with a 95\% coverage, with the concATE followed by the bootstrap approaches being the closest to the nominal 95\% level. In fact, their slightly-below-100\% coverage reflects better calibration rather than a deficiency, since a method that always reports 100\% coverage is being unnecessarily conservative. As the sample size increases, Hybrid coverage converges to 100\% as the asymptotic delta-method component improves, while remaining well above the 95\% nominal threshold throughout.

The width of the bounds reveals an entirely different story. For DGPs A--D the Manski, IM, Stoye and Bootstrap approaches all exhibit similar performances. As it is evident, in the case of DGP A, the Manski, IM and Stoye approaches all seem to converge to the same bound width of $\approx 5.6$ as the sample size increases. Similar converging phenomena is observed for DGP C, which are sampled from a normal distribution, but with a weakly dependent structure, which has increased the width. However, in the case of heavy-tailed distributions, while the results of the different techniques are similar for larger samples, the width seems to get extensively larger. This would simply be because a larger sample from the heavy-tailed DGPs implies larger deviations and hence variances.

In all instances, the width of the concATE approach shrinks with larger samples and the method provides sharper bounds on the average treatment effect. Quantitatively, the Hybrid band is 2--4 times narrower than all competing methods across DGPs A--D. For example, at $N=50$ under DGP A, the Hybrid width is 2.0 compared to 5.2 for Manski - a 62\% reduction. This efficiency gain arises because the hybrid replaces arm-specific sample extrema, which systematically overshoot the population support, with DKW-calibrated quantile endpoints that trim tail mass with controlled probability, yielding a tighter identification region without sacrificing coverage.

The advantage of the Hybrid is particularly pronounced under DGP D, which is specifically constructed so that finite samples have a tunable probability of missing the population extremes. At $N=50$, the Hybrid width is 7.9 compared to 14.1 for Manski, while both methods maintain coverage above 95\%. Even as $N$ increases to 250 and the competing methods' coverage remains at 100\%, their widths grow to 16.4 whereas the Hybrid width is only 7.5 - a gap that widens rather than closes, precisely because larger samples from heavy-tailed DGPs produce more extreme order statistics that inflate the plug-in support bounds. The Hybrid, by contrast, is insulated from this effect through its quantile-based construction.

\section{Empirical Application\label{sec:results}}

In this section, we ask ``Does gender-based board diversity causally affect firm innovation?". We begin by outlining the data and summarizing its key descriptive statistics. We then present the nonparametric bounds approach of \citet{manski1990nonparametric, manski2003partial} with simultaneous confidence bands in Proposition \ref{prop:1}, the hybrid band proposed in Proposition \ref{prop:hybrid}, and the naïve mean‐comparison framework of \citet{angrist2009mostly} (reported in Appendix \ref{sec:additionalanalysis}). The common objective is to test the null hypothesis of a zero average treatment effect of diversity on innovation, against a positive or negative alternative, when the diversity cut-off is selected endogenously (see Eq.~\eqref{eq:null}).

\subsection{Data and Descriptive Statistics\label{sec:data}}

The empirical analysis uses a panel of publicly listed firms compiled from FactSet, with quarterly observations from 2015 Q2 through 2022 Q1. The initial sample includes 945 firms, yielding a short panel of 945 cross-sectional units over 28 quarters (totalling 26,460 firm-quarter observations). 

In our analysis, we categorize the eleven Global Industry Classification Standard sectors (GICS hereafter) into five broader groups: Cyclicals (Consumer Discretionary, Materials, Industrials, Real Estate), Defensives (Health Care, Consumer Staples, Utilities), Growth \& Innovation (Information Technology, Communication Services), Financials, and Energy. This classification reflects the economic sensitivities of these sectors, as identified by Morgan Stanley Capital International (MSCI hereafter). Specifically, MSCI's Cyclical and Defensive Sectors Indexes classify sectors based on their performance correlation with the business cycle, using the OECD Composite Leading Indicator. According to MSCI, sectors like Consumer Discretionary, Materials, Industrials, Real Estate, Information Technology, Communication Services, and Financials are considered cyclical due to their positive correlation with economic expansions. Conversely, sectors such as Health Care, Consumer Staples, Utilities, and Energy are deemed defensive, exhibiting resilience during economic downturns. By adopting this grouping, we aim to capture the nuanced behaviours of these sectors in relation to macroeconomic conditions, facilitating a more informed analysis of sectoral dynamics. This classification can be found in Table \ref{tab:sector_groups}.

Following the Corporate Sustainability Reporting Directive definition (CSRD hereon) of a ‘large undertaking’ (Directive 2022/2464/EU, Art. 3 Pt 4) and the 250-employee threshold used in EU and UK gender-pay-gap statutes, we restrict the sample to firms whose time-average workforce is at least 250 employees over the sample horizon to ensure they fall under harmonized disclosure regimes. The restriction yields $n=901$ firms and a total of $N=25,228$ firm-quarter observations.  

\begin{table}[htbp]
  \centering
  \resizebox{\textwidth}{!}{%
    \begin{tabular}{lcccccccc}
      \toprule
      Variable & Min & Mean & Median & Max & Std Dev & Skewness & Kurtosis & $N$ \\
      \midrule
(\%) Women & 0.000 & 27.490 & 27.140 & 100 & 12.723 & 0.463 & 1.949 & 25,038 \\
(\%) Unknown gender & 0.000 & 0.029 & 0.023 & 0.496 & 0.031 & 2.348 & 14.420 & 25,038 \\
Tobin's $Q$ (scaled) & -0.612 & 0.445 & 0.012 & 5.047 & 1.221 & 2.133 & 4.530 & 23,085 \\
Total assets & 10.392 & 16.109 & 16.114 & 22.098 & 1.811 & 0.060 & 0.221 & 23,990 \\
Leverage & 0.000 & 0.302 & 0.288 & 3.945 & 0.230 & 3.283 & 34.292 & 23,977 \\
Total employees & 85.559 & 25707.813 & 8554.289 & 941046.440 & 54162.070 & 6.216 & 58.997 & 25,224 \\
      \bottomrule
    \end{tabular}%
  }
  \caption{Panel descriptive statistics}
  \label{tab:descriptive_stats_simple_signal}
\vspace{0.3em}
{\raggedright \footnotesize \textit{Note}:  
$N$ varies by variable because some firm-quarter observations are missing that particular item (e.g.\ Tobin’s $Q$ is reported for 23,085 of the 25,228 firm-quarters). 
The descriptive statistics are computed on all available values for each variable (“pair-wise” basis). 
For the causal analysis we use list-wise deletion, retaining only the firm-quarters for which \emph{all} diversity indicators and Tobin’s $Q$ are present. 
Tobin’s $Q$ is scaled using a robust scaler (median and interquartile range) prior to the causal analysis.
\par}
\end{table}

The key “treatment” variable is the percentage of women in senior leadership positions. To develop measures of senior management diversity, we use data from Revelio Labs. Revelio Labs is a data provider that aggregates workforce data from multiple sources, including online professional profiles (e.g., LinkedIn), job postings, company websites, government records and census data. Data are aggregated by firm and time and weighted to correct for underrepresentation in online professional profiles, such as lower-skilled or lower-paid roles, as well as to address issues such as mapping discrepancies and the presence of fake or duplicate profiles. For each firm, Revelio provides global and local headcounts and demographic characteristics of employees including gender and ethnicity, based on probabilistic estimations. From this data, we focus on the gender and ethnicity diversity of employees located in the US or the U.K., serving as a proxy for headquarters location, since the firms in our sample are listed on the MSCI US and S\&P U.K. indices. Our analysis is concentrated on senior management positions, defined as executive and senior executive roles (e.g., Managing Director, Partner, CEO, CFO).  We restrict the sample to senior management in the headquartered country as there is a growing literature indicating that executive and senior management characteristics impact management practices and firm performance \citep{flabbi2019female}. Additionally, these roles are less likely to be underrepresented in the Revelio dataset. These diversity measures are constructed using a supervised machine learning algorithm applied to senior executives’ names, which infers gender from linguistic patterns. If the algorithm cannot assign a gender with high confidence, the individual is labelled as “unknown”. Importantly, the incidence of unknown classifications is very low: on average only about 0.03\% (Table \ref{tab:descriptive_stats_simple_signal}). 

\begin{figure}[hbtp!]
\centering
\includegraphics[width=0.7\textwidth]{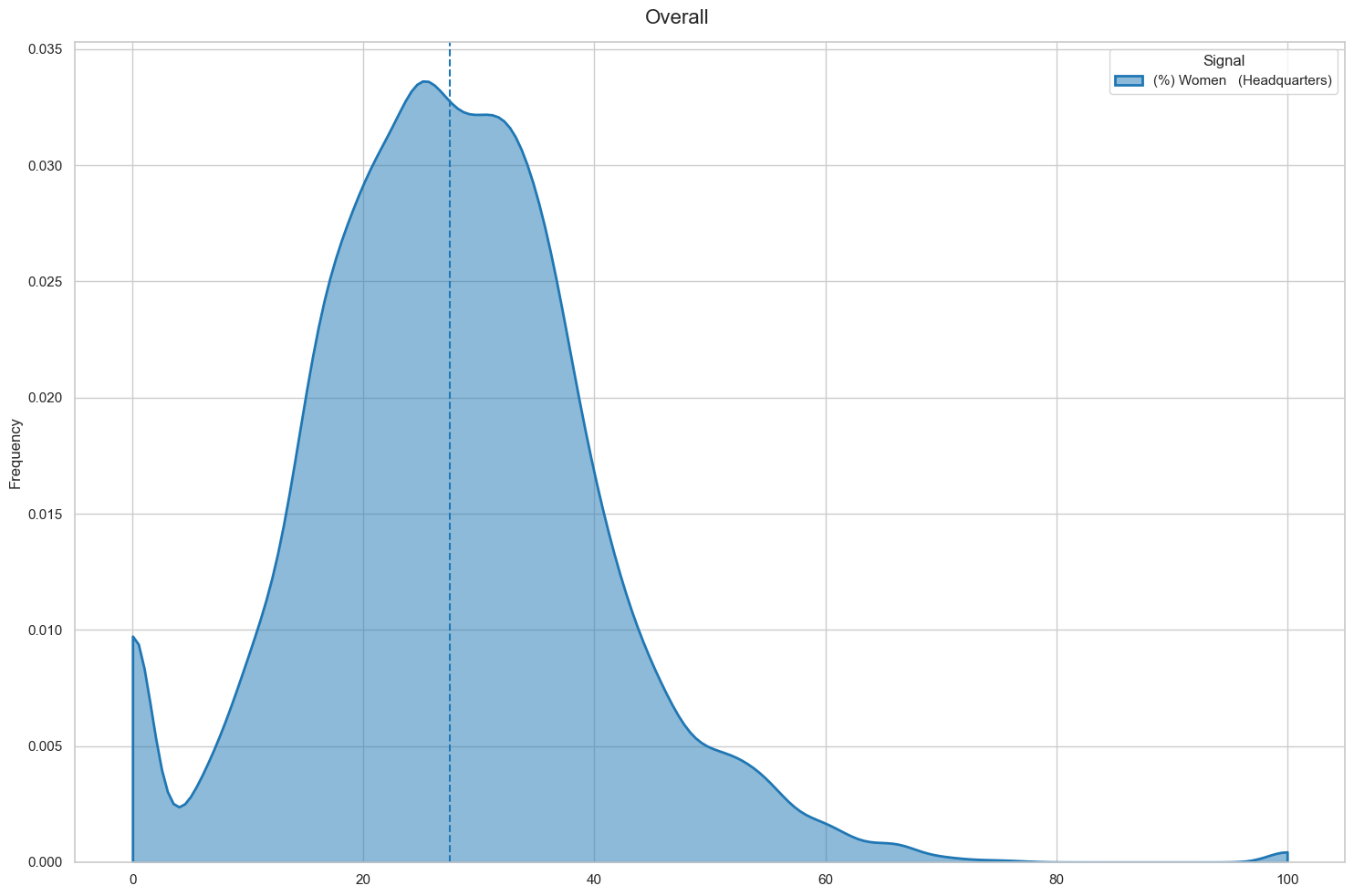}
\caption{Kernel density plot of percentage women. Scott's rule \citep{scott2015multivariate} is used to select the smoothing bandwidth parameter.}
\label{fig:density_overall}
\end{figure}

The outcome of interest is Tobin’s $Q$, defined as the ratio of the firm’s market value to the replacement cost of its assets, a standard measure of firm performance and growth opportunities \citep{tobin1969general,tobin1978monetary}. We also utilize several control variables for descriptive analysis, including firm size (log total assets), leverage (debt-to-assets ratio), and total employees. Summary statistics for all main variables are provided in Table  \ref{tab:descriptive_stats_simple_signal}. After excluding observations with missing data on key fields, the average percentage of women in senior roles is about 27.5\%. The standard deviation (around 12 percentage points for female share) indicates considerable cross-firm variation. Notably, a non-trivial subset of firm-quarters have zero diversity: roughly 4\% of observations have no women in senior leadership, at least at some point in the sample. The distribution of the diversity variables is right-skewed. Figure \ref{fig:density_overall} illustrates kernel density estimates of the percentage of female senior leaders across all firm-quarters. The distribution is skewed to the right with a primary mode around 25--35\%, and a secondary mass at 0\% corresponding to firms and periods with homogeneous leadership teams.

\begin{figure}[hbtp!]
 \centering
 \includegraphics[width=0.8\textwidth]{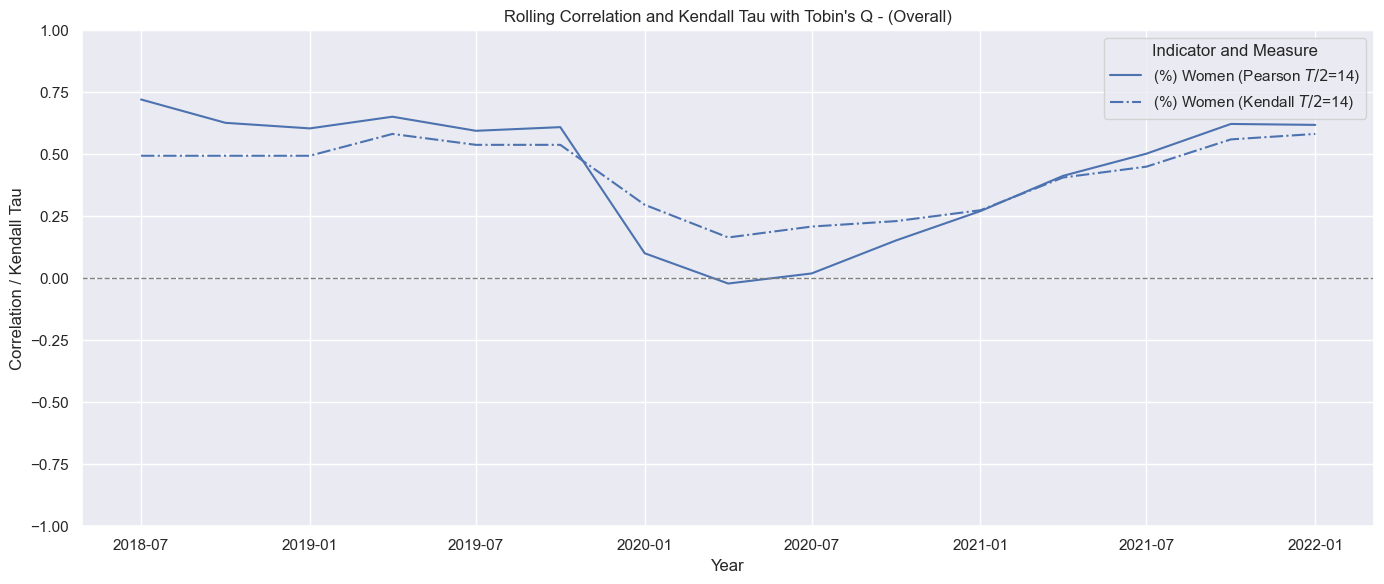}
 \caption{Rolling Pearson correlations and Kendall's $\tau$ capturing both linear and monotonic associations between Tobin's $Q$ and percentage women in senior leadership.}
\vspace{0.3em}
 {\raggedright \footnotesize \textit{Note}: The size of the rolling windows is chosen as half the length of the time dimension of the sample, i.e., $T/2$. \par}
 \label{fig:rollingcorr_overall}
\end{figure}

We next explore the raw association between gender diversity measure and firm performance. In the full sample (pooled across all sectors and time periods), there is a strong positive correlation between senior-team diversity and Tobin’s $Q$. Figure \ref{fig:rollingcorr_overall} plots rolling correlations over time, using a moving window of half the sample period ($T/2 \approx 14$ quarters) to track how the relationship evolves. The Pearson correlation between the percentage of women in leadership and Tobin’s $Q$ is in the range of +0.6 to +0.7 for most of the sample, indicating a fairly strong linear association. The Figure also reports Kendall’s $\tau$ rank correlation, which captures monotonic association; this measure corroborates the positive link while being slightly lower in magnitude, suggesting the relationship is broadly monotonic even if not perfectly linear. The association appears to strengthen from 2015 up to about 2019, consistent with increasing awareness and implementation of diversity initiatives, but then shows a noticeable drop around 2019-2020. After 2019, the rolling correlations decline, implying that the previously tight diversity-performance relationship loosened and is increasing again after 2021. One possible interpretation is that external shocks or changing market conditions (for instance, the disruptive impact of the COVID-19 pandemic or the murder of George Floyd) temporarily weakened the correlation between diversity and market valuations.

In light of this, and in addition to the sectoral group analysis, we examine the overall rolling correlations for the period preceding this drop. The corresponding Pearson correlations for this classification are reported in Figure~\ref{fig:corr_all}, while the rolling associations are shown in Figure~\ref{fig:rollingcorrs_grid}.

Several noteworthy patterns emerge. First, the Growth \& Innovation sector consistently exhibits a strong positive correlation between gender diversity measures and Tobin’s $Q$ across all years, and this sector does not experience the 2019 drop in correlation seen in the aggregate data. Second, the Energy sector shows a markedly different pattern: the percentage of women in senior positions in energy firms is actually negatively correlated with Tobin’s $Q$ in most years. These observations may reflect unique dynamics or reverse causality in the energy industry (for example, struggling firms may appoint more women to leadership roles as part of restructuring). Third, in the Financials sector, the correlation with diversity is negative in the earlier part of the sample (implying more homogenous banks were associated with slightly higher $Q$ ratios pre-2019), but this relationship reverses sign around 2019. By the end of the sample period, financial firms with more diverse leadership tend to have higher Tobin’s $Q$, indicating a possible structural change in how markets value diversity in finance or how an increase in inclusion that enabled diversity to be leveraged for business gains. 

\FloatBarrier

\subsection{Causality Analysis\label{sec:casuality}}

While the descriptive results suggest a concordance between greater senior-level gender diversity and higher firm performance, correlation alone cannot establish causality. In this section, we formally test whether increases in executive diversity causally impact Tobin’s $Q$, using the methodology developed in Sections \ref{sec:framework}-\ref{sec:inference}. Because the “treatment” (crossing a diversity threshold) is not randomly assigned, a naïve estimation of this effect risks bias from selection on unobservables. We therefore implement both a conventional point-estimation approach under strong assumptions and a robust partial-identification approach under minimal assumptions, and compare the findings. 

First, we apply an unconditional mean-comparison framework following \citet{angrist2009mostly}. For each candidate diversity threshold $\tau$ (e.g. 5\%, 10\%, …, 50\%, etc.), firms are split into a treated group (above the threshold) and a control group (below the threshold). We then estimate the difference in mean Tobin’s $Q$ between treated and control firms for that threshold. This difference-in-means is a point estimate of the ATE if one assumes mean independence (i.e. that, conditional on crossing the threshold, potential outcomes are the same for treated and control firms on average). We construct simultaneous 95\% confidence bands for these ATE estimates across all thresholds in the set $\mathcal M = {5, 10, 15, …, 90, 95}$, applying a Bonferroni or Šidák correction to account for the multiple comparisons. This yields a series of tests for the null hypothesis of no effect at each diversity level, adjusted so that the overall family-wise error rate is 5\%. It is important to note that this point-identified approach treats the threshold “treatment” as if random; in practice, firms that surpass a given diversity level could differ systematically from those that do not (for instance, more progressive or better-governed firms might both adopt diverse leadership and perform well for other reasons). As a result, the point estimates of $\delta$ may capture more than the true causal effect of diversity. We use this method as a benchmark, fully aware that its validity hinges on strong assumptions.

We next relax the strong assumptions by employing a partial identification strategy \citep{manski1990nonparametric,manski2003partial}. Instead of assuming we can precisely identify the counterfactual outcome for each firm, we derive bounds on the possible ATE. Instead of point identification, we partially identify the region in which the average treatment effect~$\Re$ lies, as characterized by Eq.\eqref{eq:nonparametricbounds}. We denote this set the identification region $\mathcal H_{\alpha_u}[\Re_u]$ for all $u$ in $\mathcal{M}$, where as noted in Section \ref{sec:tippingpoint}, $\mathcal{M}=\{5,10,15,\cdots,90,95\}$ which represents the random diversity thresholds. As previously noted, estimation of Eq.~\eqref{eq:nonparametricbounds} involves latent quantities $\E[Y_{ijt}^{(0)} \mid Z_{ijt}(\tau_u)=1]$ and $\E[Y_{ijt}^{(1)} \mid Z_{ijt}(\tau_u)=0]$, which are not observed but can be bounded by quantities $L^{(k)}$ and $U^{(k)}$. On one hand, we may acknowledge that the extrema of the latent outcomes within the finite sample may not capture the true population extrema (and consequently the true treatment effect interval), in which case we rely on the finite sample hybrid approach. On the other hand, one may argue that since using the full range of outcomes ($\min$ and $\max$) can lead to overly conservative bounds, we also construct Manksi bounds using the (5\textsuperscript{th}, 95\textsuperscript{th}) and (10\textsuperscript{th}, 90\textsuperscript{th}) quantiles of $Y_{ijt}^{(k)}$. Finally, we build a simultaneous joint 95\% confidence region for the estimated bounds to make causal inference claims.

Before turning to results, we address some practical implementation details. As noted in Section \ref{sec:estimation}, it is necessary for both the treated and control groups to be non-empty (and sufficiently large) at each threshold to estimate meaningful effects. In our panel, some extreme diversity thresholds (especially very high ones) result in very few firms in one group. We therefore discard threshold levels $\tau$ for which one of the groups contains fewer than 10 observations (approximately, we require at least 10 firm-quarters above and below the threshold). If too many high-$\tau$ values are discarded for a particular subset of the data, that subset is excluded from the threshold analysis due to lack of support. In practice, this means that for some sector-specific analyses we cannot evaluate very high diversity percentages because, for example, no firm in a given sector ever reaches 90\% female leadership. Based on this criterion, certain combinations of sector and diversity type are dropped from the causal analysis. In particular, we exclude female leadership in sectors that never approach gender parity (notably the Financials and Energy sectors). These exclusions are a matter of data availability and ensure that the identification regions for ATE do not trivially collapse to a point. All remaining sector clusters and diversity measures satisfy $0 < \Pr(Z(\tau_u)=1) < 1$ at the thresholds of interest, so both treated and control outcomes can be observed in those cases.

\begin{figure}[hbtp!] 
\centering
  \begin{subfigure}[hbtp!]{0.45\textwidth}
    \includegraphics[width=\linewidth]{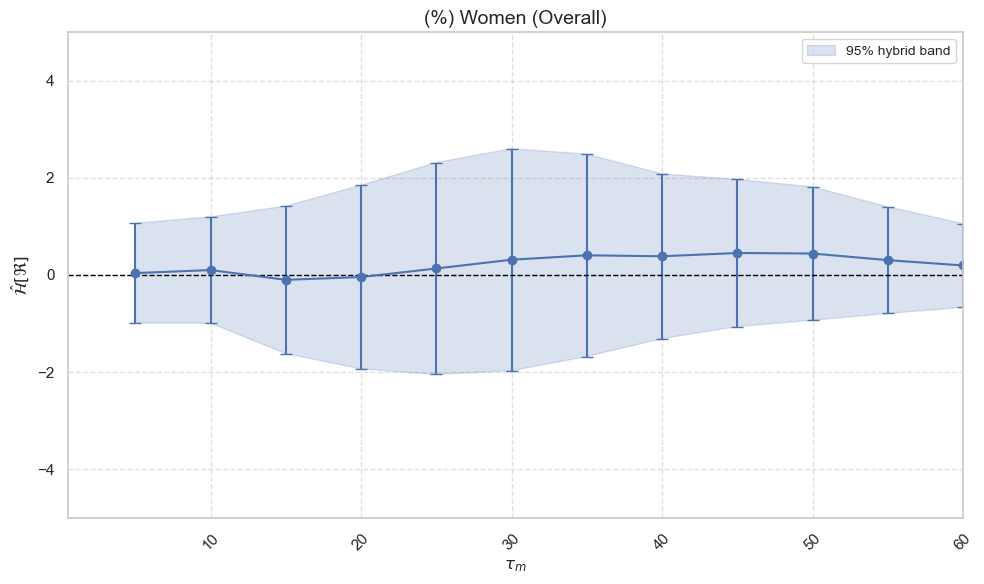}
    \caption{Hybrid: Overall (\%) women}
  \end{subfigure}
\hfill
  \begin{subfigure}[hbtp!]{0.45\textwidth}
    \includegraphics[width=\linewidth]{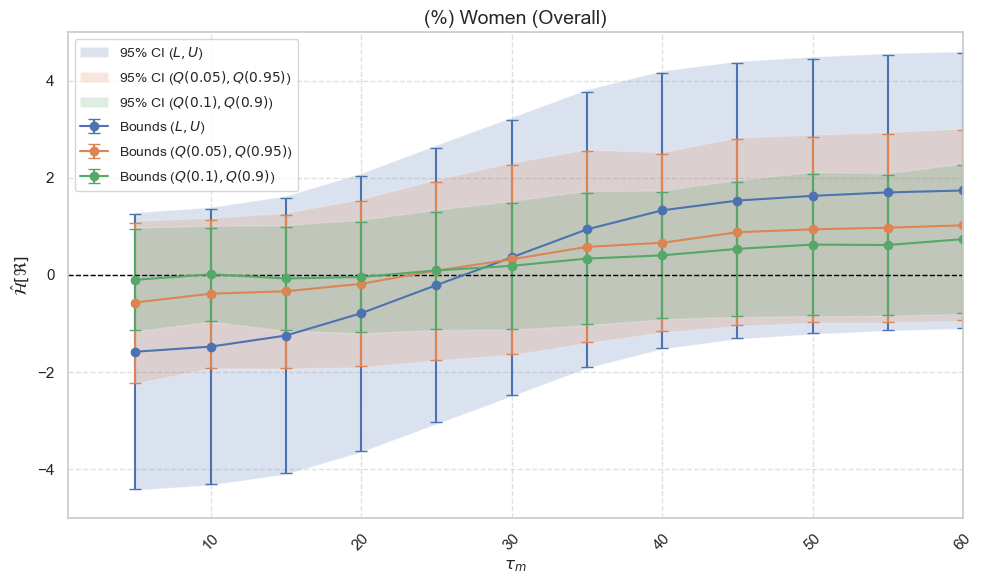}
    \caption{Manski: Overall (\%) women}
  \end{subfigure}
  \hfill
  \par\medskip
  \caption{Hybrid and Manski's Nonparametric Bounds - (Overall)}
\label{fig:overall_results}
\vspace{0.3em}
{\raggedright\footnotesize\textit{Note:} The lines in the nonparametric bounds plots represent the midpoints between the upper and lower bound estimates.\par}
\end{figure}

We first examine the impact of senior-level gender diversity on firm performance in the full sample, comparing a naive point estimate approach to our partial identification methods. Using the unconditional difference-in-means (Angrist and Pischke's approach), we find that greater female representation is associated with higher Tobin's $Q$, with an apparent ``tipping point'' at moderate diversity levels. In particular, once women comprise roughly one-third of the top management team, the naive ATE estimate becomes positive and statistically significant. For example, crossing about 30--35\% female leadership is associated with a jump in Tobin's $Q$ (see Table~\ref{tab:tipping}, which summarizes the estimated threshold levels at which the treatment effect becomes significant under each method): beyond this threshold the simple treated-control difference excludes zero at the 5\% level. This suggests that, under strong assumptions of ignorability, even a moderately gender-diverse leadership team might boost firm market value.

However, when we relax those assumptions, the evidence is less definitive. Manski's nonparametric bounds, which allow for arbitrary selection and unobserved heterogeneity, remain wide in finite samples, and their associated confidence band always includes zero. At very low diversity levels, the lower bound on the ATE is substantially negative (reflecting the worst-case scenario that ``token'' diversity could coincidentally occur in poorly performing firms). As the female share increases, this lower bound rises toward zero. We observe an inflection around 20--25\% female representation, roughly consistent with \citet{kanter1977some}'s notion of moving from tokenism to a more influential minority. Beyond that point, the worst-case impact of diversity is no longer severely negative; by around 50\% female leadership, the lower bound is near zero and the upper bound is positive. Nevertheless, without additional information about outcome limits, even at the highest diversity levels observed (e.g.\ 80--90\% female), the 95\% confidence region for the ATE still straddles zero. In other words, under minimal assumptions the data do not allow us to conclusively rule out no effect (or even a small negative effect) for the overall sample. This highlights how misleading the precise naive estimate can be: what appears as a clearly positive effect with a simple mean comparison becomes statistically ambiguous once we account for uncertainty about counterfactual outcomes.

\begin{table}[hbtp!]
  \resizebox{\textwidth}{!}{%
\begin{tabular}{llccccc}
\hline
Sector               & Signal & Hybrid & \multicolumn{3}{c}{Manski}         & Angrist \\ \cline{4-6}
                              &                 &                 & Max & 5\% & 10\% &                  \\ \hline
Overall              &                 &                 &              &              &               &                  \\
                              & (\%) Women      & -               & -            & -            & -             & 35\%             \\
Cyclicals            &                 &                 &              &              &               &                  \\
                              & (\%) Women      & -               & -            & -            & -             & 30\%             \\
Defensives           &                 &                 &              &              &               &                  \\
                              & (\%) Women      & 60\%               & -            & -            & -             & 40\%             \\
Growth \& Innovation &                 &                 &              &              &               &                  \\
                              & (\%) Women      & 55\%            & -            & -            & 55\%          & -                \\
Financials           &                 &                 &              &              &               &                  \\
                              & (\%) Women      & \multicolumn{5}{c}{N/A}                                                          \\
Energy               &                 &                 &              &              &               &                  \\
                              & (\%) Women      & \multicolumn{5}{c}{N/A}                                                          \\

\hline
\end{tabular}
}
  \caption{Random Diversity Tipping Points}
  \label{tab:tipping}
  \vspace{0.3em}
  {\raggedright \footnotesize \textit{Note}: This table presents the estimated tipping points—i.e., the random diversity thresholds at which the diversity treatment has a significantly positive effect on Tobin's~$Q$. Cells marked with a (-) indicate cases where significance is not achieved at any of the prescribed thresholds. Rows labeled ``N/A'' correspond to cases that do not meet the minimum threshold size condition of $\tau_m > 50$ discussed earlier.\par}

\end{table}

Imposing mild outcome bounds yields somewhat tighter inference. If we assume, for instance, that Tobin's $Q$ outcomes are effectively constrained within the central 90--95\% of the observed sample range (excluding extreme tail realizations), the identified ATE interval narrows. Under these plausible restrictions, the partial identification bounds move inward: the lower bound is higher (less negative) and the upper bound lower (less positive) than the unbounded Manski case. As a result, the concATE confidence band becomes more optimistic at high diversity levels. For example, using the 5\textsuperscript{th} and 95\textsuperscript{th} percentiles of $Q$ as rough bounds, we find that at very high diversity (above about 60\% women in leadership) the lower bound on the ATE is nearly zero or slightly positive. With an even tighter 10\textsuperscript{th}--90\textsuperscript{th} percentile restriction, the lower bound actually rises above zero at some thresholds. These results hint that a real positive effect may emerge once diversity is sufficiently high: with female leadership above roughly two-thirds, even the worst-case impact is likely to be zero. However, we emphasize that even under these trimmed-outcome assumptions, the joint 95\% confidence band for the ATE barely excludes zero. In the full sample, no diversity threshold produces a completely robust positive effect at the 5\% significance level unless one accepts some outcome-range assumptions. Thus, our most cautious conclusion for the overall dataset is that greater gender diversity could improve firm value, but the evidence is not statistically conclusive under minimal assumptions. The contrast between the naive point estimate (significant at $\sim$30\% diversity) and the conservative bounds (no significance without assumptions) underscores the importance of conservative, rigorous inference: smaller apparent gains may reflect unobserved biases or heavy-tailed outcomes rather than true causal effects.
 
\vspace{1em}

\noindent\textbf{Sector-Specific ``Tipping Points"} 

\vspace{1em}

We next investigate whether the diversity--performance relationship exhibits stronger effects in particular types of firms. To explore this, we apply an identical analysis within more homogeneous sector groupings. In each case we report the threshold at which the concATE band indicates a significant effect, and compare it to the naive and classical bounds results. This reveals several interesting findings.

First, in high-growth, innovation-intensive industries, we find clear evidence of a diversity tipping point. Firms in these sectors show relatively high variance in leadership composition, with some approaching gender-balanced teams. The unconditional mean comparison suggests a positive effect of diversity that becomes sizeable at upper diversity levels. However, due to the smaller sample of firms in this category, the naive threshold for significance is somewhat high: only at nearly half female representation does the simple difference in Tobin's $Q$ become significant. Our robust analysis confirms and sharpens this finding. The concATE confidence band for the average treatment effect in Growth \& Innovation firms first excludes zero at approximately 55\% female representation in senior roles. In other words, once a firm's top team is roughly half women, we can confidently assert a positive causal impact on market valuation in this sector. Below that threshold, the partial-identification interval still includes zero, meaning the effect cannot be distinguished from zero (i.e.\ while we cannot assert a positive effect, we can rule out a negative effect) with high confidence. Notably, the estimated ATE grows larger as diversity increases beyond 55\%; for firms that actually achieve gender-balanced or women-majority leadership, even the lower bound of the effect is distinctly above zero. This pattern aligns with the idea that innovative companies reap substantial benefits from diverse perspectives only after achieving a critical mass of diversity. Before that point, female voices may be too diluted to change organizational outcomes, but around parity their influence on decision-making and the innovation climate becomes strong. It is encouraging that both the naive method and the more rigorous concATE method point to a similar threshold in these sectors (around 50--55\% female): this convergence suggests the result is not merely an artefact of assumptions. In sum, for Growth \& Innovation firms we find a statistically significant positive causal effect of diversity emerging at just over half women in leadership.

Second, for Defensive sectors (Healthcare/Staples/Utilities), traditionally ``stable'' industries that historically have lower female leadership representation, fewer firms in our sample reach high diversity levels. A naive analysis indicates that even moderate diversity might help performance: the difference-in-means suggests an uptick in Tobin's $Q$ once the female share surpasses roughly 40\% in these sectors. Indeed, raw correlations in the Defensive group are positive, hinting that more diverse leadership teams tend to coincide with slightly higher $Q$ ratios. However, our robust inference reveals that the bar for significance is higher in this context. The concATE confidence band does not exclude zero until female representation reaches around 60\% or more in Defensive-sector firms. In other words, only when women form a substantial majority of top management do we find a clear positive effect on firm value with 95\% confidence. This implies that at the 60\% threshold there are true performance gains from diversity. One interpretation is that these traditional industries require a larger critical mass to overcome legacy cultures and realize the advantages of inclusion. When women remain a small minority (say 20--30\%), they may lack the influence or psychological safety needed to affect corporate strategy, yielding no measurable gain. By contrast, if a firm reaches 60\% female leadership (an uncommon achievement), it likely reflects deep organizational changes that unlock diversity's benefits (e.g.\ improved problem-solving, stakeholder alignment, or innovation even in mature markets). Thus, for Defensive sectors our findings suggest a delayed tipping point: meaningful performance improvements emerge only at a high level of representation, higher than in fast-paced growth industries. This result underscores how the required ``critical mass'' can vary by context.

The remaining sectors (Cyclical, Financial, Energy) either showed no robust threshold within our data range or could not be rigorously analysed due to limited support. Cyclical sectors (e.g.\ Consumer Discretionary, Industrials) have intermediate diversity levels. The naive analysis in cyclicals suggested a possible positive effect emerging at about 30\% female leadership (similar to the overall sample). Yet, using our more cautious approach, we found that the confidence intervals for the ATE in cyclicals still included zero at all feasible thresholds. In short, we cannot confirm a statistically significant benefit even if the point estimates are positive. We can, however, rule out a negative impact. This does not mean diversity has no effect in cyclical firms, but rather that the data do not provide high-confidence evidence of an effect under minimal assumptions. It is possible that unobserved factors or heavy-tailed outcomes obscure the impact in this mixed group of industries.

For Financial firms, we were unable to identify a tipping point because virtually none of the sampled banks or insurers exceeded 45--50\% female leadership during the study period. Since testing a threshold requires some treated and control firms on either side, the lack of any instances of very high diversity meant we could not apply our sequential threshold test in Financials. Interestingly, the correlation between diversity and Tobin's $Q$ in finance was negative in earlier years and then became positive toward the end of our sample (as noted in our descriptive analysis), suggesting a shifting dynamic. Our method would need a longer horizon or more variation to pin down where a critical mass effect might occur in finance, if at all.

The Energy sector remains an outlier. Energy companies not only had the lowest levels of female leadership (maxing out around 50\% in our data, with most far lower), but they also exhibited a negative raw correlation between diversity and performance. This negative association could reflect reverse causality or industry-specific factors. For instance, struggling energy firms might appoint more women to leadership in response to external pressures, creating a spurious negative link. In any case, our partial identification analysis did not find any significant positive effect of diversity in Energy. Even at the highest observed female share (just above 50\%), the ATE bounds encompass zero and even negative values. Thus, we find no evidence of a beneficial tipping point in Energy firms. We caution that this does not prove diversity harms performance in energy---only that, given the data and minimal assumptions, we cannot confirm any uplift. It is a reminder that the advantages of diversity may not be universal and could depend on complementary organizational changes, such as a culture of inclusion.

\begin{figure}[hbtp!]
 \centering
  \begin{subfigure}[hbtp!]{0.45\textwidth}
    \includegraphics[width=\linewidth]{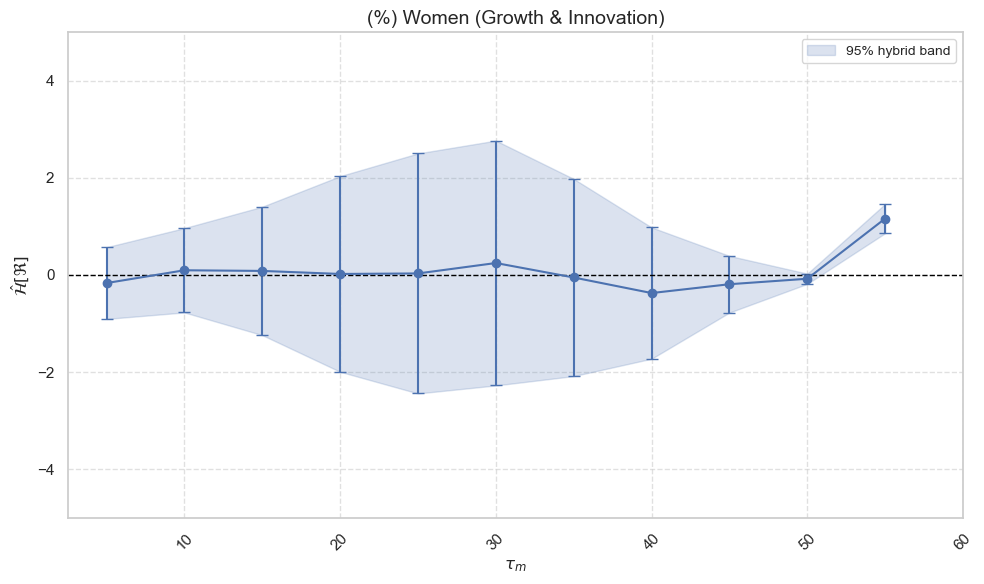}
 \caption{Hybrid: Growth \& innovation (\%) women}
  \end{subfigure}
  \hfill
  \begin{subfigure}[hbtp!]{0.45\textwidth}
    \includegraphics[width=\linewidth]{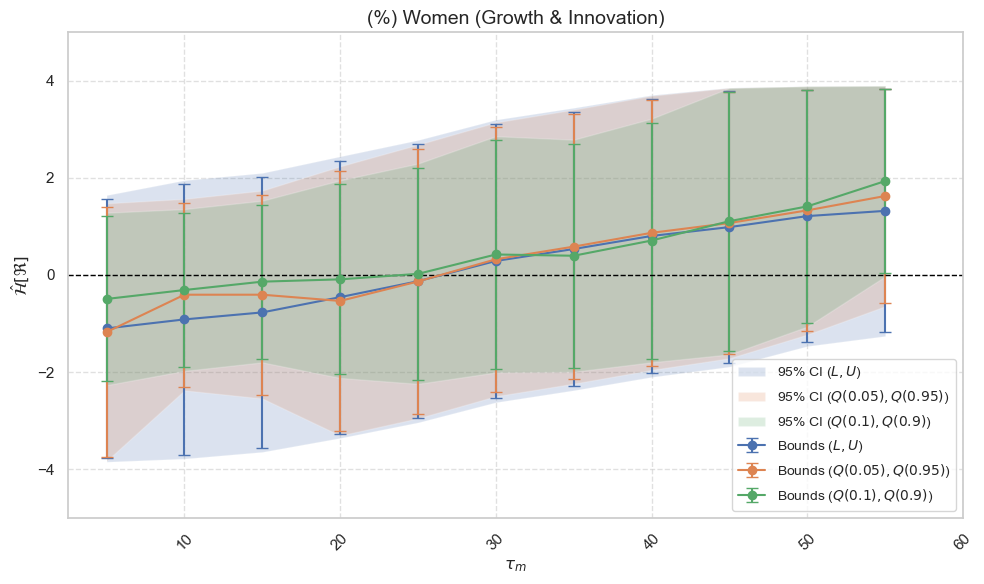}
    \caption{Manski: Growth \& innovation (\%) women}
  \end{subfigure}
  \hfill
  \caption{Hybrid and Manski Nonparametric Bounds}
\label{fig:significant_results}
\vspace{0.3em}
{\raggedright\footnotesize\textit{Note:} 
Shaded regions represent the nonparametric upper and lower bounds on the average treatment effect. Solid lines denote the midpoint between the upper and lower bounds.
\par}
\end{figure}

In summary, the causality analysis using nonparametric bounds and finite-sample confidence bands paints a more nuanced picture than the raw correlations. The data provide qualified evidence of ``tipping points'': in certain high-growth or defence sectors, reaching a critical mass of diversity (for example, women comprising about half of senior leadership roles) is associated with a reliable increase in firm value. For the remaining sectors (with the exception of the energy sector), we can conclude that diversity does not change firm value (i.e.\ it does not have any negative effects).

\FloatBarrier

\section{Concluding Remarks\label{sec:conclusion}}

This paper introduces concATE as a general framework for robust causal inference when point identification is not possible or reliable. By marrying Manski's nonparametric bounds with finite-sample concentration inequalities, concATE offers researchers a new tool to obtain ATE confidence bands without assuming away heavy-tailed outcomes or requiring strong parametric models. Our Monte Carlo results demonstrate that concATE achieves confidence bands that are 2--4 times narrower than those of \citet{imbens2004confidence}, \citet{stoye2009more}, and the nonparametric bootstrap, while maintaining valid coverage across all data generating processes considered. Moreover, the DKW-calibrated tail endpoints underlying concATE are minimax-rate-optimal among all distribution-free confidence procedures (Corollary~\ref{corr:corol2}), implying that no alternative distribution-free method can achieve a strictly faster rate of convergence while preserving coverage. The methodology's broader relevance lies in its ability to deliver valid and efficient inference under minimal assumptions (even with weakly dependent data), thereby guarding against the unnecessarily conservative intervals that can arise from plug-in methods when the outcome support is unknown or heavy-tailed.

Our empirical findings on workforce diversity illustrate the importance of such rigorous inference. While na\"ive regressions might suggest that even modest increases in female leadership yield significant gains, the concATE approach paints a more nuanced picture. We find that substantive benefits of gender diversity materialise only once a sufficient representation level is achieved. In practice, this means token diversity---for example, a lone woman or two in senior leadership---is unlikely to drive measurable performance improvement. By contrast, reaching a critical mass of women in leadership (roughly half or more in growth-oriented industries, and a clear majority in others) is associated with a reliably positive impact on firm value. These conclusions align with the critical mass hypothesis: diversity can boost performance, but only after crossing a threshold that moves an organisation beyond tokenism \citep{kanter1977some}.

By confirming this pattern under stringent inference, our study provides guidance for firms and policymakers. It emphasises that real gains from diversity require either significant numbers of women or, alternatively, substantial inclusion efforts. Indeed, an inclusive organisational culture may allow firms to reap performance gains at lower diversity levels than this critical mass. Evidence suggests that diversity alone is not sufficient and must be accompanied by inclusion to realise its full benefits \citep{nishii2013benefits,roberson2006disentangling,almeida2024diversity,josten2025makes}. Investigating the interplay between inclusion and diversity outcomes---and in particular whether inclusive cultures shift the tipping point downward---remains an important area for future research.

On the methodological side, several extensions merit investigation. First, combining the DKW-calibrated endpoints of concATE with the adaptive critical value of \citet{imbens2004confidence} could yield further efficiency gains by sharpening inference on the smooth components while retaining finite-sample tail protection. Second, practical guidance on estimating the mixing parameter $C_\upalpha$ from data, rather than relying on known dependence structures, would broaden the applicability of the method to general panel settings. Finally, our application demonstrates how concATE can be deployed in other domains, such as programme evaluation, environmental policy, or clinical trials, to uncover robust causal insights where traditional methods may produce misleadingly precise conclusions.

\section*{Acknowledgments}
This work is supported by the Gnanalingam family through their sponsorship of the G\&G Hub at The Inclusion Initiative, The London School of Economics and Political Science, and the Economic and Social Research Council (ESRC) under the ``Diversity and Productivity: from Education to Work'' (DAPEW) project [Grant Ref: ES/W010224/1].

\FloatBarrier

\bibliographystyle{apalike}
\bibliography{references}

@article{kuzmina2021gender,
  title={Gender diversity in corporate boards: Evidence from quota-implied discontinuities},
  author={Kuzmina, Olga and Melentyeva, Valentina},
  journal={Available at SSRN 3904886},
  year={2021}
}

@article{flabbi2019female,
  title={Do female executives make a difference? The impact of female leadership on gender gaps and firm performance},
  author={Flabbi, Luca and Macis, Mario and Moro, Andrea and Schivardi, Fabiano},
  journal={The Economic Journal},
  volume={129},
  number={622},
  pages={2390--2423},
  year={2019},
  publisher={Oxford University Press}
}

@article{buhlmann1998sieve,
  title={Sieve bootstrap for smoothing in nonstationary time series},
  author={B{\"u}hlmann, Peter},
  journal={The Annals of Statistics},
  volume={26},
  number={1},
  pages={48--83},
  year={1998},
  publisher={Institute of Mathematical Statistics}
}

@article{beaa45ee-b909-38c3-b8e5-43f058e8bc30,
 ISSN = {00063444},
 URL = {http://www.jstor.org/stable/2333274},
 author = {H. O. Lancaster},
 journal = {Biometrika},
 number = {1/2},
 pages = {289--292},
 publisher = {[Oxford University Press, Biometrika Trust]},
 title = {Some Properties of the Bivariate Normal Distribution Considered in the Form of a Contingency Table},
 urldate = {2026-04-03},
 volume = {44},
 year = {1957}
}

@article{doi:10.1137/1105018,
author = {Kolmogorov, A. N. and Rozanov, Yu. A.},
title = {On Strong Mixing Conditions for Stationary Gaussian Processes},
journal = {Theory of Probability \& Its Applications},
volume = {5},
number = {2},
pages = {204-208},
year = {1960},
doi = {10.1137/1105018},

URL = { 
    
        https://doi.org/10.1137/1105018
    
    

},
eprint = { 
    
        https://doi.org/10.1137/1105018
    
    

}
}

@article{bradley2005basic,
  title={Basic properties of strong mixing conditions. A survey and some open questions},
  author={Bradley, Richard C},
  year={2005}
}

@article{armstrong2021finite,
  title={Finite-Sample Optimal Estimation and Inference on Average Treatment Effects Under Unconfoundedness},
  author={Armstrong, Timothy B and Koles{\'a}r, Michal},
  journal={Econometrica},
  volume={89},
  number={3},
  pages={1141--1177},
  year={2021},
  publisher={Wiley Online Library}
}

@article{cox2023simple,
  title={Simple adaptive size-exact testing for full-vector and subvector inference in moment inequality models},
  author={Cox, Gregory and Shi, Xiaoxia},
  journal={The Review of Economic Studies},
  volume={90},
  number={1},
  pages={201--228},
  year={2023},
  publisher={Oxford University Press}
}

@article{song2024categorical,
  title={The categorical instrumental variable model: Characterization, partial identification, and statistical inference},
  author={Song, Yilin and Guo, F Richard and Chan, KC and Richardson, Thomas S},
  journal={arXiv preprint arXiv:2405.09510},
  year={2024}
}

@techreport{li2022finite,
  title={Finite Sample Inference in Incomplete Models},
  author={Li, Lixiong and Henry, Marc},
  year={2022},
  institution={Working Paper}
}

@incollection{MOLINARI2020355,
title = {Chapter 5 - Microeconometrics with partial identification},
editor = {Steven N. Durlauf and Lars Peter Hansen and James J. Heckman and Rosa L. Matzkin},
series = {Handbook of Econometrics},
publisher = {Elsevier},
volume = {7},
pages = {355-486},
year = {2020},
booktitle = {Handbook of Econometrics, Volume 7A},
issn = {1573-4412},
doi = {https://doi.org/10.1016/bs.hoe.2020.05.002},
url = {https://www.sciencedirect.com/science/article/pii/S1573441220300027},
author = {Francesca Molinari}
}

@article{d3fc5a28-c5bb-3e85-a388-e8132ffff151,
 ISSN = {00346527, 1467937X},
 URL = {http://www.jstor.org/stable/40247633},
 author = {David S. Lee},
 journal = {The Review of Economic Studies},
 number = {3},
 pages = {1071--1102},
 publisher = {[Oxford University Press, The Review of Economic Studies, Ltd.]},
 title = {Training, Wages, and Sample Selection: Estimating Sharp Bounds on Treatment Effects},
 urldate = {2026-03-30},
 volume = {76},
 year = {2009}
}

@article{bretagnolle1979estimation,
  title={Estimation des densit{\'e}s: risque minimax},
  author={Bretagnolle, Jean and Huber, Catherine},
  journal={Zeitschrift f{\"u}r Wahrscheinlichkeitstheorie und verwandte Gebiete},
  volume={47},
  number={2},
  pages={119--137},
  year={1979},
  publisher={Springer}
}

@incollection{tsybakov2008nonparametric,
  title={Nonparametric estimators},
  author={Tsybakov, Alexandre B},
  booktitle={Introduction to Nonparametric Estimation},
  pages={1--76},
  year={2008},
  publisher={Springer}
}

@article{imbens2004confidence,
  title={Confidence intervals for partially identified parameters},
  author={Imbens, Guido W and Manski, Charles F},
  journal={Econometrica},
  volume={72},
  number={6},
  pages={1845--1857},
  year={2004},
  publisher={Wiley Online Library}
}

@article{stoye2009more,
  title={More on confidence intervals for partially identified parameters},
  author={Stoye, J{\"o}rg},
  journal={Econometrica},
  volume={77},
  number={4},
  pages={1299--1315},
  year={2009},
  publisher={Wiley Online Library}
}

@article{chernozhukov2007estimation,
  title={Estimation and confidence regions for parameter sets in econometric models 1},
  author={Chernozhukov, Victor and Hong, Han and Tamer, Elie},
  journal={Econometrica},
  volume={75},
  number={5},
  pages={1243--1284},
  year={2007},
  publisher={Wiley Online Library}
}

@article{rosen2008confidence,
  title={Confidence sets for partially identified parameters that satisfy a finite number of moment inequalities},
  author={Rosen, Adam M},
  journal={Journal of Econometrics},
  volume={146},
  number={1},
  pages={107--117},
  year={2008},
  publisher={Elsevier}
}

@article{romano2008inference,
  title={Inference for identifiable parameters in partially identified econometric models},
  author={Romano, Joseph P and Shaikh, Azeem M},
  journal={Journal of Statistical Planning and Inference},
  volume={138},
  number={9},
  pages={2786--2807},
  year={2008},
  publisher={Elsevier}
}

@article{chernozhukov2013intersection,
  title={Intersection bounds: Estimation and inference},
  author={Chernozhukov, Victor and Lee, Sokbae and Rosen, Adam M},
  journal={Econometrica},
  volume={81},
  number={2},
  pages={667--737},
  year={2013},
  publisher={Wiley Online Library}
}

@article{andrews2010inference,
  title={Inference for parameters defined by moment inequalities using generalized moment selection},
  author={Andrews, Donald WK and Soares, Gustavo},
  journal={Econometrica},
  volume={78},
  number={1},
  pages={119--157},
  year={2010},
  publisher={Wiley Online Library}
}

@article{andrews2009validity,
  title={Validity of subsampling and “plug-in asymptotic” inference for parameters defined by moment inequalities},
  author={Andrews, Donald WK and Guggenberger, Patrik},
  journal={Econometric Theory},
  volume={25},
  number={3},
  pages={669--709},
  year={2009},
  publisher={Cambridge University Press}
}

@article{bugni2017inference,
  title={Inference for subvectors and other functions of partially identified parameters in moment inequality models},
  author={Bugni, Federico A and Canay, Ivan A and Shi, Xiaoxia},
  journal={Quantitative Economics},
  volume={8},
  number={1},
  pages={1--38},
  year={2017},
  publisher={Wiley Online Library}
}

@book{honore2017advances,
  title={Advances in Economics and Econometrics: Volume 2: Eleventh World Congress},
  author={Honor{\'e}, Bo and Pakes, Ariel and Piazzesi, Monika and Samuelson, Larry},
  volume={59},
  year={2017},
  publisher={Cambridge University Press}
}

@article{kaido2019confidence,
  title={Confidence intervals for projections of partially identified parameters},
  author={Kaido, Hiroaki and Molinari, Francesca and Stoye, J{\"o}rg},
  journal={Econometrica},
  volume={87},
  number={4},
  pages={1397--1432},
  year={2019},
  publisher={Wiley Online Library}
}

@article{pocock1977group,
  title={Group sequential methods in the design and analysis of clinical trials},
  author={Pocock, Stuart J},
  journal={Biometrika},
  volume={64},
  number={2},
  pages={191--199},
  year={1977},
  publisher={Oxford University Press}
}

@article{josten2025makes,
  title={What makes an individual inclusive of others? Development of the individual inclusiveness inventory},
  author={Josten, Cecily and Lordan, Grace},
  journal={Frontiers in Psychology},
  volume={16},
  pages={1473120},
  year={2025},
  publisher={Frontiers Media SA}
}

@article{nishii2013benefits,
  title={The benefits of climate for inclusion for gender-diverse groups},
  author={Nishii, Lisa H},
  journal={Academy of Management journal},
  volume={56},
  number={6},
  pages={1754--1774},
  year={2013},
  publisher={Academy of Management Briarcliff Manor, NY}
}

@article{almeida2024diversity,
  title={Diversity, equity and inclusion is not bad for business: Evidence from employee review data for companies listed in the UK and the US},
  author={Almeida, Teresa and Dayan, Yehuda and Krause, Helen and Lordan, Grace and Theodoulou, Andreas},
  year={2024},
  publisher={London School of Economics and Political Science}
}

@article{roberson2006disentangling,
  title={Disentangling the meanings of diversity and inclusion in organizations},
  author={Roberson, Quinetta M},
  journal={Group \& organization management},
  volume={31},
  number={2},
  pages={212--236},
  year={2006},
  publisher={Sage Publications Sage CA: Thousand Oaks, CA}
}

@article{o1979multiple,
  title={A multiple testing procedure for clinical trials},
  author={O'Brien, Peter C and Fleming, Thomas R},
  journal={Biometrics},
  pages={549--556},
  year={1979},
  publisher={JSTOR}
}

@article{gordon1983discrete,
  title={Discrete sequential boundaries for clinical trials},
  author={Gordon Lan, KK and DeMets, David L},
  journal={Biometrika},
  volume={70},
  number={3},
  pages={659--663},
  year={1983},
  publisher={Oxford University Press}
}

@book{siegmund2013sequential,
  title={Sequential analysis: tests and confidence intervals},
  author={Siegmund, David},
  year={2013},
  publisher={Springer Science \& Business Media}
}

@article{dvoretzky1956asymptotic,
  title={Asymptotic minimax character of the sample distribution function and of the classical multinomial estimator},
  author={Dvoretzky, Aryeh and Kiefer, Jack and Wolfowitz, Jacob},
  journal={The Annals of Mathematical Statistics},
  pages={642--669},
  year={1956},
  publisher={JSTOR}
}

@book{van2000asymptotic,
  title={Asymptotic statistics},
  author={Van der Vaart, Aad W},
  volume={3},
  year={2000},
  publisher={Cambridge university press}
}

@book{de2006extreme,
  title={Extreme value theory: an introduction},
  author={De Haan, Laurens and Ferreira, Ana},
  year={2006},
  publisher={Springer}
}

@book{van1996weak,
  title={Weak convergence},
  author={Van Der Vaart, Aad W and Wellner, Jon A and van der Vaart, Aad W and Wellner, Jon A},
  year={1996},
  publisher={Springer}
}

@article{dedecker2007empirical,
  title={The empirical distribution function for dependent variables: asymptotic and nonasymptotic results in},
  author={Dedecker, J{\'e}r{\^o}me and Merlev{\`e}de, Florence},
  journal={ESAIM: Probability and Statistics},
  volume={11},
  pages={102--114},
  year={2007},
  publisher={EDP Sciences}
}

@article{rio2000inegalites,
  title        = {Inégalités de Hoeffding pour les fonctions lipschitziennes de suites dépendantes},
  author       = {Rio, Emmanuel},
  journal      = {C. R. Acad. Sci. Paris Sér. I Math.},
  volume       = {330},
  number       = {10},
  pages        = {905--908},
  year         = {2000},
  publisher    = {Elsevier},
  doi          = {10.1016/S0764-4442(00)00290-1}
}

@incollection{merlevede2009bernstein,
  title        = {Bernstein inequality and moderate deviations under strong mixing conditions},
  author       = {Merlev{\`e}de, Florence and Peligrad, Magda and Rio, Emmanuel},
  booktitle    = {High Dimensional Probability V: The Luminy Volume},
  series       = {Institute of Mathematical Statistics Collections},
  volume       = {5},
  pages        = {273--292},
  year         = {2009},
  publisher    = {IMS},
  doi          = {10.1214/09-IMSCOLL511},
  url          = {https://projecteuclid.org/euclid.imsc/1257862903}
}

@book{white2014asymptotic,
  title={Asymptotic theory for econometricians},
  author={White, Halbert},
  year={2014},
  publisher={Academic press}
}

@book{kosorok2008introduction,
  title={Introduction to empirical processes and semiparametric inference},
  author={Kosorok, Michael R},
  volume={61},
  year={2008},
  publisher={Springer}
}

@article{hoeffding1994probability,
  title={Probability inequalities for sums of bounded random variables},
  author={Hoeffding, Wassily},
  journal={The collected works of Wassily Hoeffding},
  pages={409--426},
  year={1994},
  publisher={Springer}
}

@book{vershynin2018high,
  title={High-dimensional probability: An introduction with applications in data science},
  author={Vershynin, Roman},
  volume={47},
  year={2018},
  publisher={Cambridge university press}
}

@book{scott2015multivariate,
  title={Multivariate density estimation: theory, practice, and visualization},
  author={Scott, David W},
  year={2015},
  publisher={John Wiley \& Sons}
}

@book{casella2024statistical,
  title={Statistical inference},
  author={Casella, George and Berger, Roger},
  year={2024},
  publisher={CRC press}
}

@article{horowitz1998censoring,
  title={Censoring of outcomes and regressors due to survey nonresponse: Identification and estimation using weights and imputations},
  author={Horowitz, Joel L and Manski, Charles F},
  journal={Journal of Econometrics},
  volume={84},
  number={1},
  pages={37--58},
  year={1998},
  publisher={Elsevier}
}

@book{angrist2009mostly,
  title={Mostly harmless econometrics: An empiricist's companion},
  author={Angrist, Joshua D and Pischke, J{\"o}rn-Steffen},
  year={2009},
  publisher={Princeton university press}
}

@article{ali2011gender,
  title={The gender diversity--performance relationship in services and manufacturing organizations},
  author={Ali, Muhammad and Kulik, Carol T and Metz, Isabel},
  journal={The International Journal of Human Resource Management},
  volume={22},
  number={07},
  pages={1464--1485},
  year={2011},
  publisher={Taylor \& Francis}
}

@article{neweywest1987,
  author       = {Newey, Whitney K. and West, Kenneth D.},
  title        = {A Simple, Positive Semi-Definite, Heteroskedasticity and Autocorrelation Consistent Covariance Matrix},
  journal      = {Econometrica},
  volume       = {55},
  number       = {3},
  pages        = {703--708},
  year         = {1987},
  publisher    = {Econometric Society},
  doi          = {10.2307/1913610},
  url          = {https://www.jstor.org/stable/1913610}
}

@article{torchia2011women,
  title={Women directors on corporate boards: From tokenism to critical mass},
  author={Torchia, Mariateresa and Calabr{\`o}, Andrea and Huse, Morten},
  journal={Journal of business ethics},
  volume={102},
  pages={299--317},
  year={2011},
  publisher={Springer}
}

@article{kanter1987men,
  title={Men and Women of the Corporation Revisited.},
  author={Kanter, Rosabeth Moss},
  journal={Management Review},
  volume={76},
  number={3},
  year={1987}
}

@article{kanter1977some,
  title={Some effects of proportions on group life: Skewed sex ratios and responses to token women},
  author={Kanter, Rosabeth Moss},
  journal={American journal of Sociology},
  volume={82},
  number={5},
  pages={965--990},
  year={1977},
  publisher={University of Chicago Press}
}

@article{nathan2013cultural,
  title={Cultural diversity, innovation, and entrepreneurship: firm-level evidence from London},
  author={Nathan, Max and Lee, Neil},
  journal={Economic geography},
  volume={89},
  number={4},
  pages={367--394},
  year={2013},
  publisher={Taylor \& Francis}
}

@article{hoogendoorn2013impact,
  title={The impact of gender diversity on the performance of business teams: Evidence from a field experiment},
  author={Hoogendoorn, Sander and Oosterbeek, Hessel and Van Praag, Mirjam},
  journal={Management science},
  volume={59},
  number={7},
  pages={1514--1528},
  year={2013},
  publisher={INFORMS}
}

@article{adams2009women,
  title={Women in the boardroom and their impact on governance and performance},
  author={Adams, Ren{\'e}e B and Ferreira, Daniel},
  journal={Journal of financial economics},
  volume={94},
  number={2},
  pages={291--309},
  year={2009},
  publisher={Elsevier}
}

@article{post2015women,
  title={Women on boards and firm financial performance: A meta-analysis},
  author={Post, Corinne and Byron, Kris},
  journal={Academy of management Journal},
  volume={58},
  number={5},
  pages={1546--1571},
  year={2015},
  publisher={Academy of Management Briarcliff Manor, NY}
}

@article{ostergaard2011does,
  title={Does a different view create something new? The effect of employee diversity on innovation},
  author={{\O}stergaard, Christian R and Timmermans, Bram and Kristinsson, Kari},
  journal={Research policy},
  volume={40},
  number={3},
  pages={500--509},
  year={2011},
  publisher={Elsevier}
}

@book{manski2003partial,
  title={Partial identification of probability distributions},
  author={Manski, Charles F},
  year={2003},
  publisher={Springer Science \& Business Media}
}

@article{manski1990nonparametric,
  title={Nonparametric bounds on treatment effects},
  author={Manski, Charles F},
  journal={The American Economic Review},
  volume={80},
  number={2},
  pages={319--323},
  year={1990},
  publisher={JSTOR}
}

@article{tobin1978monetary,
  title={Monetary policies and the economy: the transmission mechanism},
  author={Tobin, James},
  journal={Southern economic journal},
  pages={421--431},
  year={1978},
  publisher={JSTOR}
}

@article{safiullah2022gender,
  title={Gender diversity on corporate boards, firm performance, and risk-taking: New evidence from Spain},
  author={Safiullah, Md and Akhter, Tanzina and Saona, Paolo and Azad, Md Abul Kalam},
  journal={Journal of Behavioral and Experimental Finance},
  volume={35},
  pages={100721},
  year={2022},
  publisher={Elsevier}
}

@article{tobin1976asset,
  title={Asset markets and the cost of capital},
  author={Tobin, James and Brainard, William C},
  year={1976}
}

@article{brainard1968pitfalls,
  title={Pitfalls in financial model building},
  author={Brainard, William C and Tobin, James},
  journal={The American economic review},
  volume={58},
  number={2},
  pages={99--122},
  year={1968},
  publisher={JSTOR}
}

@article{tobin1969general,
  title={A general equilibrium approach to monetary theory},
  author={Tobin, James},
  journal={Journal of money, credit and banking},
  volume={1},
  number={1},
  pages={15--29},
  year={1969},
  publisher={JSTOR}
}

@article{hambrick1984upper,
  title={Upper echelons: The organization as a reflection of its top managers},
  author={Hambrick, Donald C and Mason, Phyllis A},
  journal={Academy of management review},
  volume={9},
  number={2},
  pages={193--206},
  year={1984},
  publisher={Academy of Management Briarcliff Manor, NY 10510}
}

@article{massart1990tight,
  title={The tight constant in the Dvoretzky-Kiefer-Wolfowitz inequality},
  author={Massart, Pascal},
  journal={The annals of Probability},
  pages={1269--1283},
  year={1990},
  publisher={JSTOR}
}
\begin{appendices}          
\renewcommand{\thesection}{\Alph{section}}   
\section{Lemmas}
In this section we first state a lemma that justifies the quantile “sandwich” bound for the empirical distribution function used in our nonparametric bounds for the latent conditional expectations. We then collect the lemmas used in the proofs of the finite-sample propositions and corollaries. The first set of lemmas in relation to the latter corresponds to Assumption~\ref{ass:iid}, where the data is assumed to be independent and drawn from a sub-exponential distribution. The second set pertains to Assumption~\ref{ass:mixing}, which allows for weakly dependent data.

\begin{lemma}[Quantile sandwich from uniform CDF control]
\label{lem:quantilesando}
Let $F$ be a CDF and $\hat F_N$ its empirical CDF. On the event
$\sup_y|\hat F_N(y)-F(y)|\le\varepsilon$, for every $u\in[\varepsilon,1-\varepsilon]$,
\[
F^{-1}(u-\varepsilon)\ \le\ \hat F_N^{-1}(u)\ \le\ F^{-1}(u+\varepsilon),
\]
where $G^{-1}(u)=\inf\{y:\,G(y)\ge u\}$. In particular,
$\hat F_N^{-1}(u)=Y_{(\lceil uN\rceil)}$ is the $\lceil uN\rceil$th order statistic.
\end{lemma}

\subsection{Independent Data}
In what follows, we introduce the lemmas that provide the concentration inequalities for the estimators and latent quantities involved in the nonparametric bounds. The generalized Bernstein inequality for sub-exponential variables is taken from \citet{vershynin2018high}, the Hoeffding bound for Bernoulli random variables from \citet{hoeffding1994probability}, and the Dvoretzky-Kiefer-Wolfowitz inequality from \citet{kosorok2008introduction}.
\begin{lemma}[Bernstein inequality for i.i.d. data]
\label{lem:bernstein}
Let $\widetilde{Y}_{1},\ldots,\widetilde{Y}_{n}$ be independent, mean‑zero,
sub‑exponential random variables and set
\[
S_{n}:=\sum_{i=1}^{n}\widetilde{Y}_{i}.
\]
Then for every $t\ge 0$,
\begin{equation}
\Pr\left(\lvert n^{-1}S_{n}\rvert\ge t\right)
\;\le\;
2\exp\left(
        -cn\,\min\left\{
           \frac{t^{2}}
                {\left(\max_i\lVert \widetilde{Y}_i \rVert_{\psi_1}\right)^2},
           \;
           \frac{t}
                {\max_{i}\lVert \widetilde{Y}_i\rVert_{\psi_1}}
        \right\}
      \right),
\end{equation}
where $c>0$ is an absolute constant and
\[
\|X\|_{\psi_{1}}
\;:=\;
\inf\left\{s>0 : \E\exp\left(|X|/s\right)\le 2\right\}
\]
denotes the sub‑exponential (Orlicz) norm of a real random variable $X$.
\end{lemma}

\begin{lemma}[Dvoretzky-Kiefer-Wolfowitz inequality]
\label{lem:dkw}
Let $Y_{1},\ldots,Y_{n}$ be real-valued independent random variables with cumulative distribution function $F(.)$. Further denote $F_n(.)$ the empirical distribution function defined by
\begin{equation}\label{eq:empirical}
F_n(x)=\frac{1}{n}\sum\limits_{i=1}^n\mathbbm{1}_{\{X_i\leq x\}},\quad x\in\R
\end{equation}
then for every $t>0$,
\begin{equation}\label{eq:dkw}
\Pr\left(\sup_{y\in\R}\lvert F_n(y)-F(y)\rvert> t \right)\leq 2\exp\left(-2nt^2\right).
\end{equation}
\end{lemma}

\begin{lemma}[Hoeffding inequality for Bernoulli random variables]
\label{lem:hoeffing}
Let $Z_1,\cdots,Z_n$ be independent $Bernoulli(p)$ random variables with $\hat{p}=\frac{1}{n}\sum Z_i$, Since $0\leq Z_i\leq1$, Hoeffding (1963, Theorem 2)  for any $t>0$, gives
\begin{equation}
\Pr\left(\lvert\hat{p}-p\rvert\geq t\right)\leq\exp\left(-2nt^2\right).
\end{equation} 
\end{lemma}

\subsection{Weakly Dependent Data}
In this section, we present the definitions and lemmas relevant to weakly dependent data. The definition of the $\alpha$-mixing process, as well as the concentration inequalities used to derive nonparametric bounds for weakly dependent data drawn from sub-exponential distributions, are drawn from \citet{white2014asymptotic}, \citet{merlevede2009bernstein}, \citet{rio2000inegalites} and \citet{dedecker2007empirical}.

\begin{definition}[$\upalpha$-mixing process]\label{def:mixing}
Let the sequence of random variables $\widetilde{Y}_1,\cdots,\widetilde{Y}_n$ be defined on the filtered probability space $(\Omega, \mathcal{F},\left(\mathcal{F}_{t}\right)_{t\geq 0},\mathbb{P})$, where $\mathcal{F}_{t}=\sigma(\widetilde{Y}_1,\cdots,\widetilde{Y}_t)$ is the $\sigma$-field spanned by $\{\widetilde{Y}_i\}_{i=1}^t$. Additionally, let $\mathcal{G}$ and $\mathcal{H}$ be two $\sigma$-fields such that $\mathcal{G},\mathcal{H}\subset \mathcal{F}$ and define
\begin{equation}
\upalpha(\mathcal{G},\mathcal{H})=\sup_{G \in \mathcal{G},H\in \mathcal{H}}\left\{\lvert \Pr(G\cap H)-\Pr(G)\Pr(H) \rvert \right\}
\end{equation}
and define the Borel $\sigma$-field $\mathcal{B}_{1}^{m}=\sigma(\widetilde{Y}_1,\cdots,\widetilde{Y}_m)$ and the $\upalpha$-mixing coefficient $\beta(k)$ as 
\begin{equation}\label{eq:strongmixing}
\upalpha(k)\equiv \sup_m\upalpha(\mathcal{B}_1^m,\mathcal{B}_{m+k}^n)
\end{equation}
If for the sequence $\{\widetilde{Y}_t\}$, $\upalpha(k)\to 0$ as $k\to\infty$, ${\widetilde{Y}_t}$ is called $\upalpha$-mixing.
\end{definition}

\begin{lemma}[Bernstein inequality for weakly dependent data]
\label{lem:bernsteinweak}
Let $\widetilde{Y}_{1},\ldots,\widetilde{Y}_{n}$ be mean-zero,
real-valued random variables drawn from a subexponential distributions that satisfy the $\upalpha$-mixing condition with exponential decay. Moreover, for any positive $M$, let $\varphi_M(x)=(x\vee M)\wedge (-M)$ and define $V$ as,
\begin{equation}
V=\sup_{M\geq 1}\sup_{i> 0}\left(Var(\varphi_M(\widetilde{Y}_i))+2\sum_{j>1}\lvert cov(\varphi_M(\widetilde{Y}_i),\varphi_M(\widetilde{Y}_j))\rvert\right)<\infty.
\end{equation}
Further, define:
\[
S_{n}:=\sum_{i=1}^{n}\widetilde{Y}_{i}.
\]
Then for every $n \geq 4$ and $t>0$, and for positive constants $C_1$, $C_2$, $C_3$, $C_4$ depending only on $c$, $\gamma$ and $\gamma_1$, we have
\begin{align*}
\Pr\left(\lvert n^{-1} S_j\rvert \geq t\right)&\leq\Pr\left(\sup_{j\leq n}\lvert n^{-1}S_j\rvert \geq t\right)\\
&\leq n\exp\left(-\frac{(nt)^{\gamma}}{C_1}\right)+\exp\left(-\frac{(n t)^2}{C_2(1+nV)}\right)\\
&\qquad+\exp\left(-\frac{(nt)^2}{C_3n}\exp\left(\frac{(nt)^{\gamma(1-\gamma)}}{C_4(\log nt)^\gamma}\right)\right) 
\end{align*}
\end{lemma}

\begin{lemma}[Dvoretzky-Kiefer-Wolfowitz inequality for weakly dependent data]\label{lem:dkw-dep}
Let $Y_1,\cdots,Y_n$ be a strictly stationary real-valued sequence with common CDF. $F$ and assume the strong mixing coefficients $\upalpha(k)$ in \eqref{eq:strongmixing} satisfies $\sum_{k\geq 1}\upalpha(k)^{1/2}<\infty$. Define the empirical CDF as per Eq.~\eqref{eq:empirical}. Then for every $t>0$ and $n\geq 1$
\begin{equation}\label{eq:dkw-dep-proof}
P\left(\sup_{y\in\R}\lvert F_n(y)-F(y)\rvert>t\right)\leq 2\exp\left(-\frac{nt^2}{2(1+4C_{\upalpha})^2}\right)
\end{equation}
where $C_{\upalpha}=\sum\limits_{k\geq 1}\upalpha(k)^{1/2}<\infty$. In particular, if $\upalpha(k)=0$ for all $k\geq1$ (the independent case) then $C_{\upalpha}=0$ and \eqref{eq:dkw-dep-proof} reduces to $2e^{-nt^2/2}$, which is non-sharp relative to \citet{massart1990tight}.
\end{lemma}

\begin{lemma}[Hoeffding inequality for $\upalpha$-mixing Bernoulli data]\label{lem:hoeffdep}
Let $Z_1,\cdots,Z_n$ be a strictly stationary $\{0,1\}$-valued sequence with $p=\E[Z_1]$ and $\hat{p}_n=\frac{1}{n}\sum_{i=1}^nZ_i$, and strong-mixing coefficients $\upalpha(k)$. Assume $C_\upalpha =\sum_{k\ge1}\upalpha(k)^{1/2}<\infty$. Then for every $t>0$ and $n\ge1$,
\begin{equation}
\Pr\left(|\hat p_n-p|\ge t\right)\le2\exp\left(-\frac{nt^2}{2(1+4C_\upalpha)^2}\right).
\end{equation}
In particular, if $\upalpha(k)\equiv0$ then $C_\upalpha=0$ and this reduces to the usual Azuma-Hoeffding bound
$\Pr(\lvert\hat p_n-p\rvert\ge t)\le2e^{-n t^2/2}$.
\end{lemma}

\section{Proofs}

\subsection{Proof of Lemma \ref{lem:quantilesando}}
Fix $u\in[\varepsilon,1-\varepsilon]$ and $y_u:=\hat F_N^{-1}(u)$. Then $\hat F_N(y_u)\ge u$.
Hence $F(y_u)\ge \hat F_N(y_u)-\varepsilon\ge u-\varepsilon$, so
$F^{-1}(u-\varepsilon)\le y_u$. For any $y<y_u$, $\hat F_N(y)<u$, hence
$F(y)\le \hat F_N(y)+\varepsilon<u+\varepsilon$, so $F^{-1}(u+\varepsilon)\ge y_u$.

\subsection{Proof of Lemma \ref{lem:dkw-dep}}

ion of Lemma~\ref{lem:bernsteinweak} stemming from \citet{merlevede2009bernstein}.

\paragraph{\textit{(i) Pointwise application of Lemma~\ref{lem:bernsteinweak}:}} For each fixed $t$, define $\widetilde{Y}_i(t):=\mathbbm{1}\{X_i\leq t\}-F(t)$. These are mean-zero, bounded $(\lvert\widetilde{Y}_i\rvert\leq 1$, hence sub-exponential), and $\upalpha$-mixing with the same coefficients as $X_i$. The long-run variance satisfies:
\begin{equation}
V(t)=\mathrm{Var}\left(\widetilde{Y}_i(t)\right)+2\sum_{j>1}\Big\vert \mathrm{Cov}\left(\widetilde{Y}_i(t),\widetilde{Y}_j(t)\right) \Big\vert\leq \frac{1}{4}+2\sum_{k\geq 1}\upalpha(k)
\end{equation}
uniformly over $t$. Consequently, Lemma~\ref{lem:bernsteinweak} applies with a $V$ that does not depend on $t$. $\widetilde{Y}_i(t)$ is a centered Bernoulli, as such $\mathrm{Var}(\widetilde{Y}_i(t))=F(t)(1-F(t))$. The function $p(1-p)$ on $[0,1]$ is a downward parabola maximized at $p=1/2$, giving $p(1-p)=1/4$. The $\mathrm{Cov}$ term on the other hand, can be expressed as:
\begin{align}
\mathrm{Cov}(\widetilde{Y}_i(t),\widetilde{Y}_j(t))&=\E\left[\widetilde{Y}_i(t)\widetilde{Y}_j(t)\right]\\
&=\E\left[\left(\mathbbm{1}\{X_i\leq t\}-F(t)\right)\left(\mathbbm{1}\{X_j\leq t\}-F(t)\right)\right]\\
&=\E\left[\left(\mathbbm{1}\{X_i\leq t\}\right)\left(\mathbbm{1}\{X_j\leq t\}\right)\right]-F(t)^2\\
&=\Pr\left(X_i\leq t,X_j\leq t\right)-F(t)^2
\end{align}
Now looking at the $\upalpha$-mixing definition~\eqref{eq:strongmixing}, we know that $\upalpha(k)=\sup_m \upalpha(\mathcal{B}_1^m,\mathcal{B}^n_{m+k})$, where $\upalpha(\mathcal{A},\mathcal{B})=\sup_{A\in\mathcal{A},B\in\mathcal{B}}\lvert P(A\cap B)-P(A)P(B)\rvert$. The event $\{X_i\leq t\}\in \mathcal{B}_1^i$ and the event $\{X_j\leq t\} \in\mathcal{B}_j^n$. So if $j\geq i+k$,
\begin{equation}
\lvert \Pr(X_i\leq t,X_j\leq t)-P(X_i\leq t)P(X_j\leq t)\rvert \leq \upalpha(j-i).
\end{equation}
Hence, $\lvert \mathrm{Cov}(\widetilde{Y}_i(t),\widetilde{Y}_j(t))\rvert\leq \upalpha(j-i)$. Now for Lemma~\ref{lem:bernsteinweak}, since $\lvert \widetilde{Y}_i(t)\rvert\leq 1$, the truncation $\varphi_M$ is the identity for all $M\geq 1$, so:
\begin{equation}
V(t)=\sup_{i>0}\left(\mathrm{Var}\left(\widetilde{Y}_i(t)\right)+2\sum_{j>1}\lvert \mathrm{Cov}(\widetilde{Y}_i(t),\widetilde{Y}_j(t))\rvert\right)\leq \frac{1}{4}+2\sum_{k\geq 1}\upalpha(k)
\end{equation}
uniformly in $t$.

\paragraph{\textit{(ii) Discretisation of monotone functions:}} For any $\varepsilon>0$, choose points $t_0<t_1<\cdots<t_N$ such that
\begin{equation}
F(t_{j+1})-F(t_j)\leq\varepsilon \quad\text{for all } j, \qquad F(t_0)\leq\varepsilon, \qquad 1-F(t_N)\leq\varepsilon.
\end{equation}
Such a grid always exists with $N\leq\lceil 1/\varepsilon\rceil$ points, for any CDF $F$ on $\mathbb{R}$, regardless of whether $F$ admits a density or has bounded support. Since $F_n$ and $F$ are both monotone (non-decreasing), for any $t\in[t_j,t_{j+1}]$:
\begin{align}
F_n(t)-F(t) &\leq F_n(t_{j+1})-F(t_j) \nonumber\\
&= \left[F_n(t_{j+1})-F(t_{j+1})\right] + \left[F(t_{j+1})-F(t_j)\right] \nonumber\\
&\leq \lvert F_n(t_{j+1})-F(t_{j+1})\rvert + \varepsilon,
\end{align}
and similarly $F_n(t)-F(t)\geq -\lvert F_n(t_j)-F(t_j)\rvert - \varepsilon$. For $t<t_0$, both $F_n(t)\leq F_n(t_0)$ and $F(t)\leq F(t_0)\leq\varepsilon$, so $\lvert F_n(t)-F(t)\rvert\leq\lvert F_n(t_0)-F(t_0)\rvert+\varepsilon$; the case $t>t_N$ is analogous. Hence:
\begin{equation}
\sup_{t\in\mathbb{R}}\lvert F_n(t)-F(t)\rvert\leq \max_{0\leq j\leq N}\lvert F_n(t_j)-F(t_j)\rvert+\varepsilon.
\end{equation}

\paragraph{\textit{(iii) Union bound and conclusion:}} Set $\varepsilon=s/2$, where $s>0$ is the target deviation level, so that $N\leq 2/s+1$. Then
\begin{equation}
\Pr\!\left(\sup_{t\in\mathbb{R}}\lvert F_n(t)-F(t)\rvert > s\right) \leq \left(\frac{2}{s}+2\right)\cdot\sup_{t\in\mathbb{R}}\,\Pr\!\left(\lvert F_n(t)-F(t)\rvert > \frac{s}{2}\right),
\end{equation}
where the inequality is a union bound over the $N+1\leq 2/s+2$ grid points. For each fixed $t$, $F_n(t)-F(t)=n^{-1}S_n(t)$ with $S_n(t)=\sum_{i=1}^n\widetilde{Y}_i(t)$. Since $\lvert\widetilde{Y}_i(t)\rvert\leq 1$ (hence sub-exponential with $\lVert\widetilde{Y}_i(t)\rVert_{\psi_1}$ uniformly bounded) and $V(t)\leq\tfrac{1}{4}+2C_\upalpha$ uniformly in $t$ from part~(i), Lemma~\ref{lem:bernsteinweak} gives
\begin{align}
\Pr\!\left(\lvert F_n(t)-F(t)\rvert>\frac{s}{2}\right) &\leq n\exp\!\left(-\frac{(ns/2)^\gamma}{C_1}\right) + \exp\!\left(-\frac{(ns/2)^2}{C_2(1+nV)}\right) \nonumber\\
&\quad + \exp\!\left(-\frac{(ns/2)^2}{C_3\,n}\exp\!\left(\frac{(ns/2)^{\gamma(1-\gamma)}}{C_4(\log(ns/2))^\gamma}\right)\right).
\end{align}
The dominant term for moderate deviations ($s=\Theta( n^{-1/2})$) is the Gaussian term. Using $V\leq\tfrac{1}{4}+2C_\upalpha$ and the elementary inequality $1+n(\tfrac{1}{4}+2C_\upalpha)\leq\tfrac{n}{2}(1+4C_\upalpha)^2$ for $n\geq 4$, the Gaussian term satisfies
\begin{equation}
\exp\!\left(-\frac{n^2s^2/4}{C_2(1+nV)}\right) \leq \exp\!\left(-\frac{ns^2}{2C_2(1+4C_\upalpha)^2}\right).
\end{equation}
Since $\upalpha(k)\in[0,1]$ we have $\upalpha(k)\leq\upalpha(k)^{1/2}$, so $\sum_{k\geq 1}\upalpha(k)\leq\sum_{k\geq 1}\upalpha(k)^{1/2}=C_\upalpha<\infty$. For $s\geq c_0(1+4C_\upalpha)\sqrt{(\log n)/n}$ with $c_0$ chosen large enough (depending only on $c$, $\gamma$, and $\gamma_1$), the polynomial prefactor $(2/s+2)$ and the $n$-prefactor in the first term are absorbed into the exponentials, and the third term is dominated by the Gaussian term. Writing $t$ for the deviation level, we obtain: for all $t>0$ and $n\geq 4$,
\begin{equation}
\label{eq:DKWweak}
P\left(\sup_{y\in\mathbb{R}}\lvert F_n(y)-F(y)\rvert > t\right) \leq 2\exp\left(-\frac{nt^2}{2(1+4C_\upalpha)^2}\right)
\end{equation}
where $C_\upalpha=\sum_{k\geq 1}\upalpha(k)^{1/2}<\infty$. In particular, if $\upalpha(k)=0$ for all $k\geq 1$ (the independent case) then $C_\upalpha=0$ and~\eqref{eq:DKWweak} reduces to $2e^{-nt^2/2}$, which is non-sharp relative to \citet{massart1990tight}. $\hfill\square$

\subsection{Proof of Lemma~\ref{lem:hoeffdep}}

Let $Y_i:=Z_i$, where each $Z_{i}\in\{0,1\}$. Then the CDF of $Y$ is given by
\begin{equation}
F(y)=
\begin{cases}
0, &y<0,\\
1-p,&0\leq y <1,\\
1, & y\geq 1,
\end{cases}
\end{equation}
with $p=\E[Z_1]$. For the empirical CDF, we know that
\begin{equation}
\hat{F}_n(y)=\frac{1}{n}\sum_{i=1}^n \mathbbm{1}\{Z_i\leq y\}.
\end{equation}

Now evaluate at any $y\in[0,1)$. Since $Z_i$ is Bernoulli,
\begin{equation}
\1\{Z_i\leq Y\}=\1\{Z_i=0\}=1-Z_i,
\end{equation}
so,
\begin{equation}
\hat{F}_n(y)=\frac{1}{n}\sum\limits_{i=1}^n(1-Z_i)=1-\hat{p}_n.
\end{equation}
Additionally, we have $F(y)=1-p$. Hence, 
\begin{equation}
\lvert \hat F_n(y)-F(y)\rvert=\lvert (1-p_n)-(1-p)\rvert = \lvert p_n-p\rvert.
\end{equation}
Therefore, it is evident that
\begin{equation}
\lvert p_n-p\rvert \leq \sup_{y\in\R}\lvert F_n(y)-F(y)\rvert.
\end{equation}
Consequently, 
\begin{equation}
\Pr\left(\lvert \hat{p}_n-p \rvert\right)\leq \Pr\left(\sup_{y\in\R}\lvert \hat{F}_n(y)-F(y) \rvert\right)\leq 2\exp\left(-\frac{nt^2}{2(1+4C_{\upalpha})}\right),
\end{equation}
proving the Lemma.

\subsection{Proof of Proposition \ref{prop:1}}
The theory that we have laid out thus far concerns the identification problem.
However, empirical research must also be concerned with sampling variation.
Note that the empirical counterpart of the nonparametric bound \eqref{eq:nonparametricbounds} is:

\begin{equation}\label{eq:empnonparametricbounds}
\Re \in \bigg[\hat{\delta}_1 \hat{p}_1
+ L^{(1)} \hat{p}_0- U^{(0)} \hat{p}_1
- \hat{\delta}_0 \hat{p}_0, \;\hat{\delta}_1 \hat{p}_1
+ U^{(1)} \hat{p}_0- L^{(0)} \hat{p}_1
- \hat{\delta}_0 \hat{p}_0
\bigg].
\end{equation}

For $u=m_0,\cdots,m_1$, to simultaneously obtain the $(1-\alpha_u)\%$ confidence set
for both the upper and lower bounds for the identification region \eqref{eq:empnonparametricbounds},
we must first find the confidence bands with an appropriate significance level
and combine them using Bonferroni inequalities so that the combined confidence
set has $100(1-\alpha_u)\%$ coverage, or

\begin{equation}\label{eq:coveragerate1}
\Pr\left( [l(\Re),u(\Re)] \subseteq [\mathcal{L}(\hat{\bm{\theta}}),\mathcal{U}(\hat{\bm{\theta}})] \right)
\geq 1-\alpha_u,\quad \text{with}\quad \alpha_u=\frac{\alpha}{\overline{\mathcal{M}}},
\end{equation}
where $\hat{\bm{\theta}}=(\hat{\delta}_{1},\hat{\delta}_{0},\hat{p}_1,\hat{p}_0)^{\top}$
is a $4\times 1$ vector of estimators and $\mathbb{I}(\Re)=[l(\Re),u(\Re)]$.
In other words, we wish to obtain
\begin{equation}
\Pr\left(l(\Re)\leq \mathcal{L}(\hat{\theta})\right)\geq 1-\frac{\alpha_u}{2},
\quad\text{and}\quad
\Pr\left(u(\Re)\geq \mathcal{U}(\hat{\theta})\right)\geq 1-\frac{\alpha_u}{2},
\end{equation}
such that
\begin{equation}
\Pr\left(l(\Re)\leq \mathcal{L}(\hat{\theta})\cap u(\Re)\geq \mathcal{U}(\hat{\theta})\right)
\geq 1-\alpha_u.
\end{equation}
We know from Boole's inequality that:
\begin{align}\label{eq:boole}
\begin{split}
\Pr\left(l(\Re)\leq \mathcal{L}(\hat{\theta})\cap u(\Re)\geq \mathcal{U}(\hat{\theta})\right)
&\geq 1-\Pr\left(l(\Re)> \mathcal{L}(\hat{\theta})\right)
     -\Pr\left(u(\Re)< \mathcal{U}(\hat{\theta})\right)\\
&\geq 1-\frac{\alpha_u}{2}-\frac{\alpha_u}{2}\\
&= 1-\alpha_u.
\end{split}
\end{align}
Thus,
\begin{align}
\begin{split}
\mathcal{L}(\hat{\bm{\theta}})- \Phi^{-1}(1-\alpha_u/2)\,\text{S.E.}\left(\mathcal{L}(\hat{\bm{\theta}})\right)
\leq \mathcal{L}(\bm{\theta})
\leq \mathcal{L}(\hat{\bm{\theta}})+ \Phi^{-1}(1-\alpha_u/2)\,\text{S.E.}\left(\mathcal{L}(\hat{\bm{\theta}})\right)\\
\mathcal{U}(\hat{\bm{\theta}})- \Phi^{-1}(1-\alpha_u/2)\,\text{S.E.}\left(\mathcal{U}(\hat{\bm{\theta}})\right)
\leq \mathcal{U}(\bm{\theta})
\leq \mathcal{U}(\hat{\bm{\theta}})+ \Phi^{-1}(1-\alpha_u/2)\,\text{S.E.}\left(\mathcal{U}(\hat{\bm{\theta}})\right).
\end{split}
\end{align}
It remains to find the standard errors of $\mathcal{L}(\hat{\bm{\theta}})$ and
$\mathcal{U}(\hat{\bm{\theta}})$, which is a rather tedious task due to the nonlinear nature
of the estimators. Assuming relatively large sample sizes, we may rely on the delta method. 

By definition, the consistent estimator $\hat{\theta}$ converges in probability to its
true value $\theta$, and the CLT can be applied to obtain asymptotic normality, i.e.,
\begin{equation}
\sqrt{N}(\hat{\theta}-\theta)\xrightarrow{d}N(0,\Omega),
\end{equation}
for some finite covariance matrix $\Omega$. By Taylor expansion of
$\mathcal{L}(\hat{\theta})$ and $\mathcal{U}(\hat{\theta})$:
\begin{align}
\mathcal{L}(\hat{\theta})
= \mathcal{L}(\theta)
+ \nabla \mathcal{L}(\theta)^{\top}(\hat{\theta}-\theta)
+ o_p\left(\lVert \hat\theta-\theta\rVert\right),
\label{eq:lowerbound}\\
\mathcal{U}(\hat{\theta})
= \mathcal{U}(\theta)
+ \nabla \mathcal{U}(\theta)^{\top}(\hat{\theta}-\theta)
+ o_p\left(\lVert \hat\theta-\theta\rVert\right),
\label{eq:upperbound}
\end{align}
where
\begin{equation}
\nabla \mathcal{L}(\theta)
=\left(
p_1,\;-p_0,\;\delta_{1}-U^{(0)},\;L^{(1)}-\delta_{0}
\right)^{\top},
\end{equation}
with $\nabla \mathcal{U}(\theta)$ defined similarly. Since $\hat{\theta}-\theta=O_p(N^{-1/2})$,
we have $o_p(\lVert \hat \theta-\theta\rVert)=o_p(N^{-1/2})$, and hence
\begin{align}
\mathcal{L}(\hat{\theta})-\mathcal{L}(\theta)
= \nabla \mathcal{L}(\theta)^{\top}(\hat{\theta}-\theta)+o_p( N^{-1/2}),\\
\sqrt N (\mathcal{L}(\hat{\theta})-\mathcal{L}(\theta))
= \nabla \mathcal{L}(\theta)^{\top}\sqrt N(\hat{\theta}-\theta)+o_p(1).
\end{align}
Therefore,
\begin{equation}
\sqrt N (\mathcal{L}(\hat{\theta})-\mathcal{L}(\theta))
\xrightarrow{d}
N\left(0,\nabla\mathcal{L}(\theta)^\top\Omega\nabla\mathcal{L}(\theta)\right),
\end{equation}
and similarly
\begin{equation}
\sqrt N (\mathcal{U}(\hat{\theta})-\mathcal{U}(\theta))
\xrightarrow{d}
N\left(0,\nabla\mathcal{U}(\theta)^\top\Omega\nabla\mathcal{U}(\theta)\right).
\end{equation}
Consequently,
\begin{equation}
\text{Var}\left(\mathcal{L}(\hat\theta)\right)
\approx \frac{1}{N}\nabla \mathcal{L}(\theta)^{\top}\Omega\,\nabla \mathcal{L}(\theta),
\end{equation}
with $\text{Var}\left(\mathcal{U}(\hat{\theta})\right)$ defined similarly. The covariance
matrix of estimators $\hat{\bm{\theta}}$ has the form
\begin{align}
\Omega_{\hat{\bm{\theta}}}
=\begin{pmatrix}
\text{Var}(\hat{\delta}_{1}) & 0 & 0 & 0 \\
0 & \text{Var}(\hat{\delta}_{0}) & 0 & 0 \\
0 & 0 & \text{Var}(\hat{p}_1) & -\text{Var}(\hat{p}_1) \\
0 & 0 & -\text{Var}(\hat{p}_1) & \text{Var}(\hat{p}_0)
\end{pmatrix},
\end{align}
which determines $\Omega$ in the CLT after appropriate rescaling.

\subsection{Proof of Proposition \ref{prop:iid}}

Let the potential outcomes $\{Y_{ijt}^{(k)}\}_{i,j,t}$ be sub\mbox{-}exponential with $\psi_{1}$\--norm bounded by $M_k$, and assume that for each fixed threshold $\tau_u$, the observed outcomes in each arm,
\[
\{Y_{ijt} : Z_{ijt}(\tau_u)=k\}, \qquad k\in\{0,1\},
\]
are i.i.d.\ samples. We wish to show how to obtain the coverage probability
\begin{equation}\label{eq:boundapp}
\Pr\left(\forall u \in \mathcal{M},\;\Re_u\in \mathcal H_{\alpha_u}[\Re_u]\right)\geq 1-\alpha
\end{equation}
for $u=m_0,\cdots,m_1$ and some arbitrary significance level $0<\alpha<1$ when assumption \ref{ass:iid} holds. To achieve this, we first need to consider the six ``good'' events:
\begin{align*}
\e_1&:=\{\lvert \hat{\mu}_1-\mu_1\rvert \leq t_1\},\\
\e_2&:=\{\lvert \hat{\mu}_0-\mu_0\rvert \leq t_2\},\\
\e_3&:=\{\lvert \hat{p}_1-p_1\rvert \leq t_3\},\\
\e_4&:=\{\lvert \hat{p}_0-p_0\rvert \leq t_4\},\\
\e_5&:=\left\{\sup_y\bigg\lvert F_{N_1}^{(1)}(y)-F^{(1)}(y)\bigg\rvert \leq t_5\right\},\\
\e_6&:=\left\{\sup_y\bigg\lvert F_{N_0}^{(0)}(y)-F^{(0)}(y)\bigg\rvert \leq t_6\right\}.
\end{align*}
Thus, showing $
\Pr\left(\Re_u\in \mathcal H_{\alpha_u}[\Re_u]\right)\geq 1-\alpha_u$ for $u=m_0,\cdots,m_1$ is equivalent to showing that the intersection of the events, i.e., $\Pr(\bigcap\limits_{i=1}^6\e_i)\geq 1-\alpha_u$. Using De Morgan's law, it is clear that
\begin{equation}
\Pr\left(\bigcap\limits_{i=1}^6\e_i\right)=1-\Pr\left(\bigcup\limits_{i=1}^6\e^c_i\right),
\end{equation}
where $\e_i^c$ is the complement of the event $\e_i$. Furthermore, we know from Boole's inequality that
\begin{equation}
\Pr\left(\bigcup\limits_{i=1}^6\e^c_i\right)\leq \sum\limits_{i=1}^{6}\Pr\left(\e_i^c\right).
\end{equation} 
Consequently, 
\begin{align}
\Pr\left(\bigcap\limits_{i=1}^6\e_i\right)&=1-\Pr\left(\bigcup\limits_{i=1}^6\e^c_i\right)\\
&\geq 1-\sum\limits_{i=1}^{6}\Pr\left(\e_i^c\right).
\end{align}
Hence, showing that the bound \eqref{eq:boundapp} holds is equivalent to ensuring that $\sum_{i=1}^6\Pr\left(\e_i^c\right)\leq \alpha_u$ for $u=m_0,\cdots,m_1$. The only tools we need are the three inequalities in Lemmas \ref{lem:bernstein}-\ref{lem:hoeffing}.

\paragraph{\textit{(i) Means $\mu_1,\mu_0$ (events $\e_1,\e_2$).}}
Let $N_1$ (resp.\ $N_0$) be the number of observations with $Z(\tau_u)=1$ (resp.\ $Z(\tau_u)=0$).
Lemma~\ref{lem:bernstein} gives for any $t>0$
\[
\Pr\left(\lvert\hat\mu_k-\mu_k\rvert\ge t\right)
\le
2\exp\left[-c N_k
            \min\left\{t^2/M_k^{2},t/M_k\right\}\right],
\qquad k=0,1,
\]
where $M_k:=\max_{i:Z_i=k}\lVert Y_i^{(k)}-\mu_k\rVert_{\psi_1}$. Choose for each arm
\begin{equation}
t_k:=\min\left\{
         M_k\sqrt{\frac{\log(12/\alpha_u)}{cN_k}},\;
         \frac{M_k}{cN_k}\log\left(\frac{12}{\alpha_u}\right)
       \right\},
\qquad k=0,1.
\end{equation}
The first term is used when $t_k\le M_k$ — the ``quadratic'' regime;
otherwise the second, ``linear'', term is smaller. With this choice
$2\exp[-\log(12/\alpha_u)]=\alpha_u/6$,
so
$\Pr(\e_1^c)=\Pr(\e_2^c)=\alpha_u/6$.

\medskip
\paragraph{\textit{(ii) Treatment proportions $p_1,p_0$ (events $\e_3,\e_4$).}}
With $N=N_1+N_0$,
Lemma~\ref{lem:hoeffing} yields
\[
\Pr\left(\lvert\hat{p}_k-p_k\rvert \ge t\right)
\le
2\exp[-2N_k t^{2}],
\qquad k=0,1.
\]
Set
\begin{equation}
t_3=t_4:=\sqrt{\frac{\log(12/\alpha_u)}{2N}},
\end{equation}
so that
$\Pr(\e_3^c)=\Pr(\e_4^c)=\alpha_u/6$.

\paragraph{\textit{(iii) Empirical CDFs (events $\e_5,\e_6$).}}
Lemma~\ref{lem:dkw} (two-sided DKW) gives
\[
\Pr\left(\sup_y\lvert F^{(k)}_{N_k}(y)-F^{(k)}(y)\rvert>t\right)
\le
2\exp[-2N_k t^{2}],
\qquad k=0,1.
\]
Choose
\begin{equation}
t_5:=\sqrt{\frac{\log(12/\alpha_u)}{2N_1}},
\qquad
t_6:=\sqrt{\frac{\log(12/\alpha_u)}{2N_0}},
\end{equation}
so that
$\Pr(\e_5^c)=\Pr(\e_6^c)=\alpha_u/6$.

\paragraph{Step 1 Concluded.}
By construction,
$\Pr(\e_i^c)\le\alpha_u/6$ for each $i$,
hence
\begin{equation}
\Pr\left(\bigcap_{i=1}^6\e_i\right)
\ge 1-\sum_{i=1}^6\Pr(\e_i^c)
\ge 1-\alpha_u.
\label{eq:good-event}
\end{equation}

\paragraph{Step 2. From the intersection event to coverage.}
On $\cap_{i=1}^6\e_i$ we have
$\lvert\hat\mu_k-\mu_k\rvert\le t_k$,
$\lvert\hat p_k-p_k\rvert\le t_{k+2}$,
and
$\sup_y\lvert F^{(k)}_{N_k}(y)-F^{(k)}(y)\rvert\le t_{k+4}$
for $k=0,1$.
By the quantile-sandwich (Lemma \ref{lem:quantilesando}), for every $p\in[t_{k+4},\,1-t_{k+4}]$,
\[
F^{-1}(p-t_{k+4}) \le \hat F^{-1}_{N_k}(p) \le F^{-1}(p+t_{k+4}).
\]
Choose any $r_k\in\left(0,\frac12-t_{k+4}\right]$ and define the data-driven tail endpoints
\[
L^{(k)}(t_{k+4})
:=\hat F^{-1}_{N_k}(r_k+t_{k+4})
=Y^{(k)}_{\left(\lceil (r_k+t_{k+4})\,N_k\rceil\right)},
\]
and
\[
U^{(k)}(t_{k+4})
:=\hat F^{-1}_{N_k}(1-r_k-t_{k+4})
=Y^{(k)}_{\left(\lceil (1-r_k-t_{k+4})N_k\rceil\right)}.
\]
Then, on $\e_{k+4}$,
\[
F^{-1}(r_k) \le L^{(k)}(t_{k+4}),
\quad
U^{(k)}(t_{k+4}) \le F^{-1}(1-r_k),
\]
so $\left[L^{(k)}(t_{k+4}),\;U^{(k)}(t_{k+4})\right]$ are conservative surrogates for the unknown population tail quantiles $\left[F^{(k),-1}(r_k),F^{(k),-1}(1-r_k)\right]$. (When $t_{k+4}\le\frac14$ one may take $r_k=t_{k+4}$, which yields the convenient indices $\lceil 2t_{k+4}N_k\rceil$ and $\lceil(1-2t_{k+4})N_k\rceil$.)

Manski’s bounds are monotone in the support endpoints; replacing unknown support limits by $\left(L^{(k)}(t_{k+4}),\;U^{(k)}(t_{k+4})\right)$, and using the perturbed means and treatment shares from $\e_1$-$\e_4$, yields two numbers
$L_{\alpha_u}(\hat\theta)\le U_{\alpha_u}(\hat\theta)$
such that
$\Re\in[L_{\alpha_u}(\hat\theta),\;U_{\alpha_u}(\hat\theta)]$
whenever $(\hat\theta,Y)\in\cap_{i=1}^6\e_i$.
Consequently,
\[
\Pr\left(\Re_u\in \mathcal H_{\alpha_u}[\Re_u]\right)
\ge
\Pr\left(\bigcap_{i=1}^6\e_i\right)
\overset{\eqref{eq:good-event}}{\ge}1-\alpha_u,\quad\text{for}\quad{u=m_0,\cdots,m_1},
\]
and 
\[
\Pr\left(\forall u \in \mathcal M,\; \Re_u\in \mathcal H_{\alpha_u}[\Re_u]\right)
\ge1-\alpha,
\]
which establishes \eqref{eq:boundapp}.

\subsection{Proof of Corollary \ref{corol:iidtrunc}}
The argument follows Proposition~\ref{prop:iid} verbatim except that the
empirical-CDF events are now one-sided because the lower support
is the known constant~$\lambda$:

\[
\begin{aligned}
\e_{5} &:= \left\{
          \sup_{y}\left(F^{(1)}_{N_1}(y)-F^{(1)}(y)\right) \le t_5
          \right\},\\[2pt]
\e_{6} &:= \left\{
          \sup_{y}\left(F^{(0)}_{N_0}(y)-F^{(0)}(y)\right) \le t_6
          \right\}.
\end{aligned}
\]

For a one-sided Kolmogorov deviation the DKW inequality is

\[
\Pr\left(
      \sup_{y}\left[F_n(y)-F(y)\right] > t
     \right)
     \le
     \exp\left(-2N_k t^{2}\right),
\quad
\forall t\geq \sqrt{\frac{\ln 2}{2N_k}}
\]
so choosing
\[
t_{5}:=\sqrt{\frac{\log(6/\alpha_u)}{2N_1}},
\qquad
t_{6}:=\sqrt{\frac{\log(6/\alpha_u)}{2N_0}}
\]
ensures \(\Pr(\e_{5}^{c})=\Pr(\e_{6}^{c})=\alpha_u/6\).  The four
mean- and share-events \(\e_{1}\)-\(\e_{4}\) and their bounds are
unchanged, hence each still receives probability \(\alpha_u/6\).
Because the six complements jointly spend at most~\(\alpha_u\),
Boole’s inequality and the algebra in Proposition~\ref{prop:iid} give

\[
\Pr\left(\Re_u \in \mathcal H_{\alpha_u}[\Re_u]\right)\ge1-\alpha_u,
\qquad u=m_0,\dots,m_1.
\]

\subsection{Proof of Proposition \ref{prop:mixing}}

Suppose the collection $(Y_{ijt},Z_{ijt})_{i,j,t}$ is a strictly stationary 
$\upalpha$-mixing process in the sense of Definition~\ref{def:mixing}, and 
that conditional on $Z_{ijt}(\tau_u)=k$, the potential outcomes have 
sub\mbox{-}exponential tails with $\psi_{1}$-norm bounded by $M_k$.  
The proof follows the same structure as Proposition~\ref{prop:iid}: 
define the six “good’’ events $\e_1,\dots,\e_6$ and apply Boole’s inequality.  
We choose each threshold $t_i$ so that $\Pr(\e_i^{c})\le \alpha_u/6$ for 
$u=m_0,\dots,m_1$.

\paragraph{\textit{(i) Means $\mu_1,\mu_0$ (events $\e_1,\e_2$).}}
By Lemma~\ref{lem:bernsteinweak},
\[
\Pr\left(\lvert \hat\mu_k-\mu_k\rvert\ge t_k\right)
\le 
T_1(t_k)+T_2(t_k)+T_3(t_k),
\qquad k=0,1,
\]
where
\[
\begin{aligned}
T_1(t)
&= N_k \exp\!\left(-\frac{(N_k t)^\gamma}{C_1}\right), \\[2pt]
T_2(t)
&= \exp\!\left(-\frac{(N_k t)^2}{C_2(1+N_k V)}\right),\\[2pt]
T_3(t)
&= \exp\!\left(
     -\frac{(N_k t)^2}{C_3 N_k}
      \exp\!\left(
         \frac{(N_k t)^{\gamma(1-\gamma)}}{
         C_4 (\log(N_k t))^\gamma}
      \right)
     \right).
\end{aligned}
\]

To enforce $\Pr(\e_k^c)\le \alpha_u/6$, it suffices to make each term 
$\le \alpha_u/18$:
\begin{enumerate}
\item \(T_1(t)\le \alpha_u/18\) iff
   \[
   t \ge 
   t_k^{(1)}
   :=
   \frac{\left(C_1 \log(18N_k/\alpha_u)\right)^{1/\gamma}}{N_k}.
   \]

\item \(T_2(t)\le\alpha_u/18\) iff
   \[
   t \ge 
   t_k^{(2)}
   :=
   \frac{\sqrt{C_2(1+N_k V)\,\log(18/\alpha_u)}}{N_k}.
   \]

\item \(T_3(t)\le \alpha_u/18\) defines a unique positive root  
   \(t_k^{(3)}\), since the LHS is strictly increasing in \(t\).
\end{enumerate}
Set
\[
t_k:=\max\{t_k^{(1)},\,t_k^{(2)},\,t_k^{(3)}\}.
\]
Then \(\Pr(\e_k^c)\le 3\cdot(\alpha_u/18)=\alpha_u/6\).

\paragraph{\textit{(ii) Treatment proportions $p_1,p_0$ (events $\e_3,\e_4$).}}
By Lemma~\ref{lem:hoeffdep},
\[
\Pr\left(\lvert\hat p_k-p_k\rvert \ge t_k\right)
\le
2\exp\left(
-\frac{N_k t_k^2}{2(1+4C_\upalpha)^2}
\right).
\]
Solving 
\(2 \exp(-A)=\alpha_u/6\) with 
\(A=N_k t_k^2/[2(1+4C_\upalpha)^2]\)
gives
\[
t_3=t_4
=(1+4C_\upalpha)\sqrt{\frac{2\log(12/\alpha_u)}{N_k}},
\]
so that $\Pr(\e_3^c)=\Pr(\e_4^c)=\alpha_u/6$.

\paragraph{\textit{(iii) Empirical CDFs (events $\e_5,\e_6$).}}
Lemma~\ref{lem:dkw-dep} (dependent DKW) states:
\[
\Pr\left(
\sup_y \lvert F_{N_k}^{(k)}(y)-F^{(k)}(y)\rvert>t_k
\right)
\le
2\exp\!\left(
-\frac{N_k t_k^2}{2(1+4C_\upalpha)^2}
\right).
\]
Thus we may choose
\[
t_5=t_6
=(1+4C_\upalpha)\sqrt{\frac{2\log(12/\alpha_u)}{N_k}},
\]
giving $\Pr(\e_5^c)=\Pr(\e_6^c)=\alpha_u/6$.

\paragraph{Step 1 Concluded.}
By summing the six error probabilities,
\[
\Pr\left(\bigcap_{i=1}^6 \e_i\right)
\ge 1-\sum_{i=1}^6 \Pr(\e_i^c)
\ge 1-\alpha_u.
\]

\paragraph{Step 2.  From the intersection event to coverage.}
On $\bigcap_{i=1}^6 \e_i$,  
the perturbed means, shares, and CDF quantiles satisfy exactly the inequalities 
required in Lemma~\ref{lem:quantilesando}.  
Because Manski’s bounds are monotone in all these arguments, 
it follows algebraically that
\[
\Re_u \in \mathcal H_{\alpha_u}[\Re_u]
\qquad\text{whenever}\qquad 
(\hat\theta,Y)\in \bigcap_{i=1}^6\e_i.
\]
Therefore,
\[
\Pr\left(\Re_u \in \mathcal H_{\alpha_u}[\Re_u]\right)
\ge
\Pr\left(\bigcap_{i=1}^6 \e_i\right)
\ge 1-\alpha_u,
\qquad u=m_0,\dots,m_1.
\]
Finally, by the Bonferroni allocation $\sum_{u}\alpha_u=\alpha$,
\[
\Pr\left(\forall u\in\mathcal M,\;
\Re_u\in \mathcal H_{\alpha_u}[\Re_u]\right)
\ge 1-\alpha.
\]
This completes the proof.

\subsection{Proof of Proposition \ref{prop:hybrid}}
Let $(Y_{ijt},Z_{ijt})_{t=1}^T$ be strictly stationary and $\upalpha$-mixing
with $C_\upalpha=\sum_{r\ge1}\upalpha(r)^{1/2}<\infty$. To simultaneously
obtain the $100(1-\alpha)\%$ confidence set for both the upper and lower
bounds for the identification region \eqref{eq:empnonparametricbounds}, we
first construct, for each $u\in\mathcal M$, a random interval
$\mathcal H_{\alpha_u}[\Re_u]$ with marginal coverage $1-\alpha_u$, then
combine across $u$ by Bonferroni so that
\[
\Pr\left(\forall u \in \mathcal{M},\;\Re_u\in \mathcal H_{\alpha_u}[\Re_u]\right)
\geq 1-\alpha,
\qquad \sum_{u=m_0}^{m_1}\alpha_u=\alpha.
\]

For fixed $u\in\mathcal M$, we aim at
\begin{align}
\label{eq:coverageratehybrid}
\begin{split}
\Pr\Big(&
[\mathcal{L}(\hat{\bm{\theta}}),\mathcal{U}(\hat{\bm{\theta}})]
\subseteq \mathbb{I}(\Re_u)
\;\cap\;
\big\{
\sup_y\lvert F_{N_0}^{(0)}(y)-F^{(0)}(y)\rvert\le \epsilon_{0}
\big\}\\
&\cap\;
\big\{\sup_y\lvert F_{N_1}^{(1)}(y)-F^{(1)}(y)\rvert\le \epsilon_{1}\big\}
\Big)
\;\ge\; 1-\alpha_u,
\end{split}
\end{align}
where $\hat{\bm{\theta}}=(\hat{\delta}_{1},\;\hat{\delta}_{0},\;\hat{p}_1,\;\hat{p}_0)^{\top}$
is a $4\times 1$ vector of estimators and $\mathbb{I}(\Re_u)$ is an interval
$[l(\Re_u),u(\Re_u)]$.

Define the events
\begin{align*}
\e_1&=\{\mathcal{L}(\hat{\theta})\ge l(\Re_u)\},\qquad
\e_2=\{\mathcal{U}(\hat{\theta})\le u(\Re_u)\},\\
\e_3&=\left\{\sup_y\lvert F_{N_1}^{(1)}(y)-F^{(1)}(y)\rvert\le \epsilon_1\right\},\qquad
\e_4=\left\{\sup_y\lvert F_{N_0}^{(0)}(y)-F^{(0)}(y)\rvert\le \epsilon_0\right\}.
\end{align*}
By Boole's inequality,
\begin{align}
\label{eq:boole-hybrid}
\begin{split}
\Pr\left(\bigcap_{i=1}^4\e_i\right)
&\ge 1-\sum_{i=1}^{4}\Pr(\e_i^c).
\end{split}
\end{align}
If we choose the four components so that $\Pr(\e_i^c)\le\alpha_u/4$ for
$i=1,\dots,4$, then $\Pr(\cap_{i=1}^4\e_i)\ge 1-\alpha_u$.

Under $\upalpha$-mixing with $C_\upalpha<\infty$, a mixing-adjusted DKW bound
gives, for $k=0,1$,
\[
\Pr\left(\sup_y\lvert F^{(k)}_{N_k}(y)-F^{(k)}(y)\rvert>\epsilon_k\right)\le
2\exp\left(-\frac{N_k\epsilon_k^2}{2(1+4C_\upalpha)^2}\right).
\]
Thus choosing
\begin{equation}
\label{eq:epshybrid-new}
\epsilon_k:=(1+4C_{\upalpha})\sqrt{\frac{2\log(8/\alpha_u)}{N_k}},
\qquad k=0,1,
\end{equation}
ensures $\Pr(\e_{k+2}^c)\le \alpha_u/4$ for $k=0,1$.

On the event $\e_{k+2}$ the uniform control
$\sup_y\lvert F^{(k)}_{N_k}(y)-F^{(k)}(y)\rvert\le \epsilon_k$ implies the
quantile-sandwich
\[
F^{(k),-1}(p-\epsilon_k) \le \hat F^{-1}_{N_k}(p) \le F^{(k),-1}(p+\epsilon_k),
\quad p\in[\epsilon_k,1-\epsilon_k].
\]
Fix a deterministic
\[
r_k:=\min\{\epsilon_k,1/2-\epsilon_k-1/N_k\}
\]
and define the empirical quantile endpoints
\begin{equation}
\label{eq:emp-endpoints}
L^{(k)}:=\hat F^{-1}_{N_k}(r_k+\epsilon_k)
=Y^{(k)}_{\left(\lceil (r_k+\epsilon_k)N_k\rceil\right)},\qquad
U^{(k)}:=\hat F^{-1}_{N_k}(1-r_k-\epsilon_k)
=Y^{(k)}_{\left(\lceil (1-r_k-\epsilon_k)N_k\rceil\right)}.
\end{equation}
Then on $\e_{k+2}$ we have
\[
F^{(k),-1}(r_k)\le L^{(k)},\qquad
U^{(k)}\le F^{(k),-1}(1-r_k),
\]
so $[L^{(k)},U^{(k)}]$ are conservative surrogates for the unknown population
tail quantiles $\left[F^{(k),-1}(r_k),F^{(k),-1}(1-r_k)\right]$. Manski’s bounds
are monotone in the support endpoints, so using $L^{(k)},U^{(k)}$ in place of
the unknown limits preserves coverage of $\Re_u$.

For the mean and proportion components, we construct one-sided bounds
\[
\Pr\big(\mathcal{L}(\hat{\theta})\ge l(\Re_u)\big)\ge 1-\frac{\alpha_u}{4},
\qquad
\Pr\big(\mathcal{U}(\hat{\theta})\le u(\Re_u)\big)\ge 1-\frac{\alpha_u}{4},
\]
by applying a CLT and delta method to
\[
\hat{\theta}
=(\hat\delta_1,\;\hat\delta_0,\;\hat p_1,\;\hat p_0)^\top,
\qquad
\theta=(\delta_1,\;\delta_0,\;p_1,\;p_0)^\top.
\]
Under $\upalpha$-mixing, a suitable CLT yields
\begin{equation}
\sqrt{N}(\hat{\theta}-\theta)\xrightarrow{D}N(0,\Omega_{\theta}),
\end{equation}
for some positive semidefinite $4\times 4$ matrix $\Omega_{\theta}$, which can
be consistently estimated (for example) by a heteroskedasticity- and
autocorrelation-consistent estimator (see~\citet{neweywest1987}).

From Proposition~\ref{prop:1}, treating the support endpoints
$L^{(k)},U^{(k)}$ as fixed constants, the gradients of the lower and upper
functionals with respect to $\theta$ are
\begin{align}
\nabla \mathcal{L}(\theta)&=\left(p_1,\;-p_0,\;\delta_{1}-U^{(0)},\;L^{(1)}-\delta_{0}\right)^{\top},\\
\nabla \mathcal{U}(\theta)&=\left(p_1,\;-p_0,\delta_{1}-L^{(0)},\;U^{(1)}-\delta_{0}\right)^{\top}.
\end{align}
A first-order Taylor expansion gives
\[
\mathcal{L}(\hat{\theta})-\mathcal{L}(\theta)
=\nabla\mathcal{L}(\theta)^{\top}(\hat{\theta}-\theta)+o_p(N^{-1/2}),
\]
and similarly for $\mathcal{U}(\hat{\theta})$. Therefore,
\begin{align}
\sqrt{N}\left(\mathcal{L}(\hat{\theta})-\mathcal{L}(\theta)\right)
&\xrightarrow{d} N\left(0,\nabla \mathcal{L}(\theta)^{\top}\Omega_{\theta}\nabla \mathcal{L}(\theta)\right),\\
\sqrt{N}\left(\mathcal{U}(\hat{\theta})-\mathcal{U}(\theta)\right)
&\xrightarrow{d} N\left(0,\nabla \mathcal{U}(\theta)^{\top}\Omega_{\theta}\nabla \mathcal{U}(\theta)\right).
\end{align}
Let
\[
\text{S.E.}\big(\mathcal{L}(\hat{\bm{\theta}})\big)
:=\sqrt{\frac{1}{N}\nabla \mathcal{L}(\hat{\theta})^{\top}\hat\Omega_{\theta}\nabla \mathcal{L}(\hat{\theta})},
\qquad
\text{S.E.}\big(\mathcal{U}(\hat{\bm{\theta}})\big)
:=\sqrt{\frac{1}{N}\nabla \mathcal{U}(\hat{\theta})^{\top}\hat\Omega_{\theta}\nabla \mathcal{U}(\hat{\theta})},
\]
where $\hat\Omega_{\theta}$ is a consistent estimator of $\Omega_{\theta}$.
Then the one-sided Wald inequalities
\begin{align*}
\mathcal{L}(\hat{\bm{\theta}})- \Phi^{-1}(1-\alpha_u/4)\,\text{S.E.}\left(\mathcal{L}(\hat{\bm{\theta}})\right)
&\le \mathcal{L}(\bm{\theta}),\\
\mathcal{U}(\hat{\bm{\theta}})+ \Phi^{-1}(1-\alpha_u/4)\,\text{S.E.}\left(\mathcal{U}(\hat{\bm{\theta}})\right)
&\ge \mathcal{U}(\bm{\theta}),
\end{align*}
hold with probability at least $1-\alpha_u/4$ each, asymptotically. Combining
these four one-sided events with \eqref{eq:boole-hybrid} yields
\[
\Pr\left([\mathcal{L}(\hat{\bm{\theta}}),\mathcal{U}(\hat{\bm{\theta}})]
\subseteq \mathbb{I}(\Re_u)\right)\ge 1-\alpha_u.
\]
Finally, by $\sum_{u=m_0}^{m_1}\alpha_u=\alpha$ and another application of
Boole’s inequality over $u$, we obtain
\[
\Pr\left(\forall u \in \mathcal{M},\;\Re_u\in \mathcal H_{\alpha_u}[\Re_u]\right)
\ge 1-\alpha,
\]
which establishes \eqref{eq:coverageratehybrid}.

\subsection{Proof of Proposition \ref{prop:minimaxlow}}

Following the two-hypothesis reduction scheme of \citet{tsybakov2008nonparametric}, Ch. 2, let us construct two distribution in $\mathcal{F}$ whose $N$ sample laws are close in total variation, but whose upper tail endpoints differ by a controlled amount, forcing any valid bracket to be wide. 

\paragraph{\textit{(i) Two-hypothesis construction:}} For a constant $c_0>0$, define
\begin{align}
F_0&\sim N(0,1)\\
F_1&\sim N(\Delta_N,1),\quad\text{where}\quad\Delta_N:=c_0\sqrt{\frac{\log(1/\alpha')}{N}},
\end{align}
where $F_0,F_1\in \mathcal{F}$. Hence, the true upper tail endpoints differ by the additive shift:
\begin{equation}
q_{1-r}(F_0)=\Phi^{-1}(1-r),\qquad q_{1-r}(F_1)=\Phi^{-1}(1-r)+\Delta_N.
\end{equation}
\paragraph{\textit{(ii) Kullback-Leibler divergence:}} Let $P$ and $Q$ be two probability measures, such that $P\ll Q$, then the KL divergence is defined as:
\begin{equation}
KL(P,Q)=\int\log\left(\frac{dP}{dQ}\right)dP
\end{equation}
By the product-measure additivity property of KL divergence \citep[p.95 of][]{tsybakov2008nonparametric}, we know that
\begin{align}
KL\left(F_1^{\otimes N},F_0^{\otimes N}\right)&=\sum\limits_{i=1}^N KL\left(F_{i1},F_{i0}\right)\\
&=\sum\limits_{i=1}^N\int(\log f_{i1}-\log f_{i0})f_{i1}dy\\
&=\sum\limits_{i=1}^N\frac{\Delta^2_N}{2}= N\frac{\Delta^2_N}{2}=\frac{c_0^2}{2}\log(1/\alpha')
\end{align}
\paragraph{\textit{(iii) Total variation bound:}} Let us define the total variation distance between two probability measures $P$ and $Q$ as,
\begin{equation}
TV(P,Q)=\sup_{A\in \mathcal{A}}\lvert P(A)-Q(A)\rvert=\sup_{A\in \mathcal{A}}\Big\vert\int_A(p-q)dv\Big\vert.
\end{equation} 
Then according to \citet{bretagnolle1979estimation} \citep[see Lemma~2.6 of][]{tsybakov2008nonparametric}, we know that
\begin{align}
TV(F_1^{\otimes N},F_0^{\otimes N})&\leq 1-\frac{1}{2}\exp\left(-KL\left(F_1^{\otimes N},F_0^{\otimes N}\right)\right)\\
&\leq\sqrt{1-\frac{1}{2}\exp\left(-KL\left(F_1^{\otimes N},F_0^{\otimes N}\right)\right)}\\
&\leq \sqrt{ 1-\frac{1}{2}\exp\left(-\frac{c_0^2}{2}\log(1/\alpha')\right)}=\sqrt{1-\frac{1}{2}\alpha'^{c_0^2/2}}
\end{align}

\paragraph{\textit{(iv)} Le-Cam's reduction scheme:} The maximum risk can be defined in terms of bracket width. In Section~\ref{sec:inference},  we constructed two tail endpoints, $L_{\alpha_u}$ and $U_{\alpha_u}$, which we may take as $A_N:=L_{\alpha_u}$ and $B_N:=U_{\alpha_u}$ respectively. Let $(A_N,B_N)$ denote any candidate distribution-free $(1-\alpha')$-bracket for $q_{1-r}(F)$. Then, for a target width level $s>0$, define the maximum risk as
\begin{equation}
\sup_{F\in \mathcal{F}}\E_F\left[\mathbbm{1}\left( B_N-A_N\geq s\right)\right].
\end{equation}
Equivalently, since the expectation of an indicator is the probability of its event,
\begin{equation}
\sup_{F\in \mathcal{F}}\Pr_F\left(B_N-A_N\geq s\right).
\end{equation}
Hence, the minimax problem consists of finding a lower bound of the following form:
\begin{equation}
\label{eq:minimaxeq}
\liminf_{N\to\infty}\inf_{(A_N,B_N)}\sup_{F\in\mathcal{F}}\Pr\left( B_N-A_N \geq s_N\right)\geq c>0,
\end{equation}
for some sequence $s_N$. In our setting, we shall show that one may take $s_N=O(\sqrt{\log(1/\alpha')/N)}$.

Now let $(A,B)$ be any distribution-free $(1-\alpha')$-bracket over $\mathcal{F}$. Coverage under $F_1$ requires $\Pr_{F_1}^{\otimes N}(B\geq q_{1-r}(F_1))\geq 1-\alpha'$, which in our context implies
\begin{equation}
 \Pr_{F_1^{\otimes N}}\left(B\geq \Phi^{-1}(1-r)+\Delta_N\right)\geq 1-\alpha'.
\end{equation}
Using the total variation coupling property \citep[see Eq.~(2.13) of ][]{tsybakov2008nonparametric}, we know that for any measurable set $\mathcal{A}$
\begin{equation}
\label{eq:lecam}
TV(F_1^{\otimes N},F_0^{\otimes N})\geq\lvert Pr_{F_0^{\otimes N}}(\mathcal{A})-Pr_{F_1^{\otimes N}}(\mathcal{A})\rvert.
\end{equation}
Applying \eqref{eq:lecam} to $\mathcal{A}:=\{B\geq \Phi^{-1}(1-r)+\Delta_N\}$, and rearranging, yields
\begin{equation}
\label{eq:strpositive}
\Pr_{F_0^{\otimes N}}\left(B\geq \Phi^{-1}(1-r)+\Delta_N\right)\geq (1-\alpha')-\sqrt{1-\frac{1}{2}\alpha'^{c_0^2/2}}.
\end{equation}
The right-hand side of Eq.~\eqref{eq:strpositive} should be strictly positive, which requires the condition $(1-\alpha')>\sqrt{1-\frac{1}{2}\alpha'^{c_0^2/2}}$. Squaring both sides of this condition yields $(1-\alpha')^2>1-\frac{1}{2}\alpha'^{c_0^2/2}$, which is equivalent to $\frac{1}{2}\alpha'^{c_0^2/2}>(2-\alpha')\alpha' $, requiring $\alpha^{c_0^2/2}$ decaying slower than $\alpha'$ as $\alpha'\to 0$, which happens precisely when $c_0^2/2<1$, i.e., $c_0<\sqrt{2}$. Therefore, choose $c_0\in (0,\sqrt{2})$ and $\alpha'\in(0,1/4)$ sufficiently small so that the condition holds.

Under $F_0$ the true upper tail endpoint is $q_{1-r}(F_0)=\Phi^{-1}(1-r)$. Therefore, the event $\mathcal{A}$ overshoots $q_{1-r}(F_0)$ by $\Delta_n$. Since validity of the bracket under $F_0$ also requires $A\leq q_{1-r}(F_0)$, it follows that on this event,
\begin{equation}
B-A\geq B-q_{1-r}(F_0)\geq \Delta_N.
\end{equation}
Consequently,
\begin{equation}
\Pr_{F_0^{\otimes N}}\left(B-A\geq\Delta_N\right)\geq \Pr_{F_0^{\otimes N}}\left(B\geq \Phi^{-1}(1-r)+\Delta_N\right),
\end{equation}
and therefore,
\begin{equation}
\Pr_{F_0^{\otimes N}}\left(B-A\geq \Delta_N\right)\geq (1-\alpha')-\sqrt{1-\frac{1}{2}\alpha'^{c_0^2/2}}.
\end{equation}
Thus, for $S_N=\Delta_N=c_0\sqrt{\log(1/\alpha')/N}$, we have shown that 
\begin{equation}
\sup_{F\in\mathcal{F}}\Pr_{F}\left(B_N-A_N\geq s_N\right)
\end{equation}
is bounded away from zero uniformly over all valid distribution-free $(1-\alpha')$-brackets. Equivalently,
\begin{equation}
\liminf_{N\to \infty}\inf_{(A_N,B_N)}\sup_{F\in \mathcal{F}}\Pr_F\left(B_N-A_N\geq c_0\sqrt{\frac{\log(1/\alpha')}{N}}\right)\geq c
\end{equation}
for some constant $c>0$, establishing the desired minimax lower bound rate.

\subsection{Proof of Proposition~\ref{prop:minimaxupp}}
Under Assumption~\ref{ass:mixing} and given Lemma~\ref{lem:dkw-dep}, we  know that
\begin{equation}
\Pr\left(\sup_{y\in\R}\Big\lvert F_{N}(y)-F(y)\Big\rvert\leq \varepsilon\right)\geq 1-\alpha'
\end{equation}
defines the DKW event $\mathcal{E}:=\{\Delta\leq \varepsilon\}$ for  for $\varepsilon:=(1+4C_{\alpha'})\sqrt{2\log(8/\alpha')/N}$, where $\Delta:=\sup_{y\in\R}\Big\lvert F_{N}(y)-F(y)\Big\rvert$. Furthermore, for all $p\in [\varepsilon,1-\varepsilon]$, Lemma~\ref{lem:quantilesando} gives the quantile sandwich:
\begin{equation}
F^{-1}(p-\varepsilon)\leq\hat{F}^{-1}(p)\leq F^{-1}(p+\varepsilon).
\end{equation}
With $p=1-r-\varepsilon$ (which lies in $[\varepsilon,1-\varepsilon]$ when $r>0$):
\begin{align*}
U_{\alpha'}:=Y_{(\lceil (1-r-\varepsilon)N \rceil)}&= \hat{F}^{-1}(1-r-\varepsilon)\\
&\leq F^{-1}(1-r)=q_{1-r}
\end{align*}
on $\e$. Similarly, with $p=r+\varepsilon$:
\begin{equation}
L_{\alpha'}:=Y_{\left(\lceil(r+\varepsilon)N\rceil\right)}\geq F^{-1}(r)=q_r
\end{equation}
on $\e$. Hence, on $\e$ with probability $\geq 1-\alpha'$, the bracket $(L_{\alpha'},U_{\alpha'})$ brackets $(q_r,q_{1-r})$, confirming the distribution-free bracket condition in Proposition~\ref{prop:minimaxlow}.

Under the high probability excess width $\mathcal{W}$ defined in Proposition~\ref{prop:minimaxlow}, we evaluate the excess width conditional on the high-probability event $\e$. To find the worst-case width $w$ realized with probability $1-\alpha'$, we measure the outward shift in the quantile space. The endpoints in \eqref{eq:prop6q} are set at probability levels $r+\varepsilon$ and $1-r-\varepsilon$. On $\e$, the distance from the upper bound to the true quantile is bounded by the quantile shift corresponding to $\varepsilon$:
\begin{align}
\left(q_{1-r}-U_{\alpha'}\right)&= F^{-1}(1-r)-\hat{F}^{-1}(1-r-\varepsilon)\\
&\leq F^{-1}(1-r) - F^{-1}(1-r-2\varepsilon)\label{eq:fromlem1}
\end{align}
where inequality \eqref{eq:fromlem1} is due to the quantile sandwich result in Lemma~\ref{lem:quantilesando}.   

We wish to bound \eqref{eq:fromlem1}, i.e, $F^{-1}(1-r)-F^{-1}(1-r-2\varepsilon)$. Let us assume that $dF=fdy$, in other words, if the density of $F$ exists and is strictly bounded away by $f\geq m_0>0$, then the Inverse Function Theorem shows that $F^{-1}$ is differentiable with the derivative:
\begin{equation}
\frac{dF^{-1}}{dp}=\frac{1}{f(F^{-1}(p))}.
\end{equation} 
Applying the Mean-Value Theorem to $F^{-1}$ on the interval $[1-r-2\varepsilon, 1-r]$, we may show that there exists some $p^*\in(1-r-2\varepsilon, 1-r)$, such that
\begin{align}
F^{-1}(1-r)-F^{-1}(1-r-2\varepsilon)
&=(F^{-1})'(p^*)[(1-r)-(1-r-2\varepsilon)]\\
&=(F^{-1})'(p^*)\,2\varepsilon
= \frac{2\varepsilon}{f(F^{-1}(p^*))}\\
&\leq \frac{2\varepsilon}{m_0}
\end{align}

For heavy-tailed distributions without this regularity condition, i.e., $f\geq m_0$, the excess width is instead bounded directly on the probability scale. By construction, $\hat{F}(U_{\alpha'}) 
= 1-r-\varepsilon$. On $\mathcal{E}$, since $\Delta \leq 
\varepsilon$
\begin{equation}
F(U_{\alpha'}) \geq \hat{F}_(U_{\alpha'}) - \varepsilon 
= 1-r-2\varepsilon,
\end{equation}
so $U_{\alpha'}$ lies within $2\varepsilon$ of $q_{1-r}$ 
on the probability scale for any $F \in \mathcal{F}$, 
requiring no regularity on $f$. In terms of the probability-scale padding, the excess width is fundamentally bounded proportionally to $\varepsilon$
\begin{equation}
\mathcal{W}_N\left(L_{\alpha'},U_{\alpha'}\right)\leq C\varepsilon\leq C(1+4C\alpha)\sqrt{\frac{2\log(8/\alpha')}{N}}\leq C'\sqrt{\frac{\log(2/\alpha')}{N}}
\end{equation}
for some constants $C,C'>0$, establishing Proposition~\ref{prop:minimaxupp}.

\subsection{Proof of Lemma~\ref{lem:calpha_ar1}}

From Definition \ref{def:mixing}, we know that for two sub-$\sigma$-algebra $\mathcal{A},\mathcal{B}$, the $\upalpha$-mixing coefficient, is defined as
\begin{equation}
\upalpha(\mathcal{A},\mathcal{B})=\sup_{A\in \mathcal{A},B \in \mathcal{B}}\lvert P(A\cap B)-P(A)P(B)\rvert.
\end{equation}
Moreover, following \citet{bradley2005basic}, we define the $\rho$-mixing coefficient, as 
\begin{equation}
\rho(\mathcal{A},\mathcal{B})=\sup_{\substack{f\in L^2(\mathcal{A}),g\in L^2(\mathcal{B})\\Var(f)>0,Var(g)>0}}\lvert \text{Corr}(f,g)\rvert.
\end{equation} 
The $\upalpha$-mixing coefficient restricts the supremum to indicator functions ($f=\mathbbm{1}_A$,$g=\mathbbm{1}_B$). The $\rho$-mixing coefficient takes the supremum over all square-integrable functions. Moreover, we know from \citet{doi:10.1137/1105018} that for Gaussian processes
\begin{equation}
\upalpha(k)\leq \rho(k),
\end{equation}
where
\begin{equation}
\rho(k):=\rho\left(\mathcal{F}_{-\infty}^0,\mathcal{F}_{k}^{\infty}\right),
\end{equation}
with $\mathcal{F}_{-\infty}^0=\sigma(Y_s:s\leq 0)$ and $\mathcal{F}_{k}^{\infty}=\sigma(Y_s,s\geq k)$. Thus, we need to bound $\rho\left(\mathcal{F}_{-\infty}^0,\mathcal{F}_{k}^{\infty}\right)$. 
Furthermore, $Y_t$ is an AR($1$) process, and hence, a Markov process. Thus, by definition \citep[see][]{bradley2005basic}, 
\begin{equation}
\rho(\mathcal{F}_{-\infty}^0,\mathcal{F}_k^{+\infty})=\rho(\sigma(Y_0),\sigma(Y_k)).
\end{equation}
$(Y_0,Y_k)$ is bivariate Gaussian with Pearson correlation $r_k=\theta^k$. Therefore, we know from \citet{beaa45ee-b909-38c3-b8e5-43f058e8bc30} that 
\begin{equation}
\sup_{\substack{h,\ell\\Var(h(Y_0))>0,Var(\ell(Y_k))>0}}\lvert \text{Corr}(h(Y_0),\ell(Y_k))\rvert=\lvert r\rvert=\lvert\theta^k\rvert.
\end{equation}
Putting it all together:
\begin{equation}
\upalpha(k)\leq \rho(\mathcal{F}_{-\infty}^0,\mathcal{F}_{k}^{+\infty})=\rho(\sigma(Y_0),\sigma(Y_k))=\lvert\theta\rvert^k,
\end{equation}
and consequently,
\begin{equation}
C_{\upalpha}=\sum_{k\geq1}\upalpha(k)^{1/2}\leq \sum_{k\geq1}\lvert\theta\rvert^{k/2}=\frac{\lvert \theta\rvert^{1/2}}{1-\lvert\theta\rvert^{1/2}}.
\end{equation}

\newpage 

\section{Algorithms}
\begin{algorithm}[H]
\footnotesize
\caption{MC: single-threshold coverage and width comparison}
\label{alg:MC_five_methods}
\begin{algorithmic}[1]
\State \textbf{Inputs:} $B{=}2000$, $B_{\mathrm{boot}}{=}999$, $n{=}50$, $T\in\{1,2,5\}$, $\tau^\circ{=}50\%$, $\alpha{=}0.05$, $\Delta{=}4$; \\
\hspace*{1.2em} $z_M{=}\Phi^{-1}(0.975){=}1.96$, \; $z_H{=}\Phi^{-1}(1-\alpha/4){=}2.24$.

\State \textbf{Per design (once):} obtain $(a^\ast,b^\ast)$: G uses $(-5,5)$; F uses $(0,\max Y^0{+}\Delta)$; A--E use $(\min Y^0,\max Y^0{+}\Delta)$ from a large oracle draw.

\For{$b=1,\dots,B$}

\State Generate $(Y^0_{it},D_{it})$; set $Y_{it}{=}Y^0_{it}{+}\Delta D_{it}$; set $(\hat a,\hat b){=}\big(\min Y^0,\max Y^0\big)$.
\State \textbf{Analyst support} $(a,b){=}\begin{cases}
(a^\ast,b^\ast), & \text{G},\\
(0,\hat b), & \text{F},\\
(\hat a,\hat b), & \text{A--E}.
\end{cases}$
\State Split at $\tau^\circ$: arms $k\in\{0,1\}$ with sizes $N_k$, means $\bar Y^{(k)}$, shares $p_k$.

\Statex
\Statex \hspace*{0.5em}\hrulefill\; \textsc{Method 1: Plug-in Manski} \;\hrulefill
\State Point bounds $L_M^0{=}p_1\bar Y^{(1)}+p_0 a-p_1 b - p_0\bar Y^{(0)}$, \ \
$U_M^0{=}p_1\bar Y^{(1)}+p_0 b-p_1 a - p_0\bar Y^{(0)}$.
\State Delta-method SEs $(\hat\sigma_L^M,\hat\sigma_U^M)$; set
$L^{\mathrm M}{=}L_M^0{-}z_M\hat\sigma_L^M$, \;
$U^{\mathrm M}{=}U_M^0{+}z_M\hat\sigma_U^M$.

\Statex
\Statex \hspace*{0.5em}\hrulefill\; \textsc{Method 2: Imbens--Manski (2004)} \;\hrulefill
\State Set $\hat\sigma_{\max}{=}\max(\hat\sigma_L^M,\,\hat\sigma_U^M)$ and identified-set half-length
$\hat\delta_n {=} (U_M^0 - L_M^0)/(2\,\hat\sigma_{\max})$.
\State Solve for $c_n^{\mathrm{IM}}$: \;
$\Phi\!\big(c_n^{\mathrm{IM}} + 2\hat\delta_n\big) - \Phi\!\big(-c_n^{\mathrm{IM}}\big) = 1-\alpha$.
\Comment{bisection on $c_n^{\mathrm{IM}}\in[0,\,z_M{+}2]$}
\State Set
$L^{\mathrm{IM}} {=} L_M^0 - c_n^{\mathrm{IM}}\,\hat\sigma_L^M$, \;
$U^{\mathrm{IM}} {=} U_M^0 + c_n^{\mathrm{IM}}\,\hat\sigma_U^M$.

\Statex
\Statex \hspace*{0.5em}\hrulefill\; \textsc{Method 3: Stoye (2009)} \;\hrulefill
\State Set $c_n^{\mathrm{S}} {=} \max\!\big(c_n^{\mathrm{IM}},\; z_M\big)$.
\Comment{ensures uniform validity}
\State Set
$L^{\mathrm{S}} {=} L_M^0 - c_n^{\mathrm{S}}\,\hat\sigma_L^M$, \;
$U^{\mathrm{S}} {=} U_M^0 + c_n^{\mathrm{S}}\,\hat\sigma_U^M$.

\Statex
\Statex \hspace*{0.5em}\hrulefill\; \textsc{Method 4: Percentile Bootstrap} \;\hrulefill
\For{$j=1,\dots,B_{\mathrm{boot}}$}
    \State Draw $(Y^\ast,D^\ast)$ by resampling $(Y_{it},D_{it})_{i,t}$ with replacement (cluster on $i$).
    \State Compute bootstrap Manski point bounds $(L_M^{0,\ast(j)},\;U_M^{0,\ast(j)})$ using analyst support $(a,b)$.
\EndFor
\State Set
$L^{\mathrm{Boot}} {=} \widehat Q_{\alpha/2}\!\big\{L_M^{0,\ast(j)}\big\}_{j=1}^{B_{\mathrm{boot}}}$, \;
$U^{\mathrm{Boot}} {=} \widehat Q_{1-\alpha/2}\!\big\{U_M^{0,\ast(j)}\big\}_{j=1}^{B_{\mathrm{boot}}}$.
\Comment{percentile CIs}

\Statex
\Statex \hspace*{0.5em}\hrulefill\; \textsc{Method 5: Hybrid concATE} \;\hrulefill
\If{G} $(L^{\mathrm H},U^{\mathrm H}){\gets}(L^{\mathrm M},U^{\mathrm M})$
\Else
\State For each arm $k$: set
$\epsilon_k{=}\begin{cases}
\sqrt{\log(8/\alpha)/(2N_k)}, & \text{A,B,E (i.i.d.)},\\
(1{+}4C_\upalpha)\sqrt{2\log(8/\alpha)/N_k}, & \text{C,D (mixing)},\\
\sqrt{\log(4/\alpha)/(2N_k)}, & \text{F (one-sided upper)}.
\end{cases}$
and $r_k{=}\min\{\epsilon_k,\;1/2{-}\epsilon_k{-}1/N_k\}$.
\State Endpoints $(L^{(k)},U^{(k)}){=}\begin{cases}
\big(\hat F_{N_k}^{(k),-1}(r_k{+}\epsilon_k),\ \hat F_{N_k}^{(k),-1}(1{-}r_k{-}\epsilon_k)\big), & \text{A--E},\\
\big(\lambda,\ \hat F_{N_k}^{(k),-1}(1{-}r_k{-}\epsilon_k)\big),\ \lambda{=}0, & \text{F}.
\end{cases}$
\State Hybrid point bounds $L_H^0{=}p_1\bar Y^{(1)}+p_0 L^{(1)}-p_1 U^{(0)}-p_0\bar Y^{(0)}$, \ \
$U_H^0{=}p_1\bar Y^{(1)}+p_0 U^{(1)}-p_1 L^{(0)}-p_0\bar Y^{(0)}$.
\State Delta SEs with gradients $\nabla \mathcal L{=}(p_1,-p_0,\bar Y^{(1)}{-}U^{(0)},\,L^{(1)}{-}\bar Y^{(0)})$,
$\nabla \mathcal U{=}(p_1,-p_0,\bar Y^{(1)}{-}L^{(0)},\,U^{(1)}{-}\bar Y^{(0)})$; set
$L^{\mathrm H}{=}L_H^0{-}z_H\hat\sigma_L^H$, \;
$U^{\mathrm H}{=}U_H^0{+}z_H\hat\sigma_U^H$.
\EndIf

\Statex
\Statex \hspace*{0.5em}\hrulefill\; \textsc{Record} \;\hrulefill
\For{$m\in\{\mathrm{M},\,\mathrm{IM},\,\mathrm{S},\,\mathrm{Boot},\,\mathrm{H}\}$}
    \State $\mathrm{hit}_m[b] \gets \mathbbm1\{\Delta\in[L^m,\,U^m]\}$; \quad
    $\mathrm{width}_m[b] \gets U^m - L^m$.
\EndFor

\EndFor

\Statex
\State \textbf{Output:} For each method $m\in\{\mathrm{M},\,\mathrm{IM},\,\mathrm{S},\,\mathrm{Boot},\,\mathrm{H}\}$:
\Statex \hspace*{2.4em} Coverage \; $\widehat P_m{=}B^{-1}\sum_b \mathrm{hit}_m[b]$; \quad
Median width \; $\widehat W_m {=} \mathrm{median}\!\big\{\mathrm{width}_m[b]\big\}_{b=1}^B$.
\end{algorithmic}
\end{algorithm}

\section{Additional Analysis\label{sec:additionalanalysis}}

\subsection{Sector Group Classifications}

\begin{table}[H]
  \centering
  \resizebox{\textwidth}{!}{%
    \begin{tabular}{ll}
      \toprule
      \textbf{Group} & \textbf{Included Sectors} \\
      \midrule
      \textbf{Cyclicals}           & Consumer Discretionary, Materials, Industrials, Real Estate \\
      \textbf{Defensives}          & Health Care, Consumer Staples, Utilities                   \\
      \textbf{Growth \& Innovation}& Information Technology, Communication Services             \\
      \textbf{Financials}          & Financials                                                \\
      \textbf{Energy}              & Energy                                                    \\
      \bottomrule
    \end{tabular}%
  }
  \caption{Sector Group Classifications}
  \label{tab:sector_groups}
\end{table}

\subsection{Clustered Descriptive Statistics}
This section presents the descriptive statistics and kernel density plots for the sectoral groups described in Table~\ref{tab:sector_groups}, and for all companies prior to 1\textsuperscript{st} September 2019.

\begin{table}[hbtp!]
\caption{Descriptive Statistics - (Pre 01/09/2019)}
\label{tab:descriptive_stats_filtered_years}
  \resizebox{\textwidth}{!}{%
\begin{tabular}{lcccccccc}
\toprule
 Variable& Min & Mean & Median & Max & Std Dev & Skewness & Kurtosis & $N$ \\
\midrule
(\%) Women & 0.000 & 26.340 & 25.902 & 100 & 12.710 & 0.449 & 1.707 & 16066 \\
(\%) Unknown gender & 0.000 & 0.028 & 0.022 & 0.496 & 0.032 & 2.713 & 18.328 & 16066 \\
Tobin's $Q$ & -0.612 & 0.386 & 0.020 & 5.047 & 1.094 & 2.218 & 5.456 & 14466 \\
Total assets & 10.392 & 16.014 & 16.037 & 21.740 & 1.840 & 0.044 & 0.187 & 15126 \\
Leverage & 0.000 & 0.288 & 0.272 & 3.945 & 0.229 & 3.677 & 41.331 & 15119 \\
Total employees & 85.559 & 24312.371 & 7951.079 & 703268.060 & 49761.515 & 5.480 & 43.774 & 16214 \\
\bottomrule
\end{tabular}
}
\end{table}

\begin{table}[hbtp!]
\caption{Descriptive Statistics - (Cyclicals)}
\label{tab:descriptive_stats_Cyclicals}
  \resizebox{\textwidth}{!}{%
\begin{tabular}{lcccccccc}
\toprule
 Variable& Min & Mean & Median & Max & Std Dev & Skewness & Kurtosis & $N$ \\
\midrule
(\%) Women & 0.000 & 26.359 & 24.544 & 100 & 14.424 & 0.896 & 2.384 & 10429 \\
(\%) Unknown gender & 0.000 & 0.023 & 0.015 & 0.331 & 0.030 & 2.933 & 16.268 & 10429 \\
Tobin's $Q$ & -0.612 & 0.321 & 0.045 & 5.047 & 0.880 & 2.494 & 8.140 & 9833 \\
Total assets & 11.064 & 15.619 & 15.798 & 20.230 & 1.453 & -0.298 & -0.020 & 10120 \\
Leverage & 0.000 & 0.347 & 0.327 & 3.945 & 0.257 & 4.691 & 47.105 & 10118 \\
Total employees & 148.263 & 20767.166 & 8964.840 & 941046.440 & 40760.375 & 8.815 & 137.453 & 10556 \\
\bottomrule
\end{tabular}
}
\end{table}

\begin{table}[hbtp!]
\caption{Descriptive Statistics - (Defensives)}
\label{tab:descriptive_stats_Defensives}
  \resizebox{\textwidth}{!}{%
\begin{tabular}{lcccccccc}
\toprule
Variable & Min & Mean & Median & Max & Std Dev & Skewness & Kurtosis & $N$ \\
\midrule
(\%) Women & 0.000 & 32.928 & 33.008 & 67.221 & 11.497 & -0.183 & 0.653 & 4953 \\
(\%) Unknown gender & 0.000 & 0.041 & 0.036 & 0.496 & 0.037 & 2.398 & 17.184 & 4953 \\
Tobin's $Q$ & -0.612 & 0.582 & 0.103 & 5.047 & 1.263 & 1.908 & 3.286 & 4712 \\
Total assets & 10.632 & 16.308 & 16.478 & 19.347 & 1.545 & -0.500 & -0.185 & 4860 \\
Leverage & 0.000 & 0.344 & 0.337 & 2.013 & 0.177 & 0.903 & 5.015 & 4858 \\
Total employees & 99.419 & 24269.252 & 7813.385 & 430494.690 & 44319.281 & 4.504 & 28.430 & 4956 \\
\bottomrule
\end{tabular}
}
\end{table}

\begin{table}[hbtp!]
\caption{Descriptive Statistics - (Growth \& Innovation)}
\label{tab:descriptive_stats_Growth & Innovation}
  \resizebox{\textwidth}{!}{%
\begin{tabular}{lcccccccc}
\toprule
 Variable& Min & Mean & Median & Max & Std Dev & Skewness & Kurtosis & $N$ \\
\midrule
(\%) Women & 0.000 & 25.851 & 25.806 & 100 & 10.846 & 0.173 & 2.076 & 4838 \\
(\%) Unknown gender & 0.000 & 0.028 & 0.023 & 0.270 & 0.026 & 1.973 & 8.095 & 4838 \\
Tobin's $Q$ & -0.612 & 1.225 & 0.657 & 5.047 & 1.636 & 1.129 & 0.209 & 4204 \\
Total assets & 10.392 & 15.646 & 15.721 & 20.174 & 1.845 & -0.102 & -0.102 & 4419 \\
Leverage & 0.000 & 0.261 & 0.244 & 1.552 & 0.205 & 1.029 & 2.182 & 4417 \\
Total employees & 88.170 & 38170.464 & 9814.290 & 923390.810 & 85916.556 & 4.623 & 26.857 & 4844 \\
\bottomrule
\end{tabular}
}
\end{table}

\begin{table}[hbtp!]
\caption{Descriptive Statistics - (Financials)}
\label{tab:descriptive_stats_Financials}
  \resizebox{\textwidth}{!}{%
\begin{tabular}{lcccccccc}
\toprule
 Variable& Min & Mean & Median & Max & Std Dev & Skewness & Kurtosis & $N$ \\
\midrule
(\%) Women & 0.000 & 28.311 & 30.033 & 58.793 & 8.580 & -0.891 & 1.354 & 3521 \\
(\%) Unknown gender & 0.000 & 0.040 & 0.040 & 0.191 & 0.025 & 0.499 & 1.260 & 3521 \\
Tobin's $Q$ & -0.612 & -0.152 & -0.531 & 5.047 & 1.031 & 3.598 & 13.430 & 3234 \\
Total assets & 11.116 & 17.715 & 17.899 & 22.098 & 2.121 & -0.407 & -0.110 & 3356 \\
Leverage & 0.000 & 0.156 & 0.091 & 0.972 & 0.180 & 2.113 & 5.000 & 3349 \\
Total employees & 85.559 & 27831.534 & 8832.051 & 292316.720 & 49311.604 & 3.113 & 9.830 & 3552 \\
\bottomrule
\end{tabular}
}
\end{table}

\begin{table}[hbtp!]
\caption{Descriptive Statistics - (Energy)}
\label{tab:descriptive_stats_Energy}
  \resizebox{\textwidth}{!}{%
\begin{tabular}{lcccccccc}
\toprule
 Variable& Min & Mean & Median & Max & Std Dev & Skewness & Kurtosis & $N$ \\
\midrule
(\%) Women & 0.000 & 18.430 & 18.975 & 51.766 & 10.192 & 0.218 & 0.629 & 1101 \\
(\%) Unknown gender & 0.000 & 0.008 & 0.000 & 0.054 & 0.012 & 1.355 & 0.936 & 1101 \\
Tobin's $Q$ & -0.612 & -0.268 & -0.327 & 0.953 & 0.253 & 1.448 & 2.638 & 1051 \\
Total assets & 11.440 & 16.642 & 16.702 & 19.868 & 1.630 & -0.433 & 0.930 & 1088 \\
Leverage & 0.000 & 0.303 & 0.266 & 0.932 & 0.174 & 1.140 & 1.758 & 1088 \\
Total employees & 238.381 & 20174.814 & 3405.016 & 141472.060 & 32817.056 & 1.869 & 2.576 & 1120 \\
\bottomrule
\end{tabular}
}
\end{table}



\begin{figure}[hbtp!]
  \centering

  \vspace{1ex}
  \begin{subfigure}[t]{0.48\textwidth}
    \includegraphics[width=\textwidth]{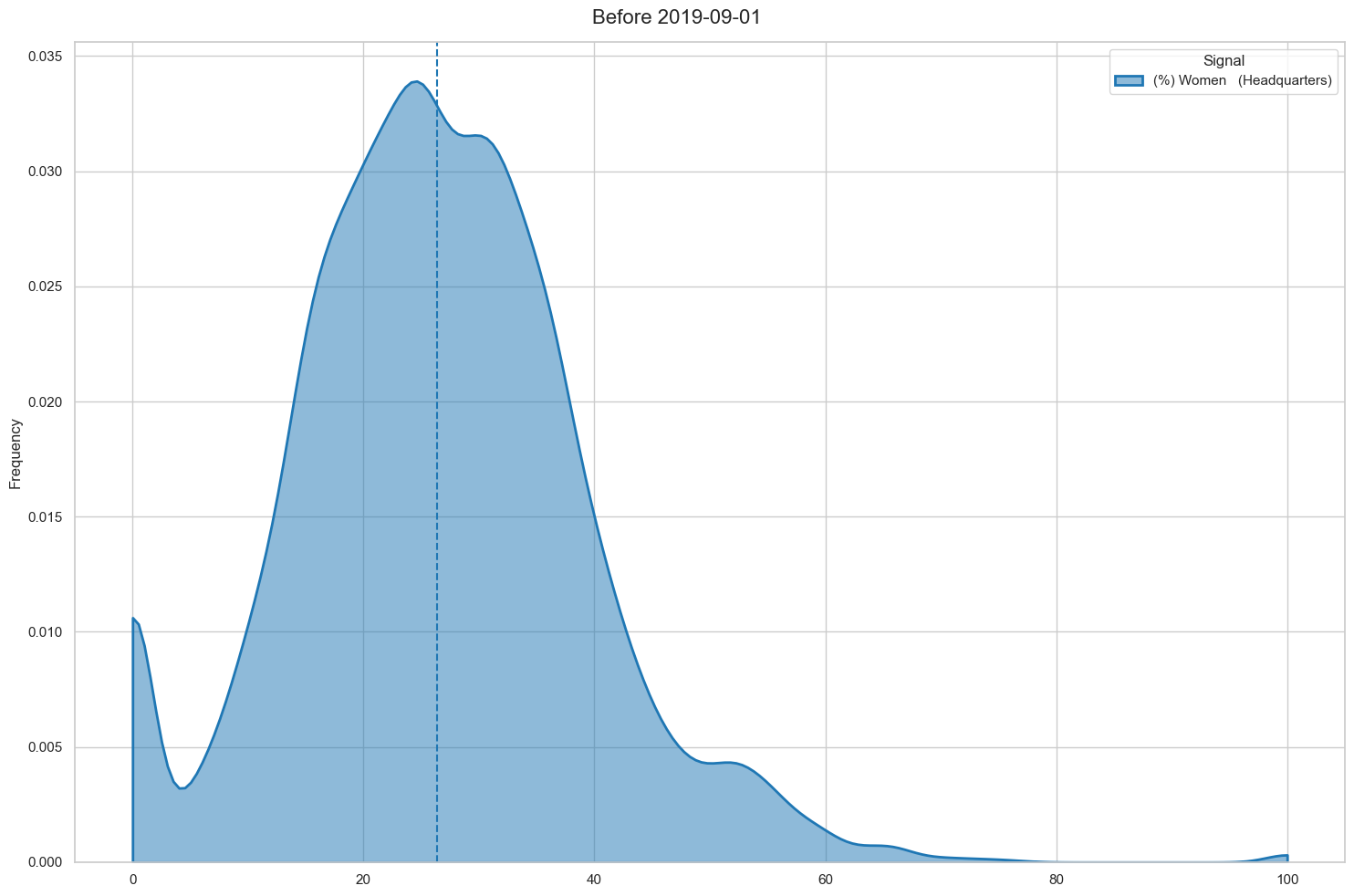}
    \caption{All firms (before 01-09-2019)}
  \end{subfigure}
  \hfill
  \begin{subfigure}[t]{0.48\textwidth}
    \includegraphics[width=\textwidth]{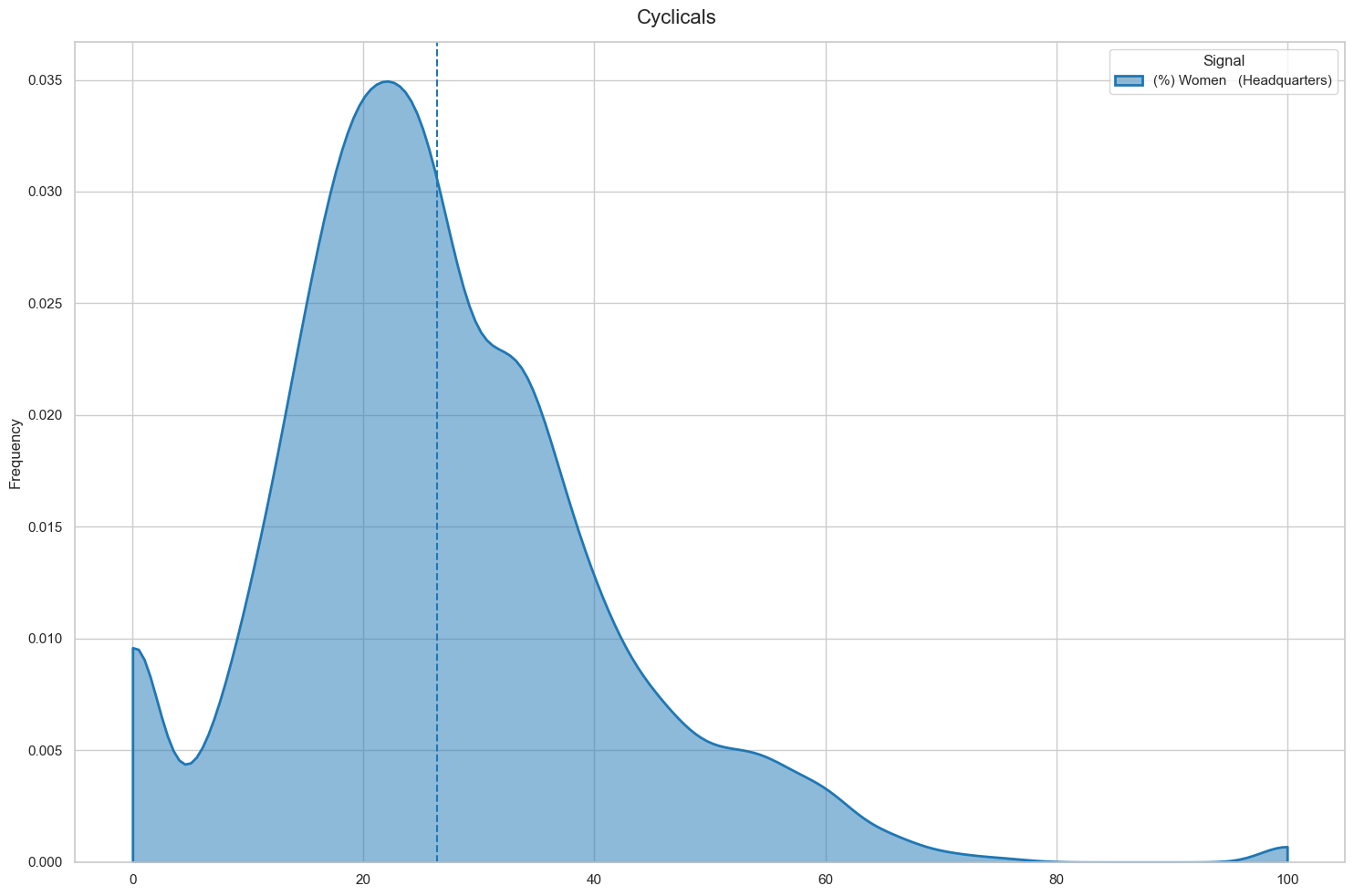}
    \caption{Cyclical sector}
  \end{subfigure}
\hfill
  \begin{subfigure}[t]{0.48\textwidth}
    \includegraphics[width=\textwidth]{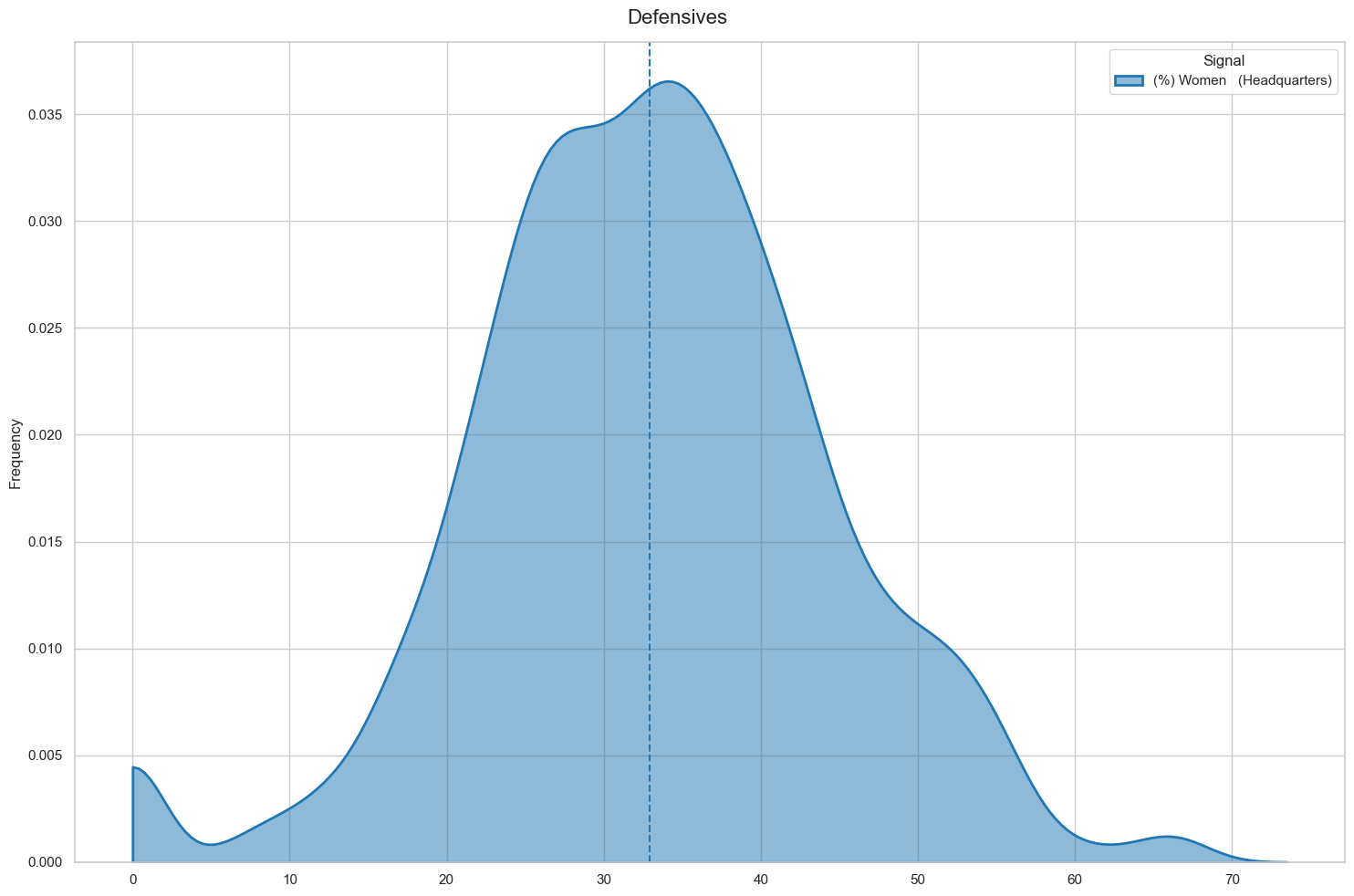}
    \caption{Defensive sector}
  \end{subfigure}
  \hfill
  \begin{subfigure}[t]{0.48\textwidth}
    \includegraphics[width=\textwidth]{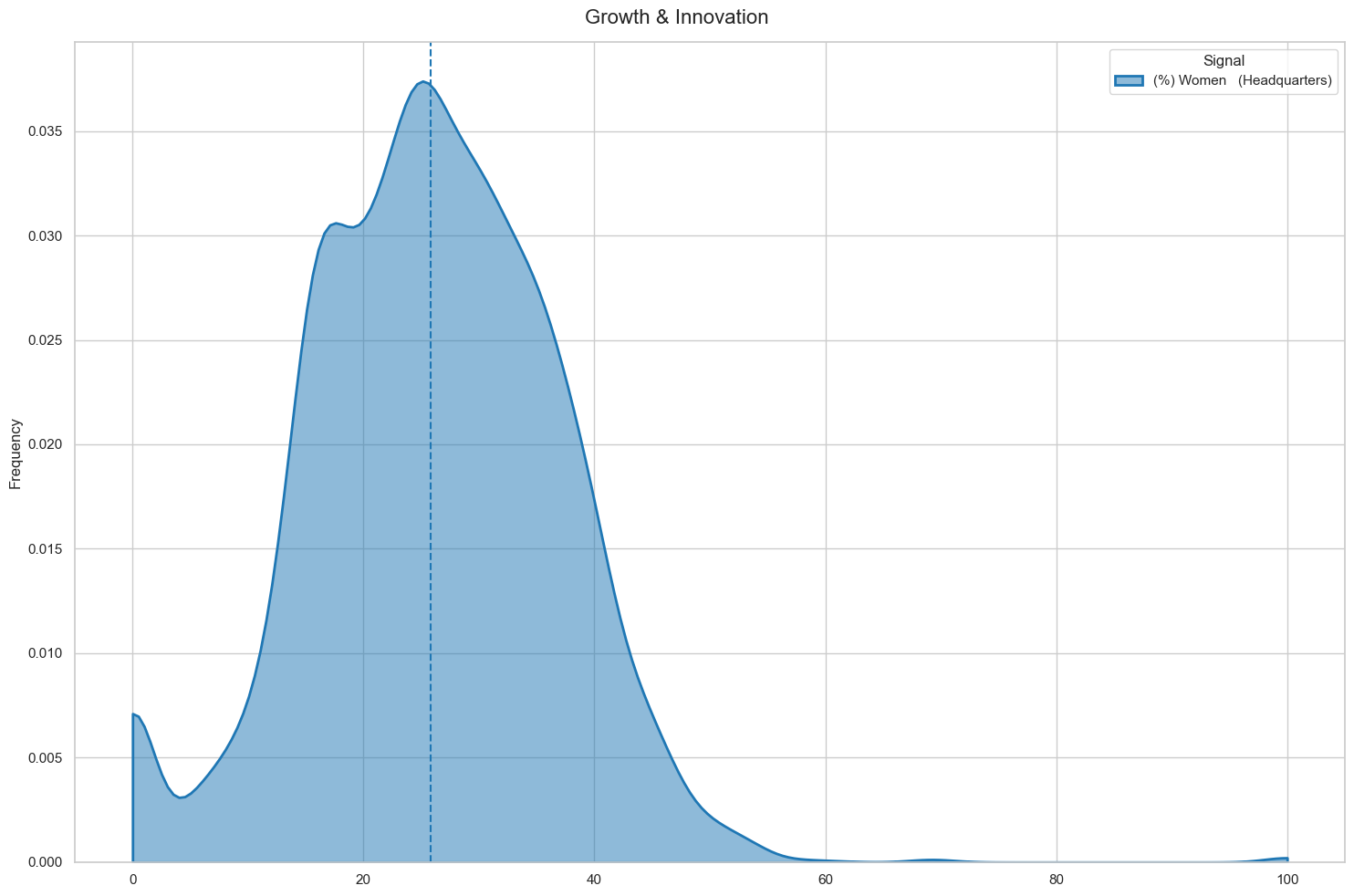}
    \caption{Growth \& Innovation sector}
  \end{subfigure}
\hfill
  \begin{subfigure}[t]{0.48\textwidth}
    \includegraphics[width=\textwidth]{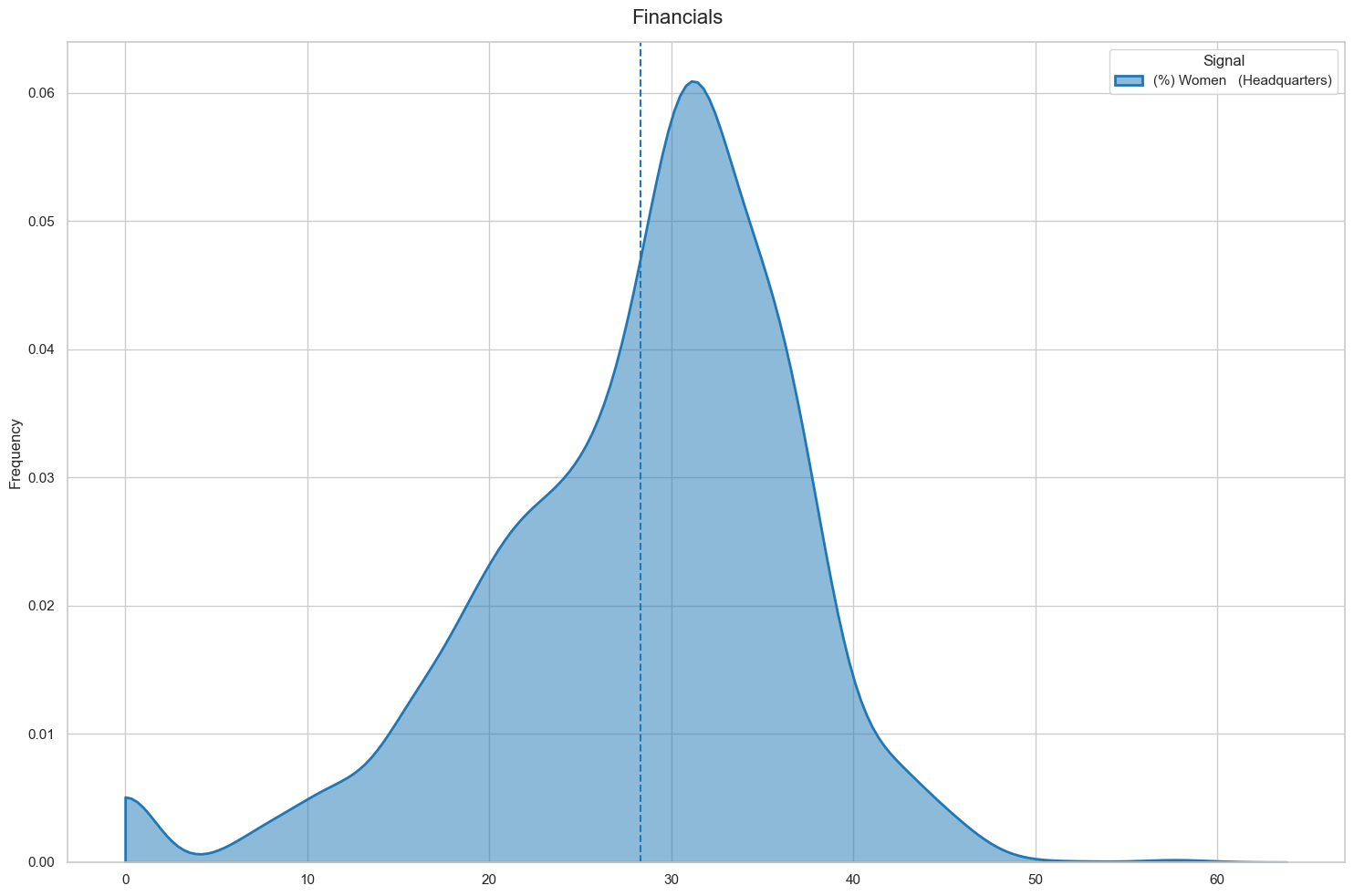}
    \caption{Financials sector}
  \end{subfigure}
  \hfill
  \begin{subfigure}[t]{0.48\textwidth}
    \includegraphics[width=\textwidth]{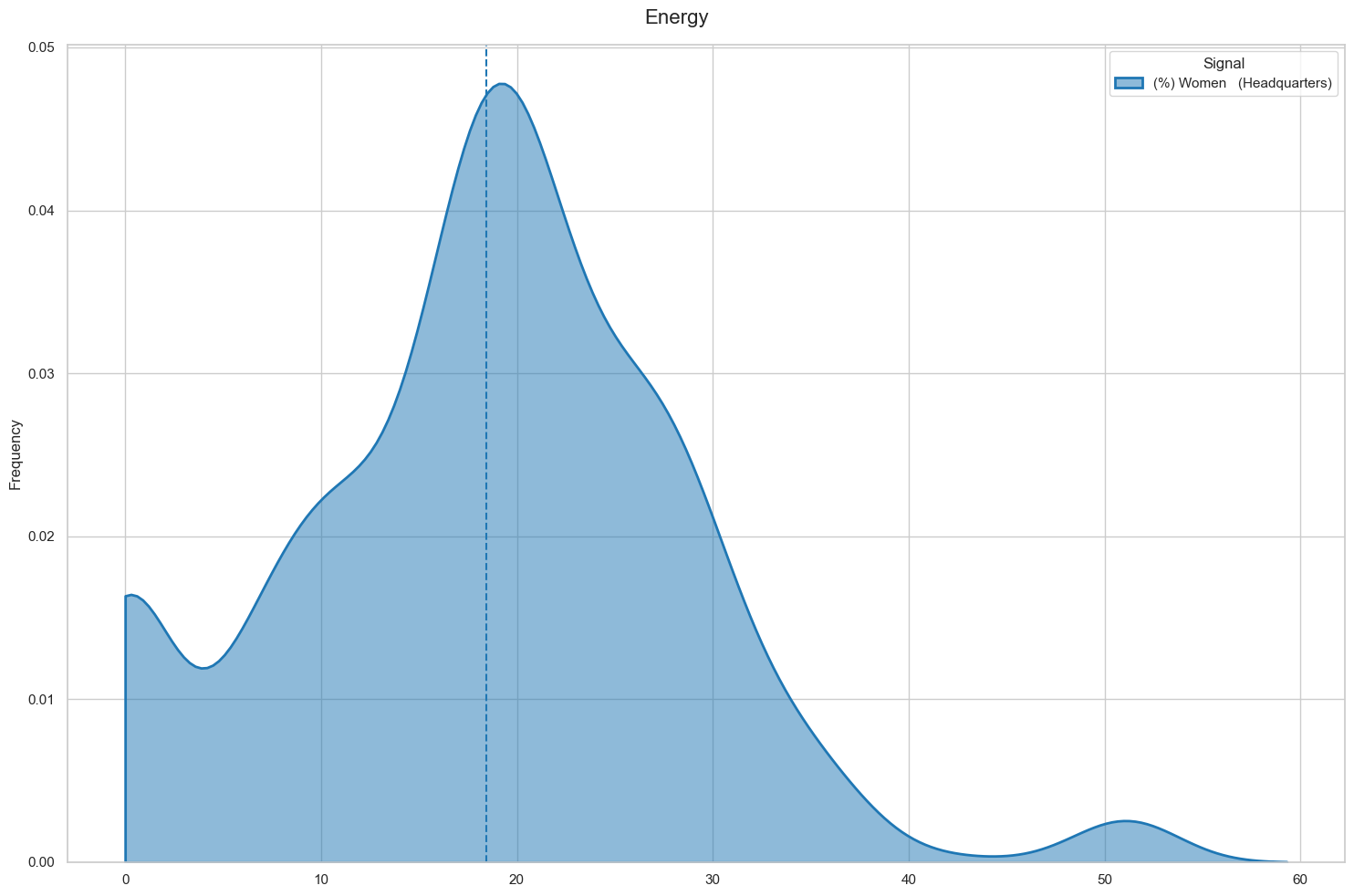}
    \caption{Energy sector}
  \end{subfigure}

  \caption{Kernel density plots of percentage women.}
  \label{fig:signalprops_grid_2}
\end{figure}
\FloatBarrier

\subsection{Correlation Analysis}

\begin{figure}[hbtp!]
\centering
\includegraphics[width=\textwidth]{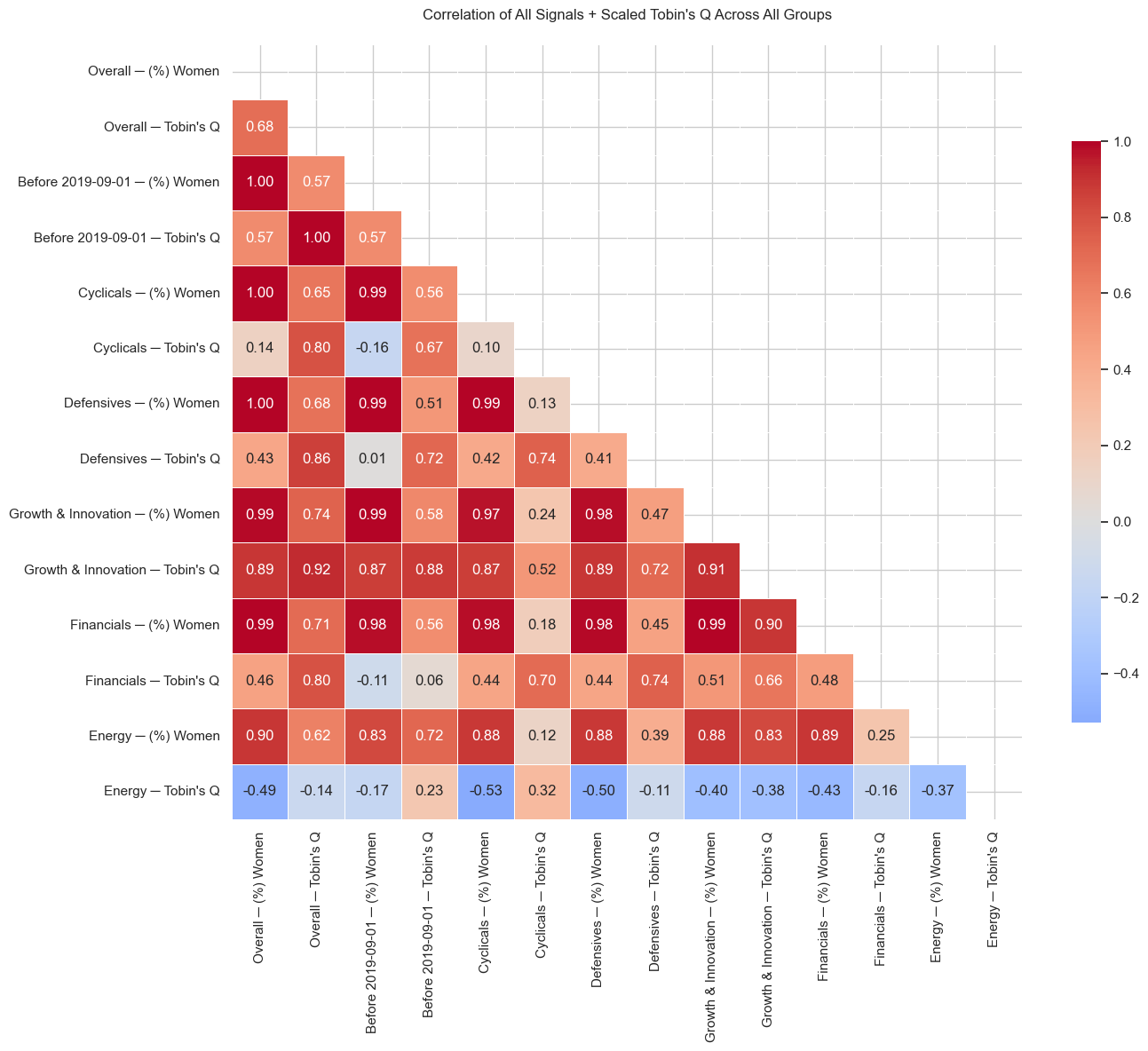}
\caption{Pearson correlation heatmap illustrating the relationships between Tobin's $Q$ and gender across all sectoral groups.}
\label{fig:corr_all}
\end{figure}

\begin{figure}[hbtp!]
 \centering

  \vspace{1ex}
  \begin{subfigure}[t]{0.48\textwidth}
    \includegraphics[width=\textwidth]{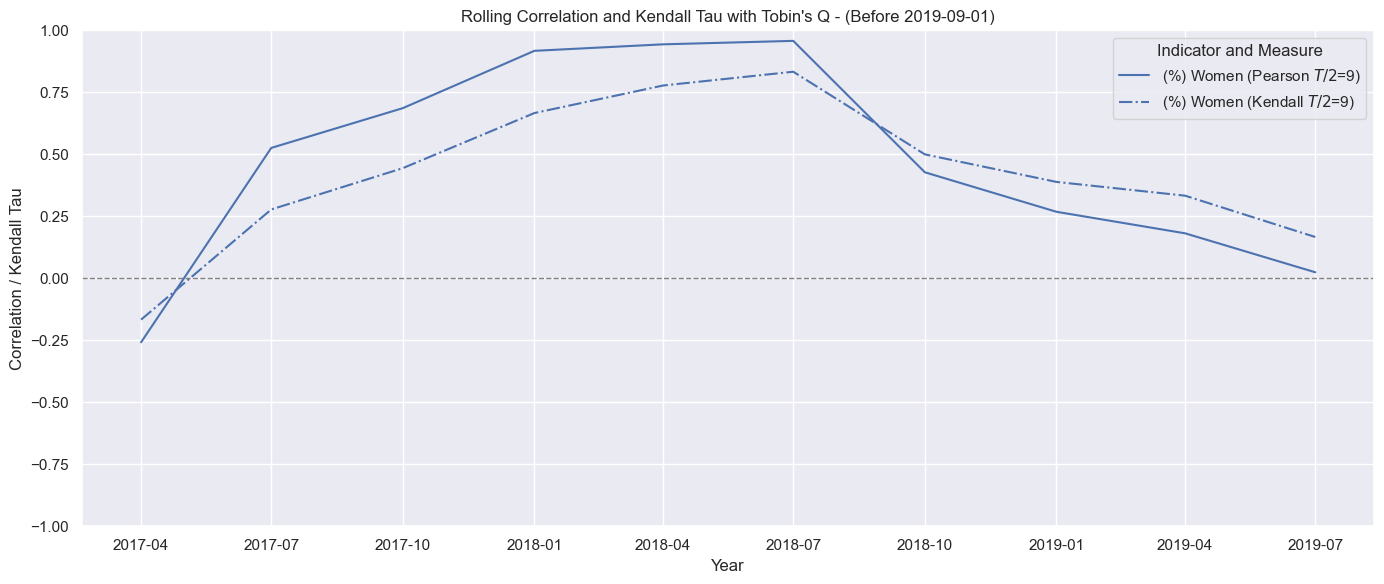}
    \caption{All firms (before 01-09-2019)}
  \end{subfigure}
  \hfill
  \begin{subfigure}[t]{0.48\textwidth}
    \includegraphics[width=\textwidth]{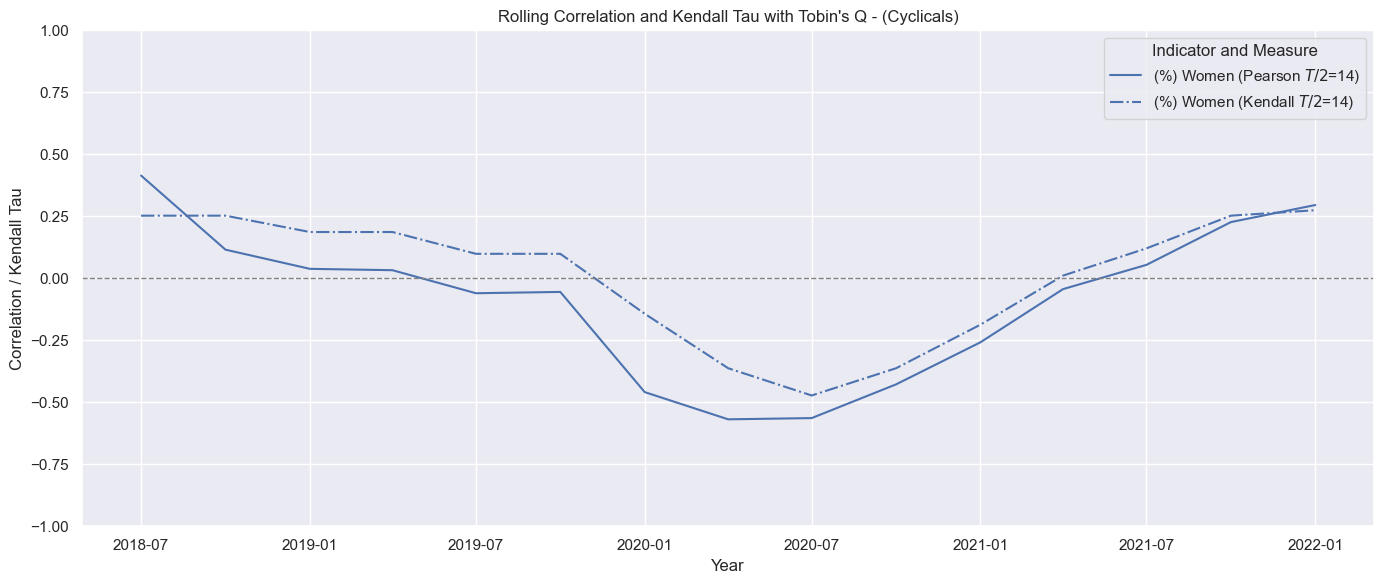}
    \caption{Cyclical sector}
  \end{subfigure}
\hfill
  \centering
  \begin{subfigure}[t]{0.48\textwidth}
    \includegraphics[width=\textwidth]{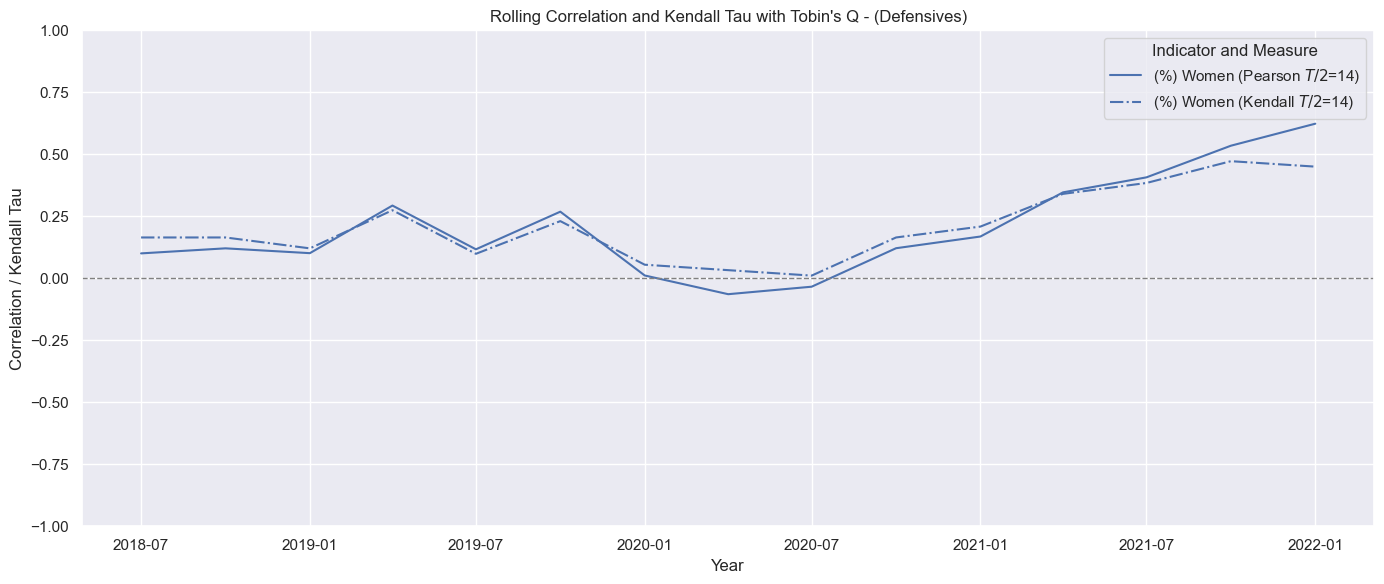}
    \caption{Defensive sector}
  \end{subfigure}
  \hfill
  \begin{subfigure}[t]{0.48\textwidth}
    \includegraphics[width=\textwidth]{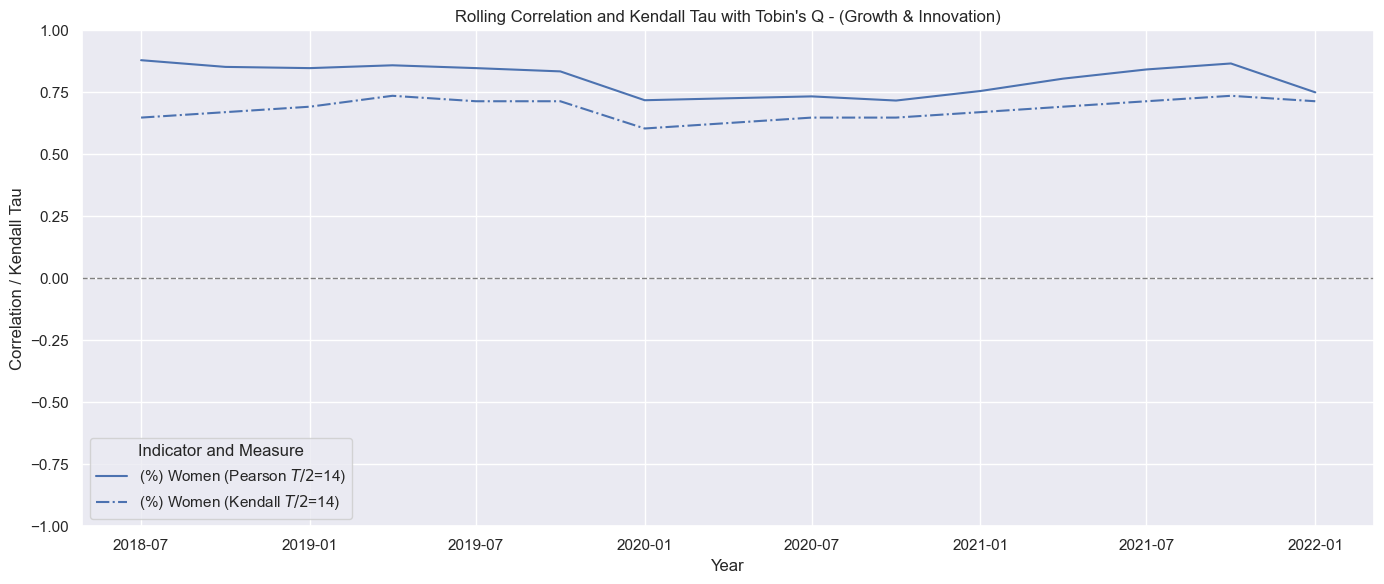}
    \caption{Growth \& Innovation sector}
  \end{subfigure}

  \vspace{1ex}
  \begin{subfigure}[t]{0.48\textwidth}
    \includegraphics[width=\textwidth]{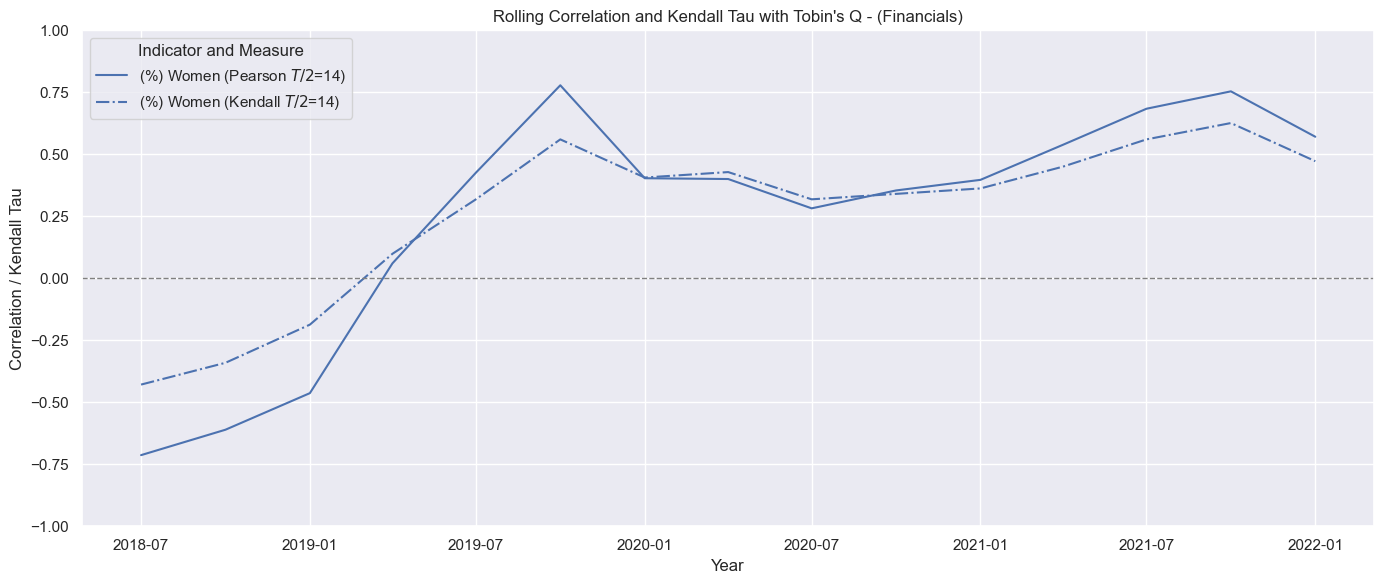}
    \caption{Financials sector}
  \end{subfigure}
  \hfill
  \begin{subfigure}[t]{0.48\textwidth}
    \includegraphics[width=\textwidth]{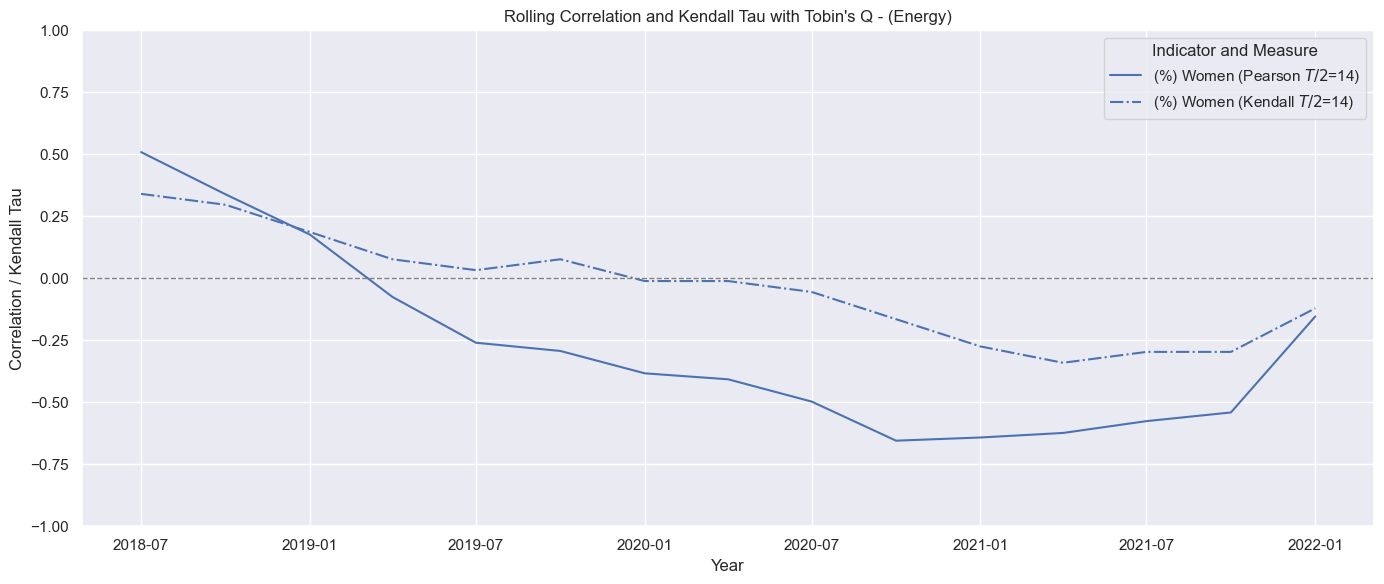}
    \caption{Energy sector}
  \end{subfigure}

  \caption{Rolling correlations plots of gender with Tobin's $Q$.}
  \label{fig:rollingcorrs_grid}
\end{figure}

\FloatBarrier

\subsection{Data Clustering by Threshold}

\begin{table}[hbtp!]
\centering
\caption{Counts by Threshold and Group -- (\% Women)}
\footnotesize\textit{Note}: Grey-shaded cells indicate clusters with $N_k < 10$ for $k=0,1$. \\
\vspace{0.5em}
\label{tab:counts_by_tau_women}
\resizebox{\linewidth}{!}{%
\begin{tabular}{ccccccccccccccc}
\toprule
(\%) Women & \multicolumn{2}{c}{Overall} & \multicolumn{2}{c}{Pre 01/04/2019} & \multicolumn{2}{c}{Cyclicals} & \multicolumn{2}{c}{Defensives} & \multicolumn{2}{c}{Growth \& Innovation} & \multicolumn{2}{c}{Financials} & \multicolumn{2}{c}{Energy} \\
 & $N_0$ & $N_1$ & $N_0$ & $N_1$ & $N_0$ & $N_1$ & $N_0$ & $N_1$ & $N_0$ & $N_1$ & $N_0$ & $N_1$ & $N_0$ & $N_1$ \\
\midrule
$\tau_{5}$ & 1298 & 23930 & 1004 & 15214 & 734 & 9822 & 119 & 4837 & 199 & 4645 & 110 & 3446 & 136 & 984 \\
\cline{1-15}
$\tau_{10}$ & 1965 & 23263 & 1484 & 14734 & 1079 & 9477 & 158 & 4798 & 317 & 4527 & 155 & 3401 & 238 & 882 \\
\cline{1-15}
$\tau_{15}$ & 3607 & 21621 & 2759 & 13459 & 2030 & 8526 & 246 & 4710 & 652 & 4192 & 273 & 3283 & 368 & 752 \\
\cline{1-15}
$\tau_{20}$ & 6886 & 18342 & 5004 & 11214 & 3674 & 6882 & 505 & 4451 & 1432 & 3412 & 578 & 2978 & 635 & 485 \\
\cline{1-15}
$\tau_{25}$ & 10902 & 14326 & 7685 & 8533 & 5538 & 5018 & 1102 & 3854 & 2272 & 2572 & 1068 & 2488 & 853 & 267 \\
\cline{1-15}
$\tau_{30}$ & 14969 & 10259 & 10219 & 5999 & 6976 & 3580 & 1965 & 2991 & 3130 & 1714 & 1784 & 1772 & 1012 & 108 \\
\cline{1-15}
 $\tau_{35}$ & 18937 & 6291 & 12634 & 3584 & 8167 & 2389 & 2850 & 2106 & 3872 & 972 & 2830 & 726 & 1074 & 46 \\
\cline{1-15}
 $\tau_{40}$ &  21853 & 3375 & 14332 & 1886 & 9027 & 1529 & 3708 & 1248 & 4441 & 403 & 3397 & 159 & 1099 & 21 \\
\cline{1-15}
$\tau_{45}$ & 23330 & 1898 & 15163 & 1055 & 9525 & 1031 & 4279 & 677 & 4703 & 141 & 3526 & 30 & 1101 & 19 \\
\cline{1-15}
 $\tau_{50}$ & 24102 & 1126 & 15549 & 669 & 9855 & 701 & 4592 & 364 & 4799 & 45 & {\cellcolor{lightgray}} 3552 & {\cellcolor{lightgray}} 4 & 1108 & 12 \\
\cline{1-15}
 $\tau_{55}$ & 24651 & 577 & 15886 & 332 & 10113 & 443 & 4836 & 120 & 4833 & 11 & {\cellcolor{lightgray}} 3553 & {\cellcolor{lightgray}} 3 & {\cellcolor{lightgray}} 1120 & {\cellcolor{lightgray}} 0 \\
\cline{1-15}
$\tau_{60}$ & 24952 & 276 & 16080 & 138 & 10335 & 221 & 4910 & 46 & {\cellcolor{lightgray}} 4835 & {\cellcolor{lightgray}} 9 & {\cellcolor{lightgray}} 3556 & {\cellcolor{lightgray}} 0 & {\cellcolor{lightgray}} 1120 & {\cellcolor{lightgray}} 0 \\
\cline{1-15}
$\tau_{65}$ & 25081 & 147 & 16135 & 83 & 10445 & 111 & 4928 & 28 & {\cellcolor{lightgray}} 4836 & {\cellcolor{lightgray}} 8 & {\cellcolor{lightgray}} 3556 & {\cellcolor{lightgray}} 0 & {\cellcolor{lightgray}} 1120 & {\cellcolor{lightgray}} 0 \\
\cline{1-15}
$\tau_{70}$ &  25158 & 70 & 16175 & 43 & 10491 & 65 & {\cellcolor{lightgray}} 4956 & {\cellcolor{lightgray}} 0 & {\cellcolor{lightgray}} 4839 & {\cellcolor{lightgray}} 5 & {\cellcolor{lightgray}} 3556 & {\cellcolor{lightgray}} 0 & {\cellcolor{lightgray}} 1120 & {\cellcolor{lightgray}} 0 \\
\cline{1-15}
$\tau_{75}$ &  25177 & 51 & 16191 & 27 & 10510 & 46 & {\cellcolor{lightgray}} 4956 & {\cellcolor{lightgray}} 0 & {\cellcolor{lightgray}} 4839 & {\cellcolor{lightgray}} 5 & {\cellcolor{lightgray}} 3556 & {\cellcolor{lightgray}} 0 & {\cellcolor{lightgray}} 1120 & {\cellcolor{lightgray}} 0 \\
\cline{1-15}
$\tau_{80}$ &  25182 & 46 & 16195 & 23 & 10515 & 41 & {\cellcolor{lightgray}} 4956 & {\cellcolor{lightgray}} 0 & {\cellcolor{lightgray}} 4839 & {\cellcolor{lightgray}} 5 & {\cellcolor{lightgray}} 3556 & {\cellcolor{lightgray}} 0 & {\cellcolor{lightgray}} 1120 & {\cellcolor{lightgray}} 0 \\
\cline{1-15}
$\tau_{85}$ & 25182 & 46 & 16195 & 23 & 10515 & 41 & {\cellcolor{lightgray}} 4956 & {\cellcolor{lightgray}} 0 & {\cellcolor{lightgray}} 4839 & {\cellcolor{lightgray}} 5 & {\cellcolor{lightgray}} 3556 & {\cellcolor{lightgray}} 0 & {\cellcolor{lightgray}} 1120 & {\cellcolor{lightgray}} 0 \\
\cline{1-15}
$\tau_{90}$ &  25182 & 46 & 16195 & 23 & 10515 & 41 & {\cellcolor{lightgray}} 4956 & {\cellcolor{lightgray}} 0 & {\cellcolor{lightgray}} 4839 & {\cellcolor{lightgray}} 5 & {\cellcolor{lightgray}} 3556 & {\cellcolor{lightgray}} 0 & {\cellcolor{lightgray}} 1120 & {\cellcolor{lightgray}} 0 \\
\cline{1-15}
$\tau_{95}$ &  25182 & 46 & 16195 & 23 & 10515 & 41 & {\cellcolor{lightgray}} 4956 & {\cellcolor{lightgray}} 0 & {\cellcolor{lightgray}} 4839 & {\cellcolor{lightgray}} 5 & {\cellcolor{lightgray}} 3556 & {\cellcolor{lightgray}} 0 & {\cellcolor{lightgray}} 1120 & {\cellcolor{lightgray}} 0 \\
\cline{1-15}
\bottomrule
\end{tabular}
} 
\end{table}

\begin{table}[H]
\centering
\begin{tabular}{llrr}
\toprule
Cluster & Diversity Signal & Min & Max \\
\midrule
Overall & (\%) women & 5 & 95 \\
Pre 01/09/2019 & (\%) women & 5 & 95 \\
Cyclicals & (\%) women & 5 & 95 \\
Defensives & (\%) women & 5 & 65 \\
Growth \& Innovation & (\%) women & 5 & 55 \\
\rowcolor{lightgray} Financials & (\%) women & 5 & 45 \\
\rowcolor{lightgray}Energy & (\%) women & 5 & 50 \\
\bottomrule
\end{tabular}
\caption{Summary of Minimum and Maximum Permissible Values of $\tau_m$ Across Clusters.}
\label{tab:summary_table}
\vspace{0.3em}
{\raggedright\footnotesize\textit{Note:} The greyed rows are eliminated from the analysis due to small maximum $\tau_m$ values, which render them unsuitable for further analysis.\par}
\end{table}

\subsection{Partial and Point Identification of the Treatment Effect}

\begin{figure}[H]
  \centering
  \par\medskip
  \begin{subfigure}[b]{0.30\textwidth}
    \includegraphics[width=\linewidth]{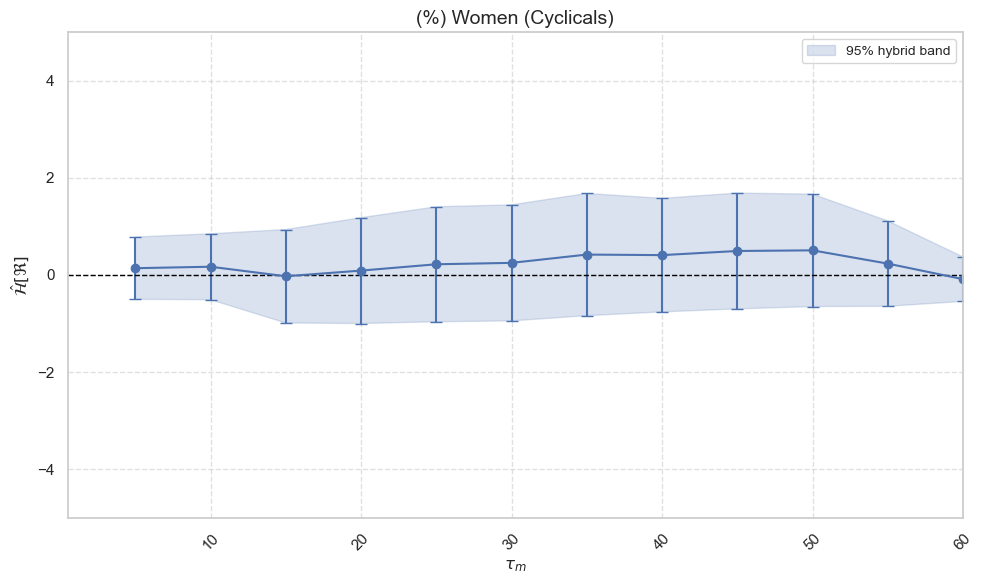}
    \caption{Hybrid: Cyclicals (\%) women}
  \end{subfigure}
  \hfill
  \begin{subfigure}[b]{0.30\textwidth}
    \includegraphics[width=\linewidth]{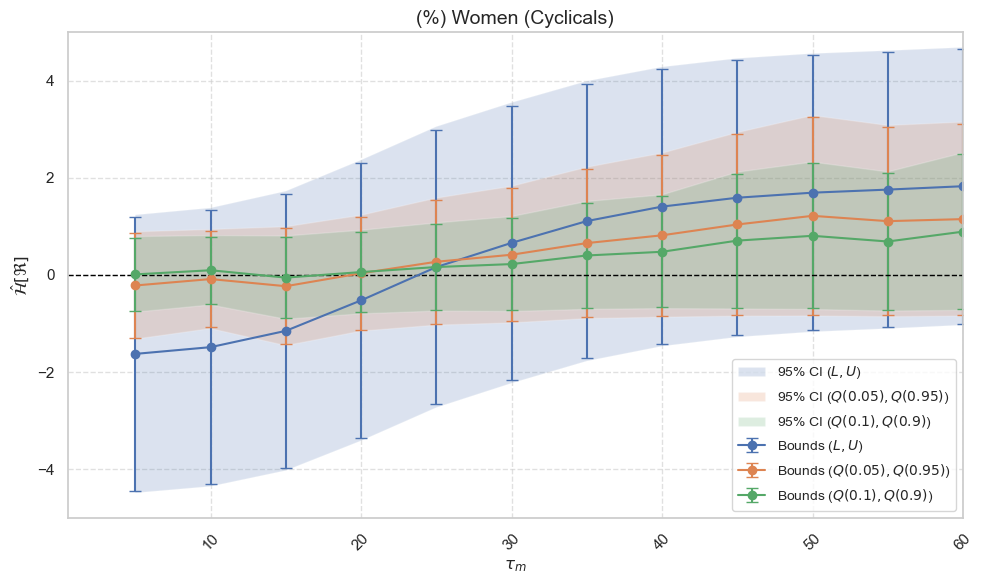}
    \caption{Nonparametric: Cyclicals (\%) women}
  \end{subfigure}
  \hfill
  \begin{subfigure}[b]{0.30\textwidth}
    \includegraphics[width=\linewidth]{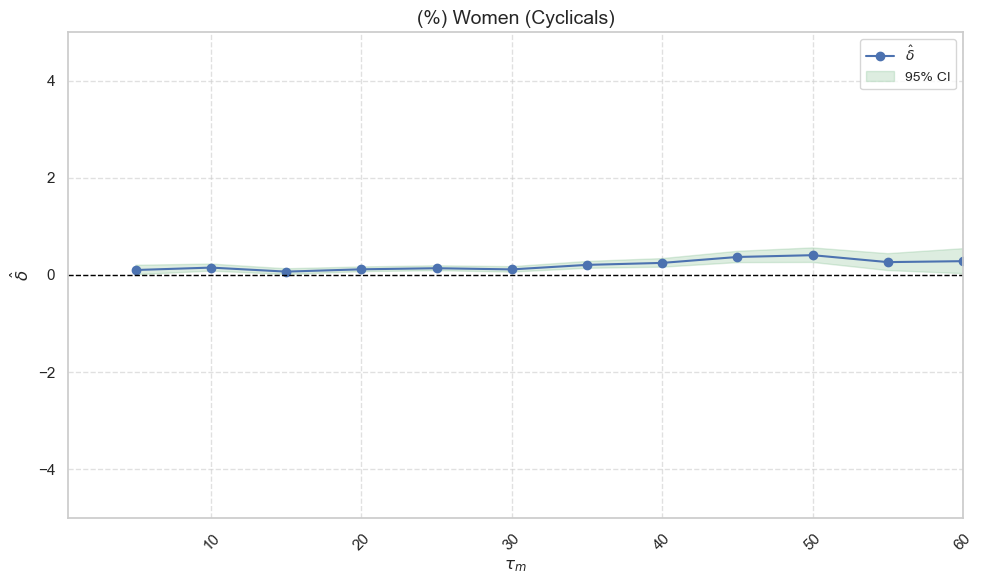}
    \caption{Angrist: Cyclicals (\%) women}
  \end{subfigure}
\par\medskip
  \centering
  \begin{subfigure}[b]{0.30\textwidth}
    \includegraphics[width=\linewidth]{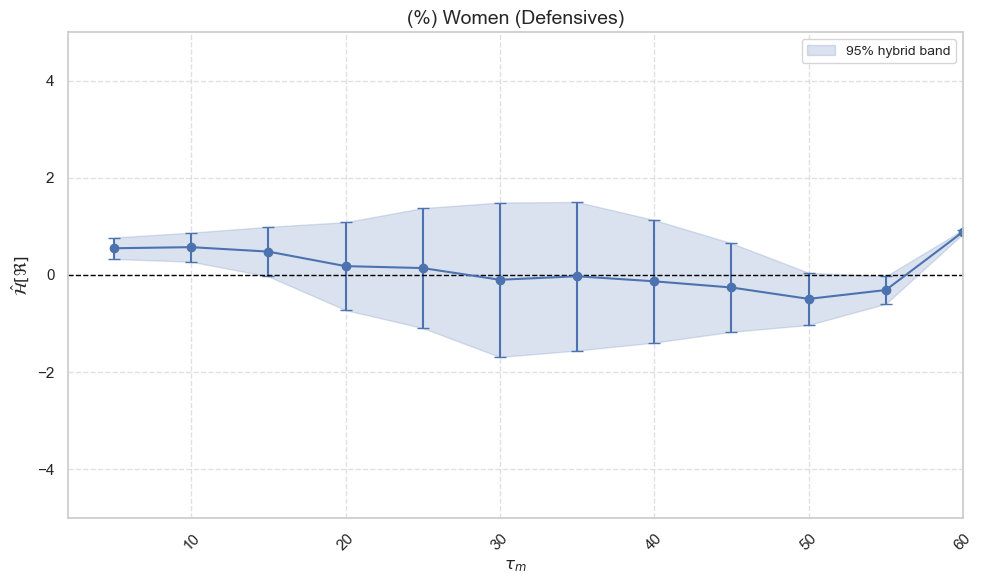}
    \caption{Hybrid: Defensives (\%) women}
  \end{subfigure}
  \hfill
  \begin{subfigure}[b]{0.30\textwidth}
    \includegraphics[width=\linewidth]{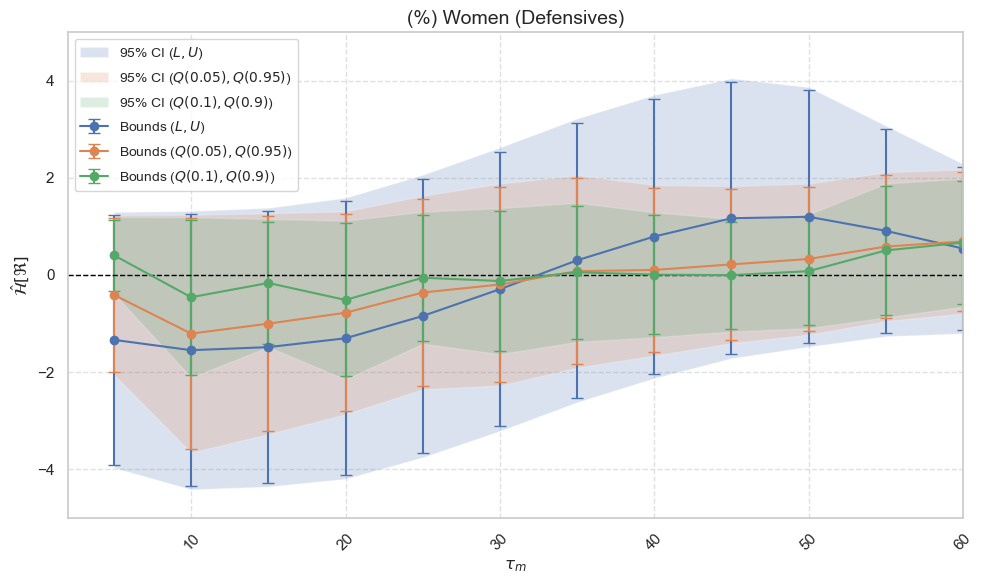}
    \caption{Nonparametric: Defensives (\%) women}
  \end{subfigure}
  \hfill
  \begin{subfigure}[b]{0.30\textwidth}
    \includegraphics[width=\linewidth]{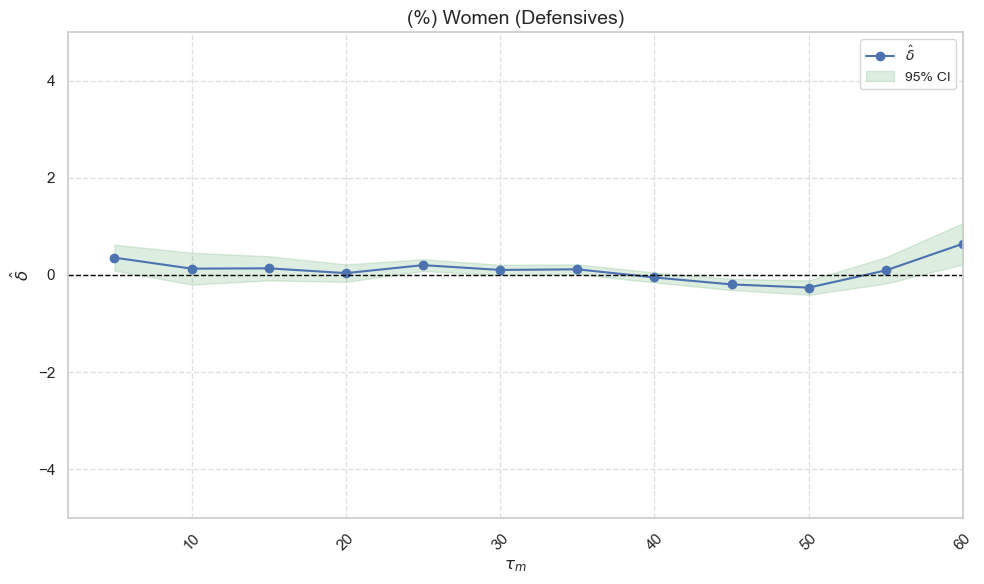}
    \caption{Angrist: Defensives (\%) women}
  \end{subfigure}
  \par\medskip
  \begin{subfigure}[b]{0.30\textwidth}
    \includegraphics[width=\linewidth]{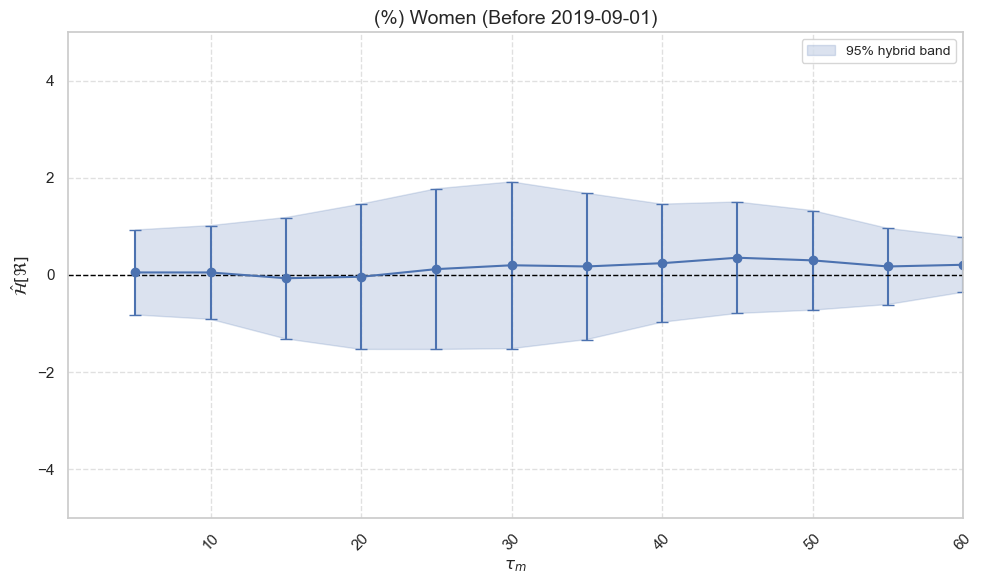}
    \caption{Hybrid: Pre 01/04/2019 (\%) women}
  \end{subfigure}
  \hfill
  \begin{subfigure}[b]{0.30\textwidth}
    \includegraphics[width=\linewidth]{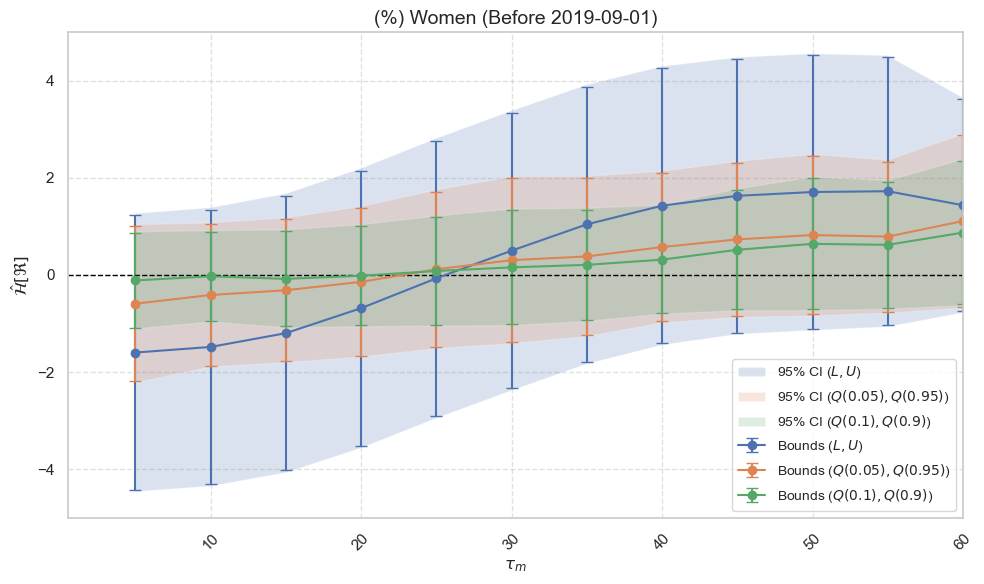}
    \caption{Nonparametric: Pre 01/04/2019 (\%) women}
  \end{subfigure}
  \hfill
  \begin{subfigure}[b]{0.30\textwidth}
    \includegraphics[width=\linewidth]{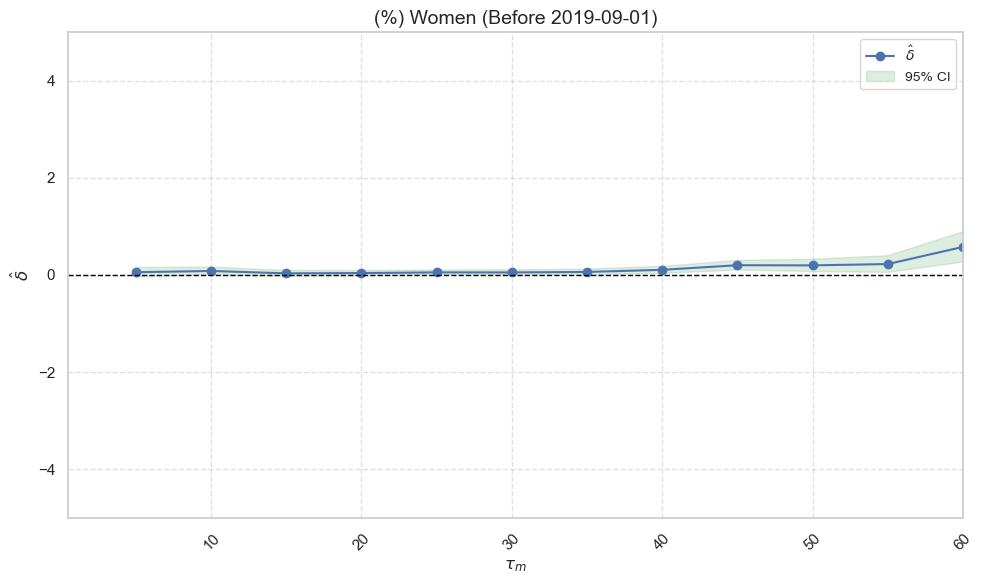}
    \caption{Angrist: Pre 01/04/2019 (\%) women}
  \end{subfigure}
  \caption{Hybrid, Manski nonparametric and Angrist estimates}
\end{figure}

\begin{figure}[H]
  \centering
  \begin{subfigure}[b]{0.30\textwidth}
    \includegraphics[width=\linewidth]{Nonparametric_Bounds_hybrid_Growth_and_Innovation_p_women_senior_top_hq.png}
    \caption{Hybrid: Growth \& innovation (\%) women}
  \end{subfigure}
  \hfill
  \begin{subfigure}[b]{0.30\textwidth}
    \includegraphics[width=\linewidth]{Nonparametric_Bounds_Growth_and_Innovation_p_women_senior_top_hq.png}
    \caption{Nonparametric: Growth \& innovation (\%) women}
  \end{subfigure}
  \hfill
  \begin{subfigure}[b]{0.30\textwidth}
    \includegraphics[width=\linewidth]{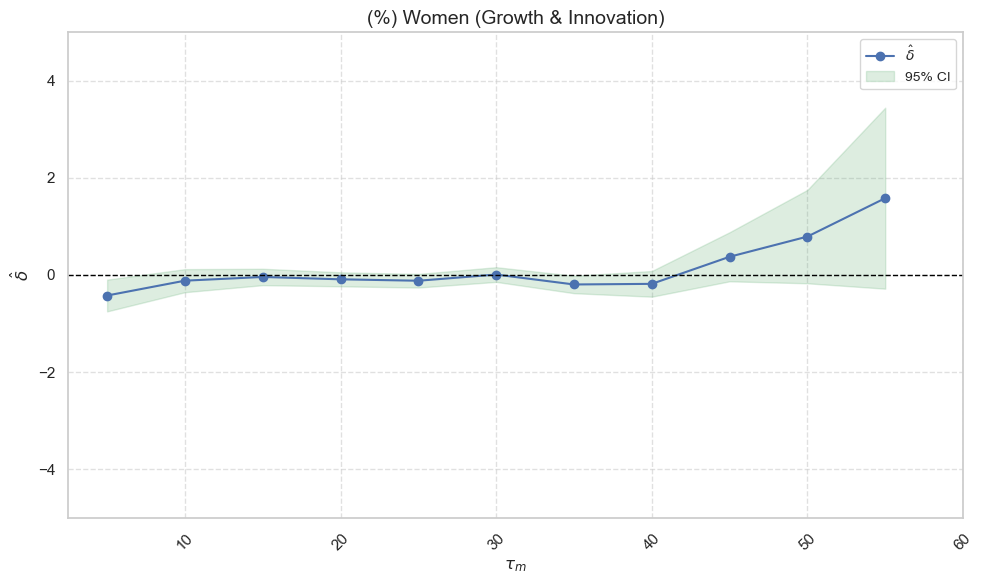}
    \caption{Angrist: Growth \& innovation (\%) women}
  \end{subfigure}
  \par\medskip
  \begin{subfigure}[b]{0.30\textwidth}
    \includegraphics[width=\linewidth]{Nonparametric_Bounds_hybrid_simple_signal_p_women_senior_top_hq.png}
    \caption{Hybrid: Overall (\%) women}
  \end{subfigure}
  \hfill
  \begin{subfigure}[b]{0.30\textwidth}
    \includegraphics[width=\linewidth]{Nonparametric_Bounds_simple_signal_p_women_senior_top_hq.png}
    \caption{Nonparametric: Overall (\%) women}
  \end{subfigure}
  \hfill
  \begin{subfigure}[b]{0.30\textwidth}
    \includegraphics[width=\linewidth]{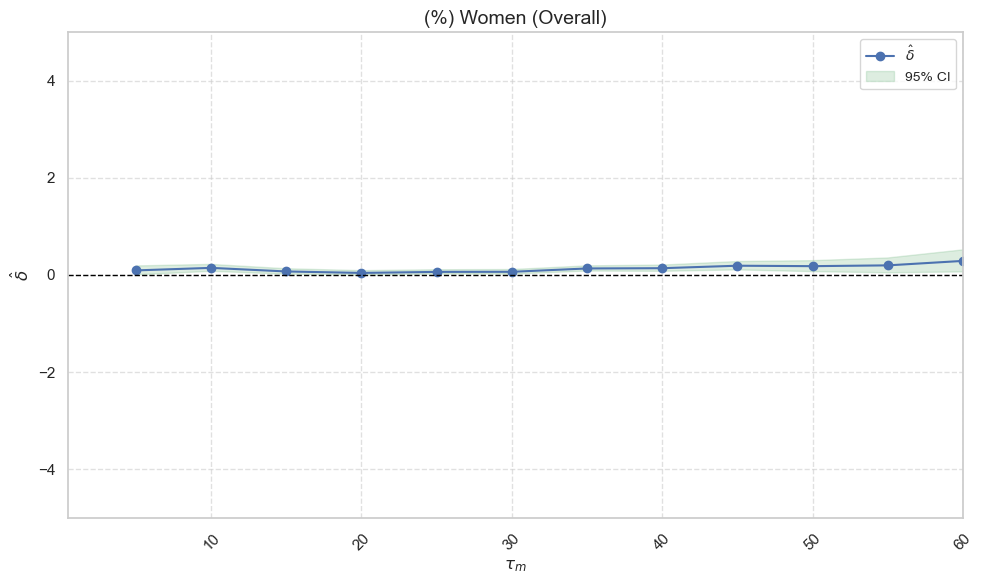}
    \caption{Angrist: Overall (\%) women}
  \end{subfigure}
  \caption{Hybrid, Manski nonparametric and Angrist estimates}
\label{fig:overall_women}
\end{figure}

\end{appendices}
    \end{document}